\documentclass{elsart}
\usepackage{graphicx}
\usepackage{amssymb}
\usepackage{epsfig}
\def\Journal#1#2#3#4{{#1} {#2} (#3) #4}
\def\NPA{Nucl. Phys. A}

\def\PLB{Phys. Lett.  B}
\def\PRL{Phys. Rev. Lett.}
\def\PRC{Phys. Rev. C}
\def\PRD{Phys. Rev. D}

\def\ZPC{Z. Phys. C}

\def\EPJC{Eur. Phys. J. C}
\def\JPG{J. Phys. G}

\newcommand{\ud}{\mathrm{d}}
\newcommand{\be}{\begin{equation}}
\newcommand{\ee}{\end{equation}}

\newcommand{\AmS}{{\protect\the\textfont2
  A\kern-.1667em\lower.5ex\hbox{M}\kern-.125emS}}

\begin{document}

\begin{frontmatter}

\title{Hadron production in central nucleus-nucleus collisions at 
chemical freeze-out}

\author[gsi]{A.\,Andronic},
\author[gsi]{P.\,Braun-Munzinger},
\author[hei]{J.\,Stachel}

\address[gsi]{Gesellschaft f\"ur Schwerionenforschung,
D-64291 Darmstadt, Germany}
\address[hei]{Physikalisches Institut der Universit\"at Heidelberg,
D-69120 Heidelberg, Germany}


\begin{abstract}
We analyze the experimental hadron yield ratios for central nucleus-nucleus
collisions in terms of thermal model calculations over a broad energy range, 
$\sqrt{s_{NN}}$=2.7-200 GeV. 
The fits of the experimental data with the model calculations provide the 
thermal parameters, temperature and baryo-chemical potential at chemical
freeze-out.
We compare our results with the values obtained in other studies and 
also investigate more technical aspects such as a potential bias in the 
fits when fitting particle ratios or yields.
Using parametrizations of the temperature and baryonic chemical potential
as a function of energy, we compare the model calculations with data for 
a large variety of hadron yield ratios.
We provide quantitative predictions for experiments at LHC energy, 
as well as for the low RHIC energy of 62.4 GeV.
The relation of the determined parameters with the QCD phase boundary is
discussed.

\end{abstract}


\end{frontmatter}

\section{Introduction}

The success of the statistical (thermal) model \cite{rev} in describing 
the ratios of hadron yields produced in nucleus-nucleus collisions 
is remarkable.
The thermal model was initially used for the AGS and SPS data \cite{pbm1}
and was subsequently employed to describe data at SIS \cite{cle1,ra}, 
SPS \cite{pbm2} and more recently at RHIC \cite{pbm3,bro1,kan,star}.
An analysis of the energy dependence of the thermal parameters extracted 
from fits of the experimental data, temperature ($T$) and baryo-chemical 
potential ($\mu_b$), established the "line of chemical freeze-out"
\cite{qm97}.
These data were subsequently interpreted in terms of an universal condition 
for chemical freeze-out \cite{1gev}.
Remarkably, it appears that at the top SPS energy the ($T$,$\mu_b$) values
reach the phase boundary between the hadronic world and the quark-gluon 
plasma (QGP) \cite{rev,qm97}, as calculated solving Quantum Chromo-Dynamics 
(QCD) on the lattice \cite{lqcd,lqcd1,lqcd2}.
In this context, it has been argued that the QGP itself and the 
"deus ex machina" of phase space filling during hadronization
are playing the crucial roles in achieving thermalization in high energy 
nucleus-nucleus collisions \cite{stock}.
More recently it was demonstrated that passing through the phase transition
leads to multiparticle scattering of Goldstone bosons which drives even
(multi)strange baryons rapidly into equilibration \cite{pbmx},
providing a natural explanation for the observation that the
chemical freeze-out line reaches the phase boundary for small values
of $\mu_b$. The situation is less well understood for $\mu_b>$400 MeV
and needs further investigation.

Despite its simplicity, the thermal model allows to extract essential  
properties of the hot and dense fireball produced in high energy 
nucleus-nucleus collisions at a given stage in its evolution, namely
when the inelastic collisions cease (chemical freeze-out).
This makes the model a unique tool in the attempt to quantify from the 
experimental side the features of the phase diagram of hadronic matter 
\cite{phase,fk,op}.
Recent analyses over a broad energy range \cite{bec1,raf1} have contributed
in the on-going efforts to understand the chemical freeze-out criteria 
\cite{rev,qm97,1gev,pbm5,taw,cle2b} in the phase diagram.
It has been pointed out within a thermal model analysis that scanning 
the energy one encounters a transition from baryon- to meson-dominated
freeze-out \cite{cle2}, with its associated fingerprint on the characteristics
of hadron yields, also evidenced earlier \cite{pbm4}.

\vspace{-1.cm}
\begin{figure}[hbt]
\begin{tabular}{lr} \begin{minipage}{.66\textwidth}
\hspace{-.5cm}
\includegraphics[width=1.18\textwidth]{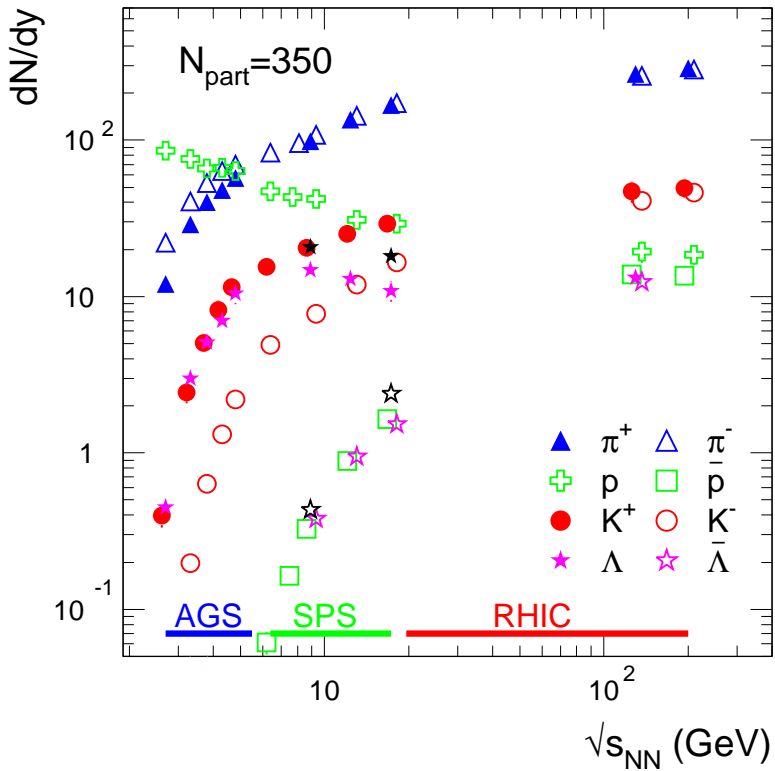}
\end{minipage} &\begin{minipage}{.3\textwidth}
\caption{The energy dependence of experimental hadron yields at
mid-rapidity for various species produced in central nucleus-nucleus
collisions.
The energy regimes for various accelerators are marked.
Note that, for SPS energies, there are two independent measurements available 
for the $\Lambda$ hyperon yields. For more details see text (also in
Section~\ref{fits}).} 
\end{minipage}
\end{tabular}
\label{fig0}
\end{figure}

It is important to emphasize that experimental data on hadron yields
are now available over a broad collision energy range and for a large sample 
of hadron species. 
A compilation of measurements of yields at mid-rapidity for
the most abundant hadron species is shown in Fig.~\ref{fig0} for 
central nucleus-nucleus (Au-Au or Pb-Pb) collisions. 
As the centrality selection differs between various measurements, we have
scaled the data at the same number of participating nucleons, 
$N_{part}$=350.
The contribution from feeding due to weak decays has been removed
whenever the case (see Section~\ref{fits}).
The main properties concerning the chemical composition of the fireball
(at mid-rapidity) can be derived from the yields plotted in Fig.~\ref{fig0}
without the need of any model.
At lower energies ($\sqrt{s_{NN}}\le$5 GeV), measured at Brookhaven's
Alternating Gradient Synchrotron (AGS), the fireball is dominated
by the incoming nucleons, while the yield of the produced pions, with a 
strong energy dependence, dominates at larger energies.
The importance of isospin at lower energies is reflected in the different
yields of $\pi^+$ and $\pi^-$.
The decreasing yield of protons points to an increasing transparency 
of the incoming nuclei as a function of energy. 
Beyond $\sqrt{s_{NN}}\simeq$100 GeV, the newly produced protons become 
dominant.
The yield of strange hadrons shows a sharp rise at AGS energies, 
with characteristic features for various species, determined by their
quark content. The yields of $K^+$ and $\Lambda$ (both with only the 
strange quark newly produced) are larger compared to $K^-$, which has 
two newly produced quarks.
The remarkable similarity of the yields of $K^+$ and $\Lambda$, despite
their large mass difference, is determined chiefly by their (anti)strange
quark content, leading to their associated production, and less by the 
abundant presence of light quarks from stopped incoming nucleons.
The yield of antiprotons and antihyperons (containing three newly produced 
quarks) is very similar and with a strong energy dependence (onset of 
production) at energies of CERN's Super Proton Synchrotron (SPS).
At the Relativistic Heavy Ion Collider (RHIC), due to a rather small 
net baryon content of the fireball, these differences almost disappear.
The data are further discussed in Section \ref{fits} in comparison
with thermal model calculations. 
We notice here that, in general, there is good consistency between 
overlapping data sets measured by different experiments. One exception
are the hyperon yields at SPS energies, illustrated for $\Lambda$ hyperons
in Fig.~\ref{fig0}.

In view of these considerations it is important to assess with precision 
potential uncertainties in the extraction of thermal parameters with a
consistent, 2nd generation, analysis of all available experimental data 
over a broad range of collision energies ($\sqrt{s_{NN}}$=2.7-200 GeV).
This includes a comparison of the results from fits of hadron yield ratios
and from absolute yields and a discussion on the influence of systematic 
errors on the fit parameters.
The next section contains a brief description of the model.
In Section \ref{fits} we perform fits of model calculations to the data.
We discuss the energy dependence of the resulting $T$ and $\mu_b$ 
parameters, along with a critical assessment of various sources of 
uncertainties, in Section \ref{edep}.
Section \ref{sdep} is devoted to the comparison of model and data
in terms of excitation functions of various hadron yield ratios, which 
we extend up to the LHC energy.
The relation of the extracted parameters with the predicted QCD phase 
boundary, calculated solving QCD on the lattice \cite{lqcd,lqcd1,lqcd2},
is discussed in Section \ref{spha}. The Appendix contains a discussion 
of more technical aspects in the fit procedures.

\section{Model description}

We restrict ourselves here to the basic features and essential
results of the statistical model approach. A complete survey of 
the assumptions and results, as well as of the relevant references, 
is available in ref.~\cite{rev}.

The basic quantity required to compute the thermal composition of
hadron yields  measured in heavy ion collisions is the
partition function $Z(T,V)$. In the grand canonical (GC) ensemble,
the partition function for species $i$ is ($\hbar=c=1$):
\be 
\ln Z_i ={{Vg_i}\over {2\pi^2}}\int_0^\infty \pm p^2\ud p \ln [1\pm 
\exp (-(E_i-\mu_i)/T)],
\label{eq:part}
\ee

from which the density is then calculated according to:
\be
n_i=N_i/V=-\frac{T}{V}\frac{\partial\ln Z_i}{\partial\mu}=\frac{g_i}{2 \pi^2} 
\int_0^\infty \frac{p^2 \ud p}{\exp[(E_i-\mu_i)/T] \pm 1},
\ee
where $g_i=(2J_i+1)$ is the spin degeneracy factor, $T$ is the temperature 
and $E_i =\sqrt {p^2+m_i^2}$ is the total energy. 
The (+) sign is for fermions and (--) is for bosons.
For hadron $i$ of baryon number $B_i$, third component
of the isospin $I_{3i}$, strangeness $S_i$, and charmness $C_i$,  
the chemical potential is 
$\mu_i = \mu_b B_i+\mu_{I_3} I_{3i}+\mu_S S_i+\mu_C C_i$.  
The chemical potentials related to baryon number ($\mu_b$), 
isospin ($\mu_{I_3}$), strangeness ($\mu_S$) and charm ($\mu_C$) ensure 
the conservation (on average) of the respective quantum numbers: 
i) baryon number: $V\sum_i n_i B_i = N_B$;  
ii) isospin:  $V \sum_i n_i I_{3i} = I_{3}^{tot}$;
iii) strangeness: $V \sum_i n_i S_i = 0$;
iv) charm: $V \sum_i n_i C_i = 0$.
The (net) baryon number $N_B$ and the total isospin $I_{3}^{tot}$ of the
system are input values which need to be specified according to the
colliding nuclei studied.
The degree of stopping of the colliding nuclei, which is energy dependent
and cannot be precisely determined experimentally, brings some uncertainty
in the choice of $N_B$ and $I_{3}^{tot}$. 
In our case, as we study central collisions of heavy nuclei (Au or Pb), 
but focus on data at mid-rapidity, we have chosen $N_B$=200 and 
$I_{3}^{tot}$=-20.
The sensitivity of the hadron ratios, which are the calculated 
"observables" to be compared with the experimental data, on $N_B$ and 
$I_{3}^{tot}$ is rather small.
Taking into account the conservation laws i)-iv), $T$ and the baryo-chemical
potential $\mu_b$ are the only parameters of the model, which will be
obtained from fits to experimental data. 
When fitting hadron yields rather than ratios of yields, the fireball 
volume appears as additional parameter. We discuss its physical significance
in Section~\ref{edep} and in the Appendix.

We want to emphasize that our model does not contain any strangeness 
suppression factor, $\gamma_S$, as used e.g. in ref. \cite{bec1,raf1}.
Whenever $\gamma_S$ is used in thermal model calculations\footnote{We recall
that the usage of $\gamma_S$ implies that the thermal density of any given 
hadron carrying strangeness has a "suppression" factor $\gamma_S$ for 
every strange or antistrange quark.}, it is meant 
to account for non-equilibration in the strangeness sector.
Departure from equilibrium is expected for elementary collisions or for 
peripheral nucleus-nucleus collisions (as well as for light nuclei).
The introduction of a $\gamma_S$ factor is an attempt to model this
situation (see e.g. \cite{bec3,bec1}).
It is established that $\gamma_S\simeq 1$ for central collisions 
at RHIC \cite{kan,star}. At the lower end of the energy range, namely at 
SIS energies (up to 2 AGeV), there are indications \cite{cle1} that 
$\gamma_S$ is not needed either.
As we investigate central Au-Au or Pb-Pb collisions we prefer to stay 
within the thermodynamically well defined equilibrium model.
Nevertheless we investigate, in the Appendix, the influence on our results
as well as the statistical significance of the introduction of $\gamma_S$ 
into the fit procedure.  This leads to the conclusion that our main results,
i.e. the determination of $T$ and $\mu_b$ from fits of ratios of yields
at mid-rapidity, are little changed by this additional fit parameter.

The following hadrons are included in the calculations:
i) mesons: non-strange (37), strange (28), charm (15), bottom (16);
ii) baryons: non-strange (30), strange (33), charm (10);
iii) "composites" (nuclei up to $^4$He and $K^-$-clusters \cite{ksys}, 18).
The corresponding anti-particles are of course also included.
Their characteristics, including a rather complete set of decay channels,
(all strong decays and a suitable fraction of the weak decays contribution
matching the experimental conditions)
are implemented according to the most recent PDG compilation \cite{pdg}.
Studies of QCD thermodynamics using lattice results have demonstrated
that the partition function of a hadron and resonance gas model, 
which is a sum over all individual partition functions (Eq.~\ref{eq:part}), 
describes the lattice QCD data after suitably rescaling hadron masses 
to match lattice QCD \cite{lat0}.

We have used vacuum masses for all hadrons.
The effect of hadron mass (and width) modification was analyzed 
in the context of the thermal model for the top SPS energy  
in ref. \cite{mich}. 
The conclusion is that at SPS a 10\% reduction of masses leads to a 
slightly better thermal fit, with a corresponding 10\% reduction of the 
resulting $T$ and $\mu_b$, 
A thermal analysis of the RHIC data using hadron masses derived from
chiral models \cite{zsc} found no preference for any scenario of in-medium
masses compared to the non-interacting gas model.
Note that at AGS (Si+Au) the nominal masses are preferred \cite{rev}.
It was argued \cite{br,vk1} that changes of masses in-medium
in connection to chiral symmetry restoration can lead to larger values 
of the chemical potentials at lower energies.

The finite widths of resonances are taken into account 
in the density calculation by an additional integration, over 
the particle mass, with a Breit-Wigner distribution as a weight:
\be
n_i= \frac{g_i}{2\pi^2} \frac{1}{N_{BW}} \int_{M_0}^\infty \ud m\int_0^\infty 
\frac{\Gamma_i^2}{(m-m_i)^2+\Gamma_i^2/4} \cdot
\frac{p^2 \ud p}{\exp[(E_i^m-\mu_i)/T]\pm 1},
\ee
where $m_i$ is the nominal mass and $\Gamma_i$ is the width of particle $i$.
The energy is then calculated for every value of $m$ in the integration 
step, $E_i^m =\sqrt {p^2+m^2}$. 
Here, $N_{BW}$ is the normalization of the Breit-Wigner distribution 
and $M_0$ is the threshold for the dominant decay channel.
As this procedure significantly increases the computing time, for the
fits it was used only for the AGS energies (see below), where in particular
the widths of nucleon resonances are expected to play a significant role
\cite{wdel}.
At higher energies the width of the resonances plays little role in the 
determination of the thermal parameters.

The interaction of hadrons and resonances is usually included by 
implementing a hard core repulsion of Van der Waals--type via
an excluded volume correction. This is implemented in an iterative
procedure according to ref. \cite{exv,exv2}:
\be P^{excl.} (T,\mu)= P^{id.gas}(T,\hat{\mu}); \qquad 
\hat{\mu} = \mu - V_{eigen} P^{excl.}(T,\mu) \ee
where $V_{eigen}$ is calculated for a radius of 0.3~fm, considered identical
for all hadrons \cite{pbm2}.
This correction influences the particle densities, but has little ($<$5\%) 
effect on ratios.
The effect on absolute densities is very significant, however: at top SPS 
energies and above the excluded volume procedure leads to a reduction in
densities by about a factor of 2, leading to a corrspondingly larger 
volume at chemical freeze-out.

\begin{figure}[hbt]
\centering\includegraphics[width=.65\textwidth,height=.59\textwidth]{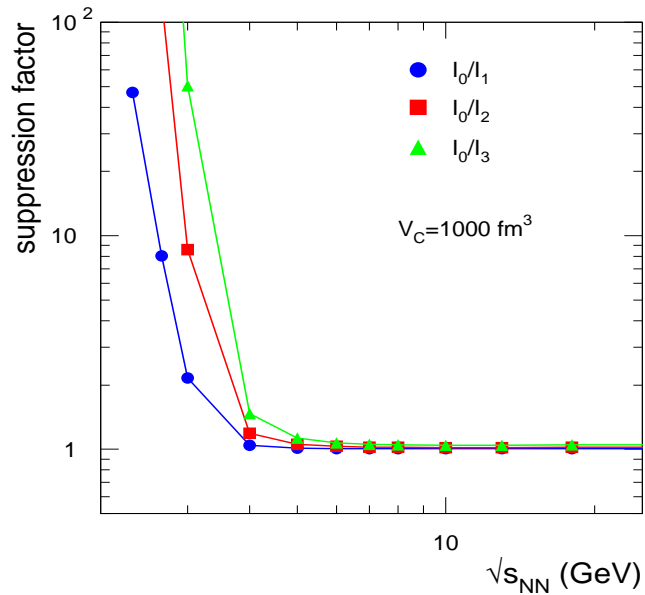}
\caption{The energy dependence of the canonical suppression factor for 
strangeness, calculated for the canonical volume $V_C$=1000 fm$^3$.} 
\label{fig1}
\end{figure}

The grand canonical ensemble is the simplest realization of a statistical 
approach and is suited for systems with a large number of produced 
hadrons.
However, for small systems (or peripheral nucleus-nucleus collisions) and
for low energies in case of strangeness production, a canonical ensemble (C)
treatment is mandatory \cite{bec0}. 
It leads to a phase space reduction for particle production (so-called 
``canonical suppression'').
It has been shown \cite{cle0,rev} that the density of particle $i$ with 
strangeness $S$ calculated in the canonical approach, $n_{i}^{C}$, 
is with a good approximation related to the grand canonical value, 
$n_{i}^{GC}$, as: 
$n_{i}^{C} = n_{i}^{GC} / F_S$, with $F_S={I_{0}(x)}/{I_S(x)}$.
The argument of the Bessel function of order $S$ is the total yield of 
strange and antistrange hadrons.
The suppression factor is shown in Fig.~\ref{fig1} for a canonical 
volume $V_C$=1000 fm$^3$, which we employ in our calculations 
(see next section).
In this case, the canonical suppression is negligible for all strange 
hadron species already for the highest AGS energy 
($\sqrt{s_{NN}}\simeq$5~GeV) but is sizeable for the lower energies. 
Whenever the canonical suppression is needed, $V_C$, becomes another 
parameter of the model and needs to be evaluated carefully (see below). 

We mention that several similar codes are now available for thermal model
calculations: 
SHARE \cite{share}, THERMUS \cite{thermus}, Therminator \cite{term}.

\section{Fits to experimental data} 
\label{fits}

To extract the parameters of the model we perform fits of the 
experimental data with model calculations.
In order to employ a minimal number of model parameters, namely $T$ and 
$\mu_b$, the temperature and the baryonic chemical potential at chemical 
freeze-out (ceasing of inelastic collisions), a common practice is to fit
hadron yield ratios \cite{rev,bro1,kan,star}.
It is important to recognize that one is faced with a choice when selecting
$N-1$ statistically independent ratios from the total number $N(N-1)/2$ of
ratios which can be constructed from $N$ experimentally measured hadron 
yields.
Although the number of independent hadron yields is typically rather large 
($N>$10) one may argue that fitting a particular choice of (statistically 
independent) particle ratios may introduce a bias for the
extracted $T$ and $\mu_b$ whenever a given experimental yield 
used in several ratios is subject to a systematic error .
To minimize such a possible bias, we use ratios constructed from all 
measured hadron species avoiding, as much as possible, the repetition of 
any particular yield.
Another possibility is to fit the measured hadron yields \cite{bec1,raf1}, 
which implies an extra parameter, the fireball volume.
This method is free of the bias discussed above, but is subject of
another bias, arising, for instance, whenever two measured yields (i.e.
of a particle and its antiparticle) have a similar systematic error,
which is cancelled in the ratio. Also, ratios typically have a smaller 
systematic error than absolute yields.
Based on this, our main analysis focusses on ratios of hadron yields.
As mentioned in Section~\ref{edep} and demonstrated in detail in the Appendix, 
the difference of fit parameters resulting from the two different fit 
procedures is always smaller than the uncertainty in the overall method, 
given the current status of measured yields and corresponding errors.

We briefly discuss below the fit procedures. The best fit is obtained by 
minimizing the distribution of $\chi^2$. 
To get an estimate of the systematic error of the fits, we additionally
consider the quadratic deviation, $\delta^2$.
The two quantities are defined as:
\begin{equation}
\chi^2=\sum_i \frac{(R_i^{exp}-R_i^{therm})^2}{\sigma_i^2}, \quad 
\delta^2=\sum_i \frac{(R_i^{exp}-R_i^{therm})^2}{(R_i^{therm})^2}, 
\end{equation}
where $R_i^{exp}$ is the measured value of either the yield or the ratio 
of hadron yields\footnote{We denote a yield ratio of two particles by 
the ratio of their respective symbols.}
with its uncertainty $\sigma_i$ and $R_i^{therm}$ is the value from 
the model calculations.
For the experimental errors we have quadratically added the statistical 
and the systematic errors of the measured yields.
When the systematic errors of the measurements were not available
we have assigned values of 10\%, unless otherwise stated.
The sum runs over the number of hadron ratios available experimentally.
We focus on central collisions of Au or Pb nuclei.
Ratios are calculated from the measured hadron yields.
Whenever needed, we scale the measured yields with the number of participants
\cite{misko} to account for different centrality classes.

We focus on mid-rapidity data ($\ud N/\ud y$), for which the bulk of the 
published hadron yields is available, but we also analyze data integrated 
over the full solid angle whenever available. 
While at low energies the consideration of hadron yields integrated over 
4$\pi$ is optimal, this changes as, with increasing beam 
energy, the nuclei become transparent and, besides a central fireball, 
fragmentation regions develop. Once the stopping is not complete, 
necessarily the baryo-chemical potential depends on rapidity and a 
consideration of 4$\pi$ data with one value of $\mu_b$ would not be 
appropriate. In that case the statistical model should be compared to data 
over a central region near mid-rapidity. One may consider in separate 
analyses more forward or backward fireballs with different parameters.
The possible presence of a finite net strangeness at mid-rapidity, as 
suggested by a recent transport model calculation \cite{ble}, could pose 
a special difficulty; this is not accounted for in our model.
However, the effect is expected to be small enough to not cause a systematic
bias of the extracted thermal parameters.

An important aspect in the comparison of calculations with measurements
is the contribution of feed-down from weak decays, mainly in the yields 
of pions and (anti)protons. In our model, the fraction of those hadrons
originating from weak decays is adjusted.
In general, this contribution is subtracted in the experiments, either 
explicitly (via simulations) or implicitly (due to specific reconstruction 
methods). 
We have tried to model in detail the contributions from weak decays,
following the information in the relevant experimental papers.
In the Appendix we demonstrate the importance to determine as precisely as 
possible the size of this contribution.

\subsection{AGS}

At the AGS, collisions of Au nuclei at beam kinetic energies of 2 to 10.7 
AGeV, corresponding to $\sqrt{s_{NN}}$=2.70-4.85 GeV were studied.
For the beam energies of 2, 4, 6, and 8 AGeV, the yields of protons 
\cite{e895p,e917p}, pions \cite{e895pi,e866pik} and kaons \cite{e866pik} 
are available (there are no $K^-$ data at 2 AGeV). 
The $\Lambda$ yields are only available integrated over 4$\pi$ \cite{e895l}
and for the energy of 6 AGeV the integrated $\Xi^-$ yield was also measured
\cite{e895xi}.
In these cases we have derived the mid-rapidity values assuming the same
ratio between 4$\pi$ and $\ud N/\ud y$ as for protons \cite{e895p}.

\vspace{-.7cm}
\begin{figure}[hbt]
\begin{tabular}{lr} \begin{minipage}{.49\textwidth}
\centering\includegraphics[width=1.1\textwidth]{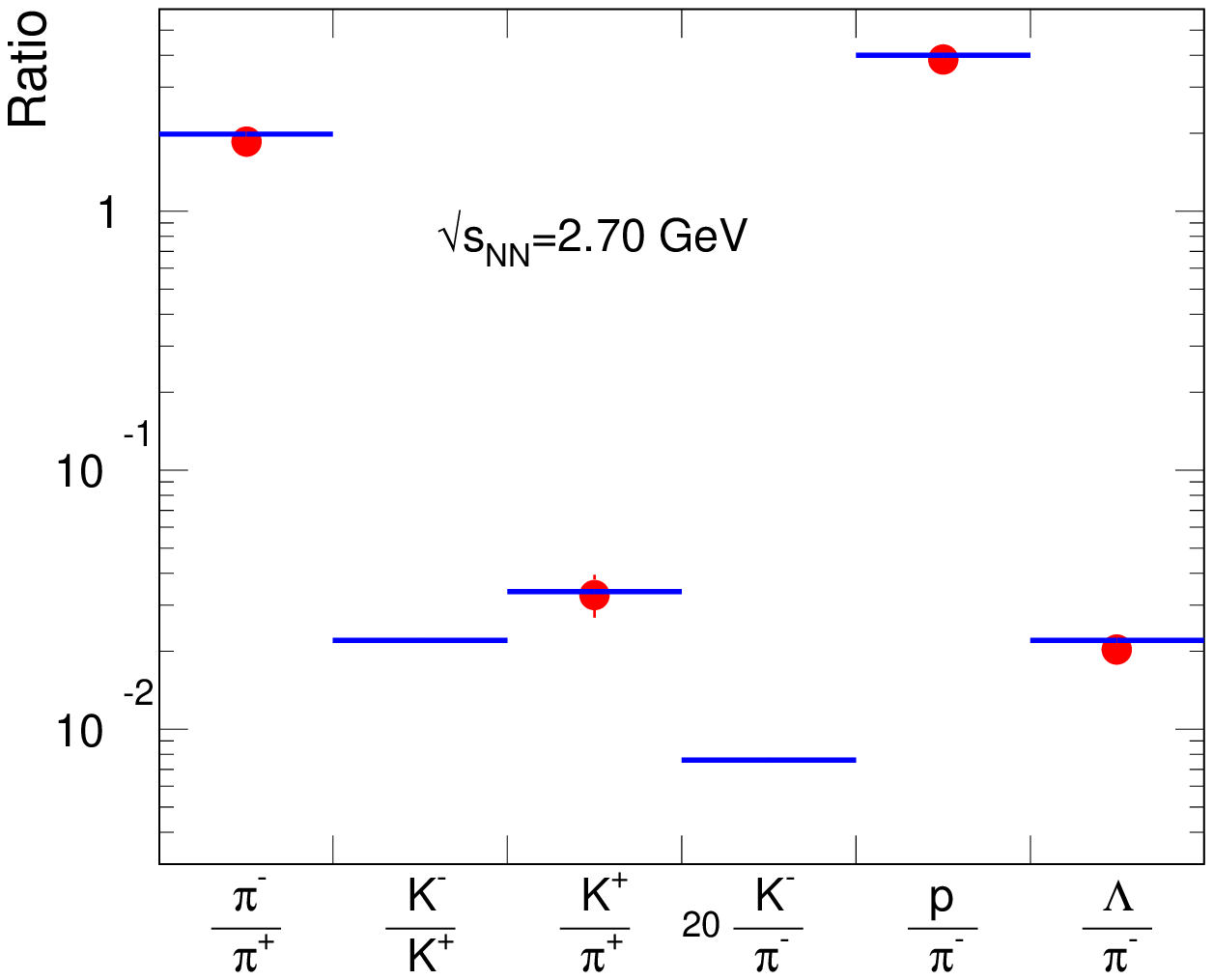}
\end{minipage} &\begin{minipage}{.49\textwidth}
\centering\includegraphics[width=1.1\textwidth]{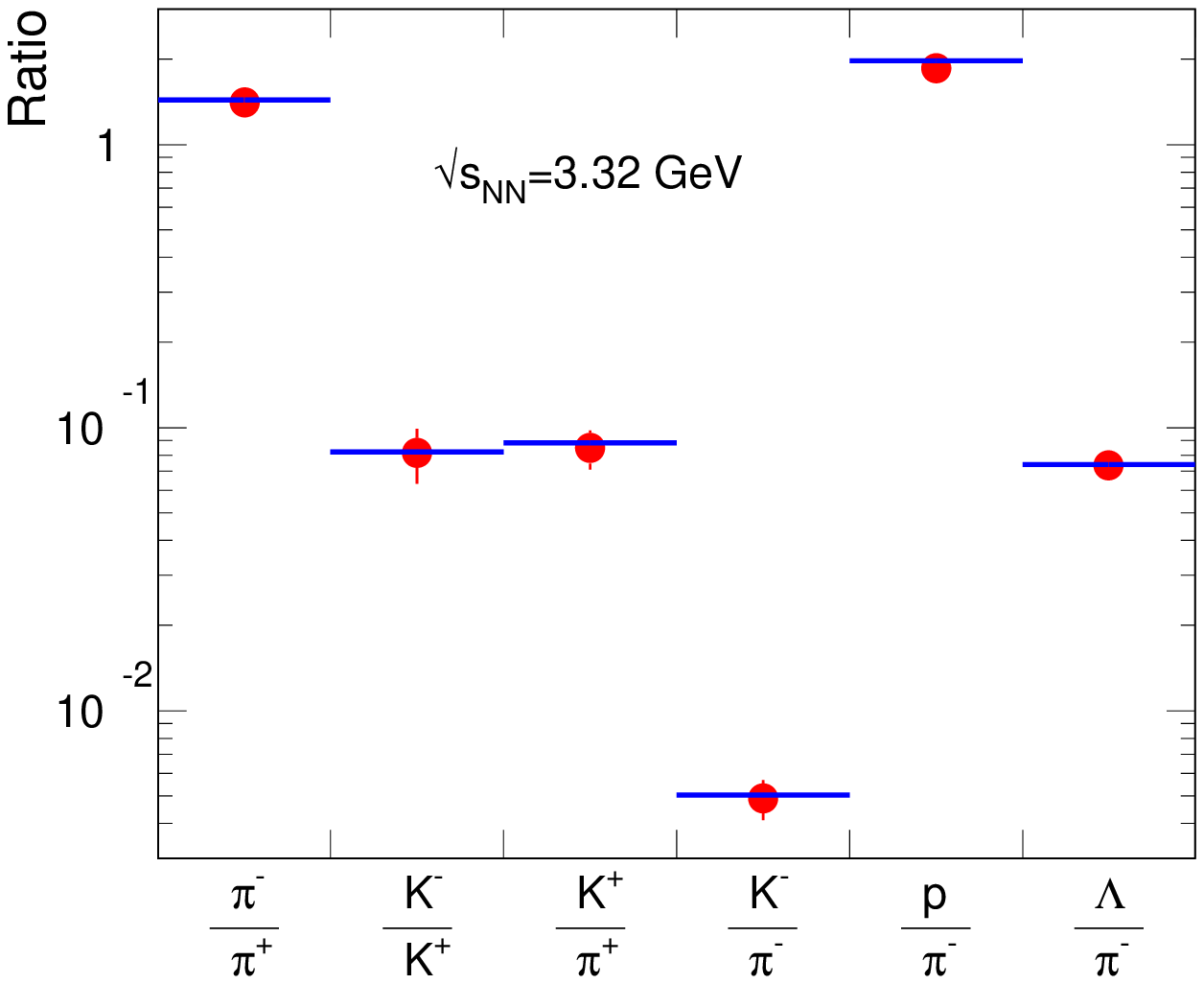}
\end{minipage} \\
\begin{minipage}{.49\textwidth}
\vspace{-1cm}
\centering\includegraphics[width=1.1\textwidth]{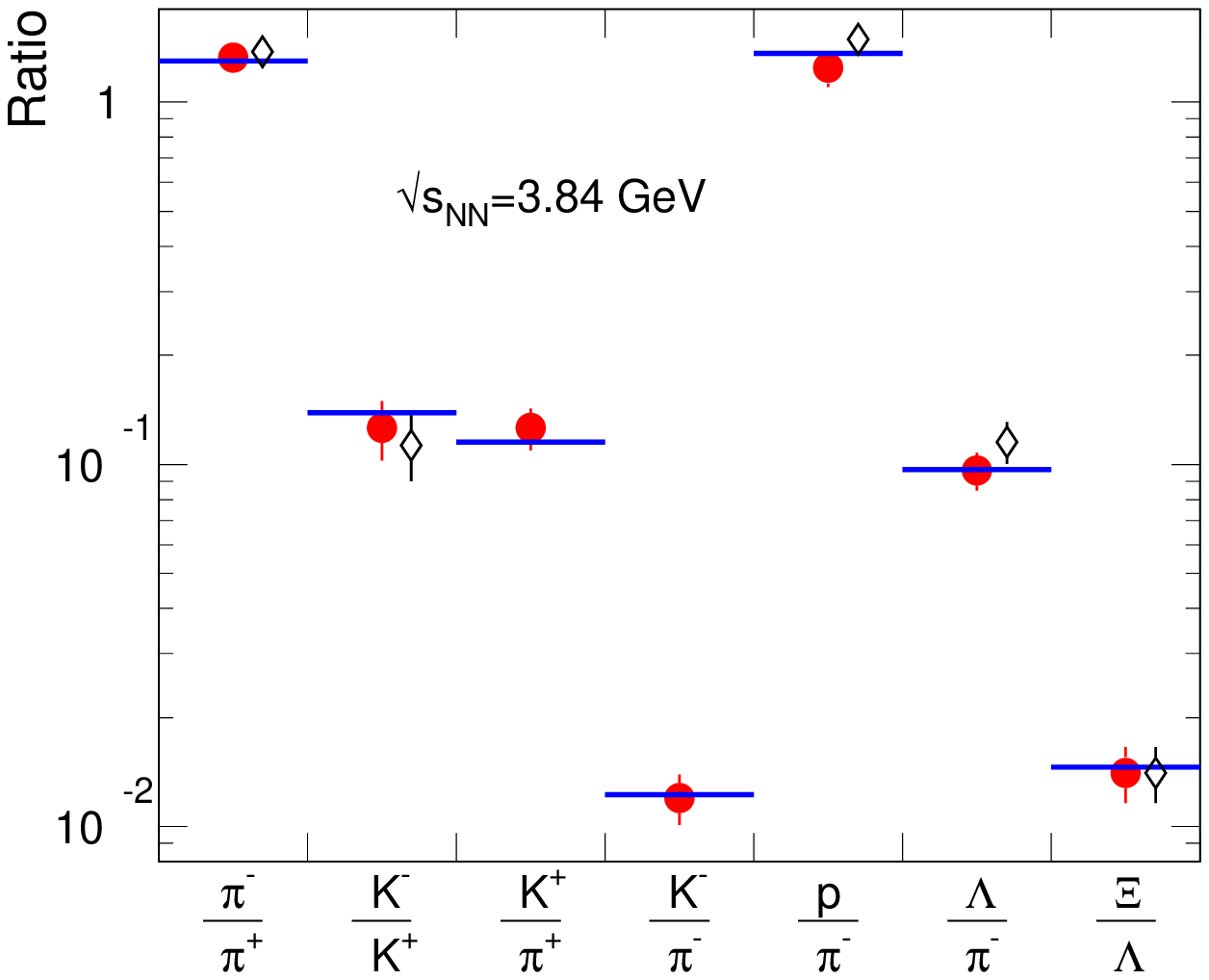}
\end{minipage} &\begin{minipage}{.49\textwidth}
\vspace{-1cm}
\centering\includegraphics[width=1.1\textwidth]{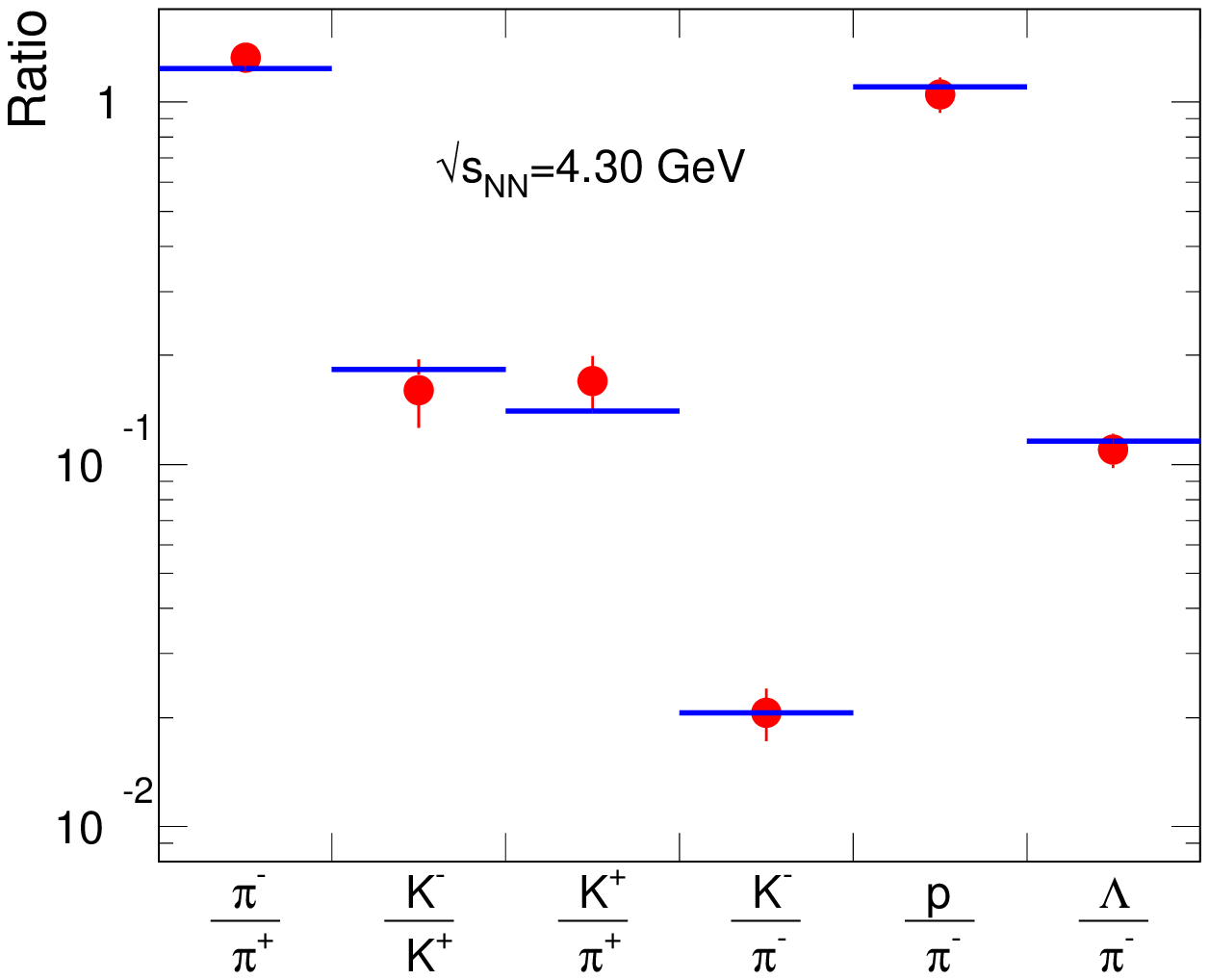}
\end{minipage} \end{tabular} 
\vspace{-.5cm}
\caption{Measured hadron yield ratios compared to the thermal model 
calculations for the AGS beam energies of 2, 4, 6, and 8 AGeV.
The symbols are data, the lines are model calculations corresponding
to the best fit ($T$, $\mu_b$) values of (64,760), (78,670), (86,615),
and (93,580) MeV, respectively. For the beam energy of 6 AGeV the
ratios of 4$\pi$ yields are plotted as diamonds.}  
\label{fig2}
\end{figure}

The comparison of the measured and calculated ratios for the best fit 
($\chi^2$ minimization) is presented in Fig.~\ref{fig2} for the four energies.
Here and in the following plots we show both the $K^+/\pi^+$ and the 
$K^-/\pi^-$ ratios, but in the fit either one or the other is included 
(as only one is statistically independent from the ratios $\pi^-/\pi^+$ 
and $K^-/K^+$), depending on the minimum value of $\chi^2$. 
The resulting values of $T$ and $\mu_b$ are between 64 and 93 MeV
and 760 and 580 MeV, respectively.
The model describes the data very well; the $\chi^2$ values per
number of degrees of freedom, $\chi^2/N_{df}$ are 1.0/2, 0.43/3, 1.15/4, 
1.14/3 for 2, 4, 6, and 8 AGeV, respectively. 
Although this could be accidental, as the number of data points in the fits
is very small, these low values could imply that the systematic errors 
of the measurements are actually overestimated.
We mention that if we do not include in the calculations the finite widths 
of resonances the corresponding values of $\chi^2/N_{df}$ are
slightly worse, but the values of $T$ and $\mu_b$ change only marginally.
For both cases, the $\chi^2$ distributions are narrow around the minimum
and there is no correlation between $T$ and $\mu_b$ as is observed for 
higher energies (see below). In some cases, close-by minima are identified 
and we have included this feature into the errors of the parameters.

\begin{figure}[hbt]
\begin{tabular}{lr} \begin{minipage}{.42\textwidth}
\hspace{-.7cm}
\includegraphics[width=1.24\textwidth]{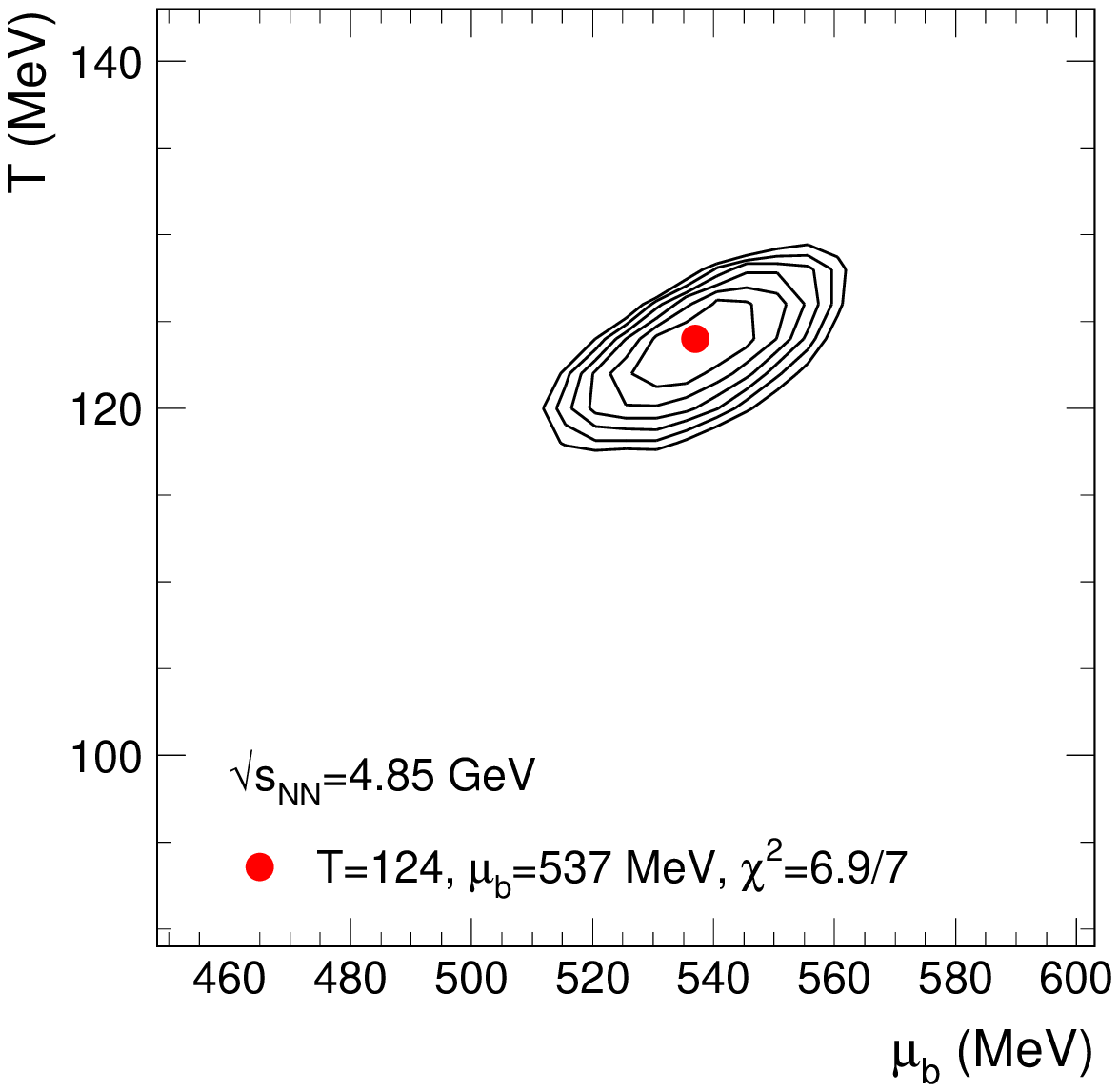}
\end{minipage} &\begin{minipage}{.58\textwidth}
\hspace{-.7cm}
\includegraphics[width=1.15\textwidth,height=.93\textwidth]{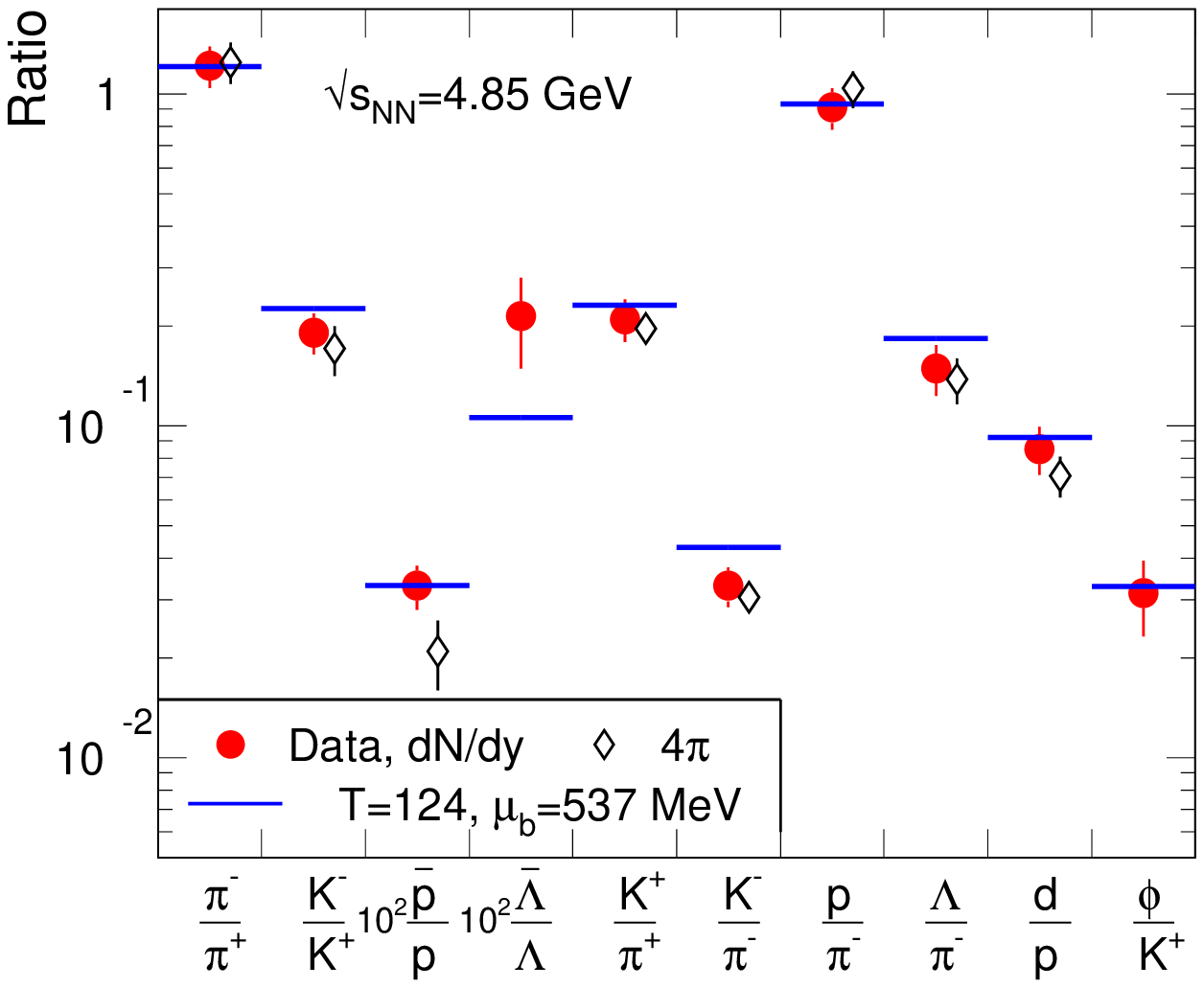}
\end{minipage} \end{tabular} 
\caption{The $\chi^2$ contours (in steps of 1 from the minimum, marked by 
the dot) and yield ratios with best fit at mid-rapidity for the top AGS beam 
energy of 10.7 AGeV. The diamonds represent ratios of yields integrated
over 4$\pi$.
The ratio $K^-/\pi^-$ was not included in the fit. Note the scaling factor 
of 100 for the ratios $\bar{p}/p$ and $\bar{\Lambda}/\Lambda$.}
\label{fig3}
\end{figure}

\begin{figure}[h]
\begin{tabular}{lr} \begin{minipage}{.49\textwidth}
\centering\includegraphics[width=1.1\textwidth]{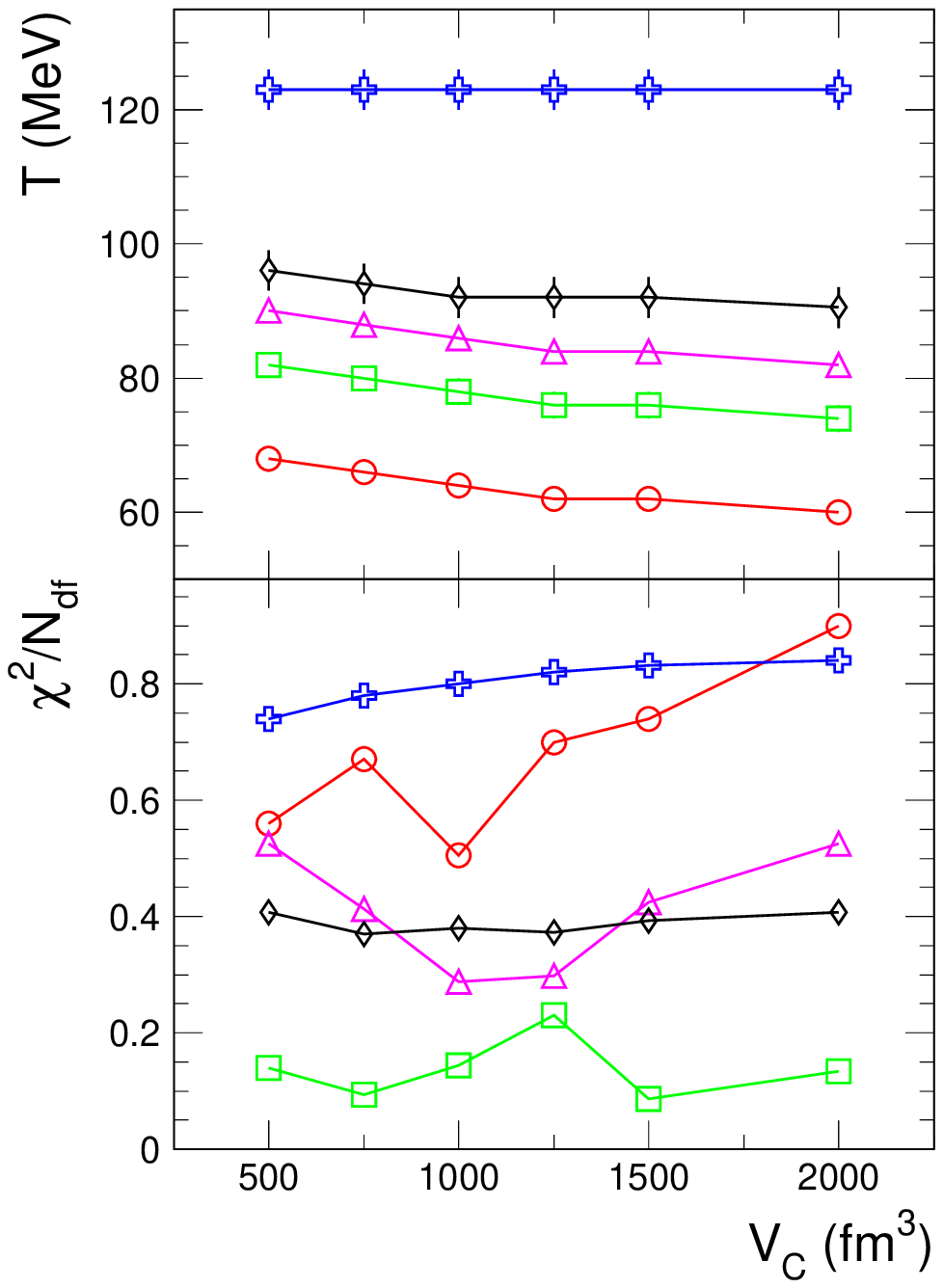}
\end{minipage} &\begin{minipage}{.49\textwidth}
\centering\includegraphics[width=1.1\textwidth]{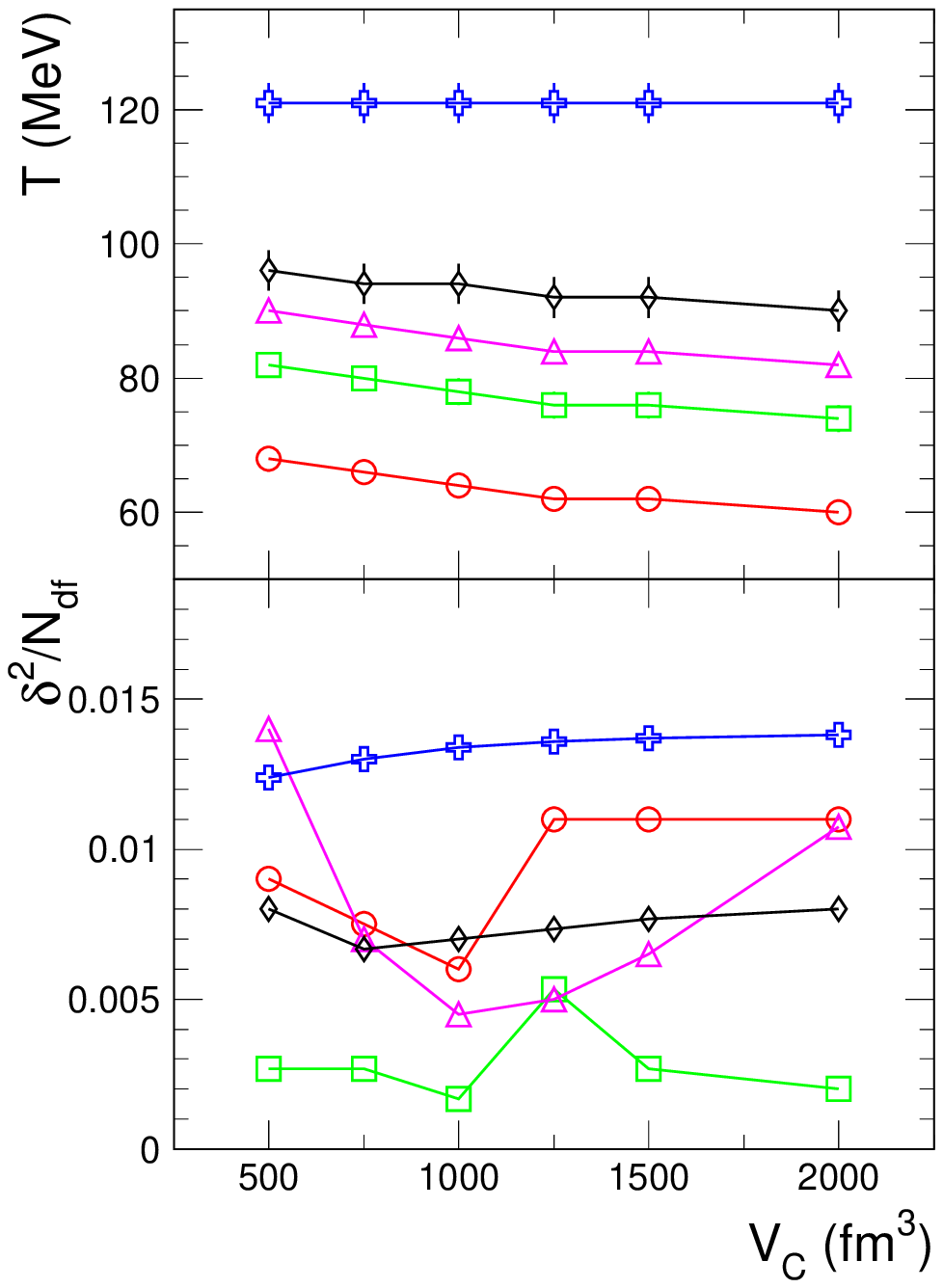}
\end{minipage} \end{tabular} 
\caption{The canonical volume dependence of temperature extracted from 
thermal fits for the AGS beam kinetic energies of 2, 4, 6, 8, and 
10.7 AGeV (in order of increasing temperature). 
Left panel is for minimizing $\chi^2$, right panel is for $\delta^2$. 
The corresponding $\chi^2/N_{df}$ and $\delta^2/N_{df}$ are also shown.}
\label{fig3x}
\end{figure}

At the top AGS beam kinetic energy of 10.7 AGeV ($\sqrt{s_{NN}}$=4.85 GeV) 
there is a large set of hadron yields experimentally available:
$p$ \cite{e877p,e802p,e917p},
$d$ \cite{e802p},
$\pi$ \cite{e877p,e866pik},
$K$ \cite{e866pik,e802k},
$\Lambda$ \cite{e896l},
$\phi$ \cite{e917phi},
$\bar{\Lambda}$ \cite{e917l}.
The thermal fit to the data yields $T=124\pm 3$ MeV and 
$\mu_b=537\pm 10$ MeV, corresponding to a minimum of $\chi^2/N_{df}$=6.9/7
(the errors correspond to 1 $\sigma$).
The model fits the data very well, as evident from the narrow $\chi^2$ 
distribution and from the comparison of the ratios shown in Fig.~\ref{fig3}.
Not included in the fit is the ratio $K^-/\pi^-$,which is not statistically
independent from the ratios already included in the fit. 
The ratio $\bar{\Lambda}/\Lambda$ deviates from the model calculations
substantially ($\bar{\Lambda}$ "anomaly" \cite{e917l}).
The experimental $\bar{\Lambda}/\Lambda$ is a factor of 2 higher than 
the model value and as such has a significant contribution to the $\chi^2$
(if $\bar{\Lambda}/\Lambda$ is not included in the fit, the resulting 
values are: $T=123\pm 3$ MeV, $\mu_b=538\pm 9$ MeV, $\chi^2/N_{df}$=4.0/6).

The fit without the ratios $d/p$, $\bar{p}/p$, $\bar{\Lambda}/\Lambda$, and 
$\phi/K^+$ gives $T=108\pm 9$, $\mu_b=555\pm 18$ MeV, with 
$\chi^2/N_{df}$=1.3/3.
We note that, in this case, $T$ and $\mu_b$ are not anymore correlated
(as seen in Fig.~\ref{fig3}), but rather loosely anticorrelated. 
We have performed this fit to test a possible bias in the extracted thermal
parameters at the lower energies due to the smaller number of ratios
experimentally available.
We estimate in this way that the extracted temperature can be biased 
by up to 15\%, while for $\mu_b$ the bias can be up to 3\%.
These systematic (upper limits) errors are used in the following.

To investigate the sensitivity of the extracted ($T$,$\mu_b$) values on 
the canonical volume $V_C$ (which enters via the canonical suppression 
factor), we have performed the fits for all AGS energies for a range 
of $V_C$ values.
The results are presented in Fig.~\ref{fig3x}, where we show the dependence 
of $T$ on the canonical volume parameter $V_C$ for two cases: minimizing 
$\chi^2$ and $\delta^2$.
Both cases result in very similar freeze-out temperatures. 
As expected, $T$ slightly grows with decreasing $V_C$. 
The effect is gradually reduced towards higher energies and totally vanishes 
at the top AGS energy.
Based on the dependence of $\chi^2/N_{df}$ and $\delta^2/N_{df}$ on 
$V_C$ (also shown in Fig.~\ref{fig3x}) one cannot constrain the canonical
volume parameter. As a consequence, and since the canonical volume dependence
is a rather small effect, we have chosen the value $V_C$=1000 fm$^3$.
We mention that $\mu_b$ is unaffected by the canonical volume choice.

\subsection{SPS}

At the CERN SPS Pb-Pb collisions are measured in fixed target experiments, 
with beam momenta ($\approx$energies) of 20 to 158 AGeV 
($\sqrt{s_{NN}}$=6.27-17.3 GeV).
For the time being only the data collected at 40, 80 and 158 AGeV are final. 
The mid-rapidity yields of pions, kaons and (anti)protons and deuterons 
\cite{na44,na49pik,na49pd,na49pbar} are available for these energies.
The yields of the complete set of strange (anti)hyperons 
\cite{na49l,na49xi,na49o,na57,na57b} and of $\phi$ \cite{na49phi} are 
published for the top energy and at 40 AGeV.
Since the deuteron yield is measured over a reduced range in momentum,
we have decided to not include the deuterons in the fitting procedure
for all the SPS energies, although the ratio $d/p$ will be shown in the
following plots.
Also shown in the plots are the ratios $\Xi/\pi^-$ and $\Omega/\pi^-$,
which are not included in the fits because they are not independent
from the ratios $\Xi/\Lambda$ and $\Omega/\Xi$, respectively, which are
used because they result in a better fit.

\begin{figure}[hbt]
\hspace{-.7cm}
\begin{tabular}{lr} \begin{minipage}{.37\textwidth}
\centering\includegraphics[width=1.26\textwidth]{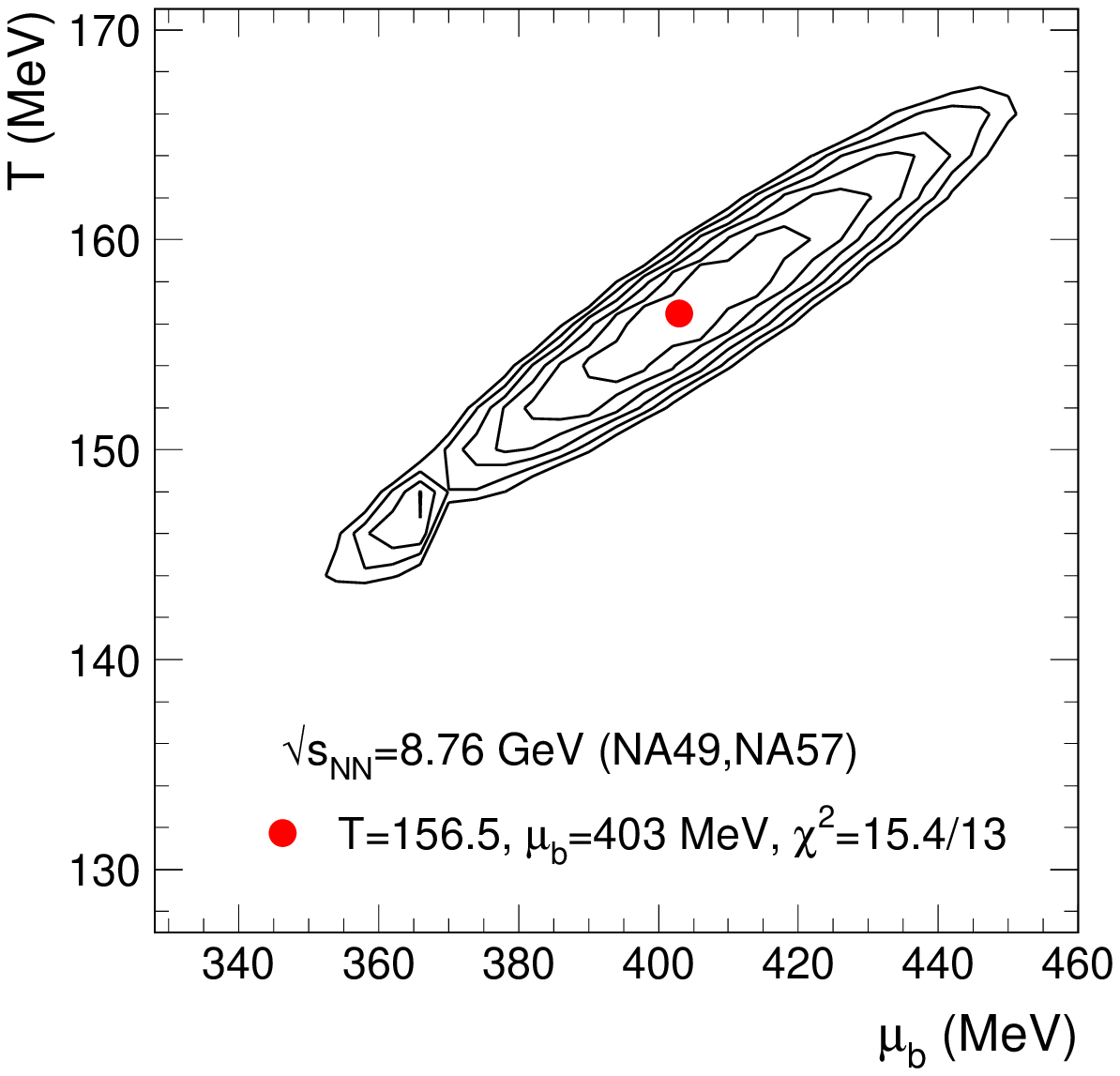}
\end{minipage} &\begin{minipage}{.58\textwidth}
\centering\includegraphics[width=1.17\textwidth]{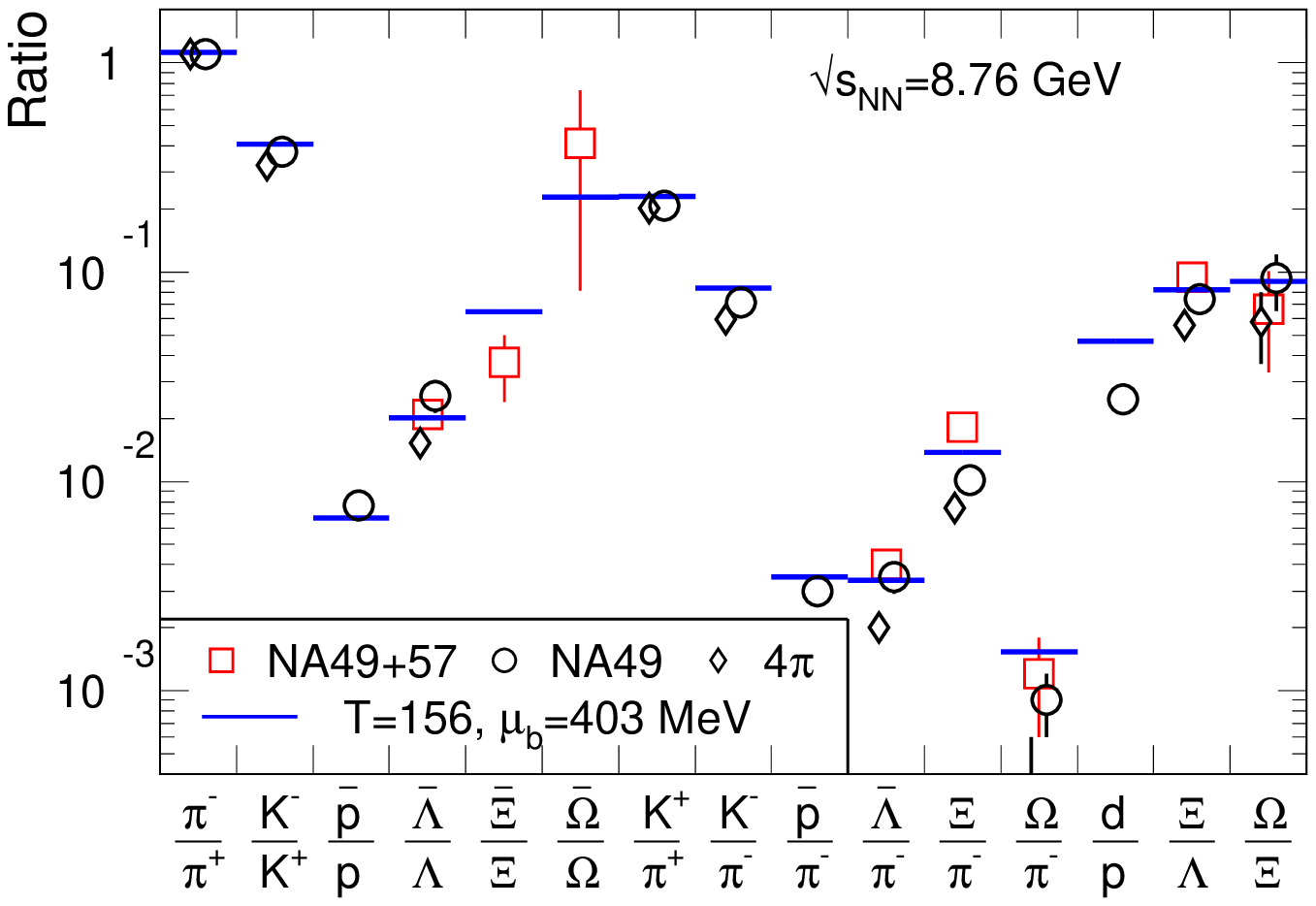}
\end{minipage} \end{tabular} 
\caption{Distribution of $\chi^2$ and hadron yield ratios with best fit at 
the SPS beam energy of 40 AGeV using the NA49 and NA57 mid-rapidity data
(the ratios $K^-/\pi^-$, $d/p$, $\Xi/\pi^-$ and $\Omega/\pi^-$ are not 
included in the fits).
The NA49 4$\pi$ data are plotted for comparison (diamonds).
For the ratios $\Omega/\pi^-$ and $\Omega/\Xi$, the $\Omega$ yield includes
both $\Omega$ and $\bar{\Omega}$.} 
\label{fig4}
\end{figure}

At the beam energy of 40 AGeV ($\sqrt{s_{NN}}$=8.76 GeV), a combined fit 
to all data gives $T$=156$^{+4}_{-3}$ MeV, $\mu_b$=403$^{+18}_{-14}$ MeV, 
with $\chi^2/N_{df}$=15.4/13
(from the quadratic deviation analysis, $T$=166 MeV, $\mu_b$=434 MeV, 
$\delta^2$=0.58). 
The $\chi^2$ distribution and the comparison of the measured and
calculated ratios for the best fit are shown in Fig.~\ref{fig4}.
A fit of NA49 data alone leads to $T$=166$\pm$2 MeV, $\mu_b$=438$\pm$7 MeV, 
with $\chi^2/N_{df}$=6.2/7
($T$=162 MeV, $\mu_b$=414 MeV, $\delta^2$=0.10).

\begin{figure}[hbt]
\begin{tabular}{lr} \begin{minipage}{.42\textwidth}
\hspace{-.7cm}
\includegraphics[width=1.24\textwidth]{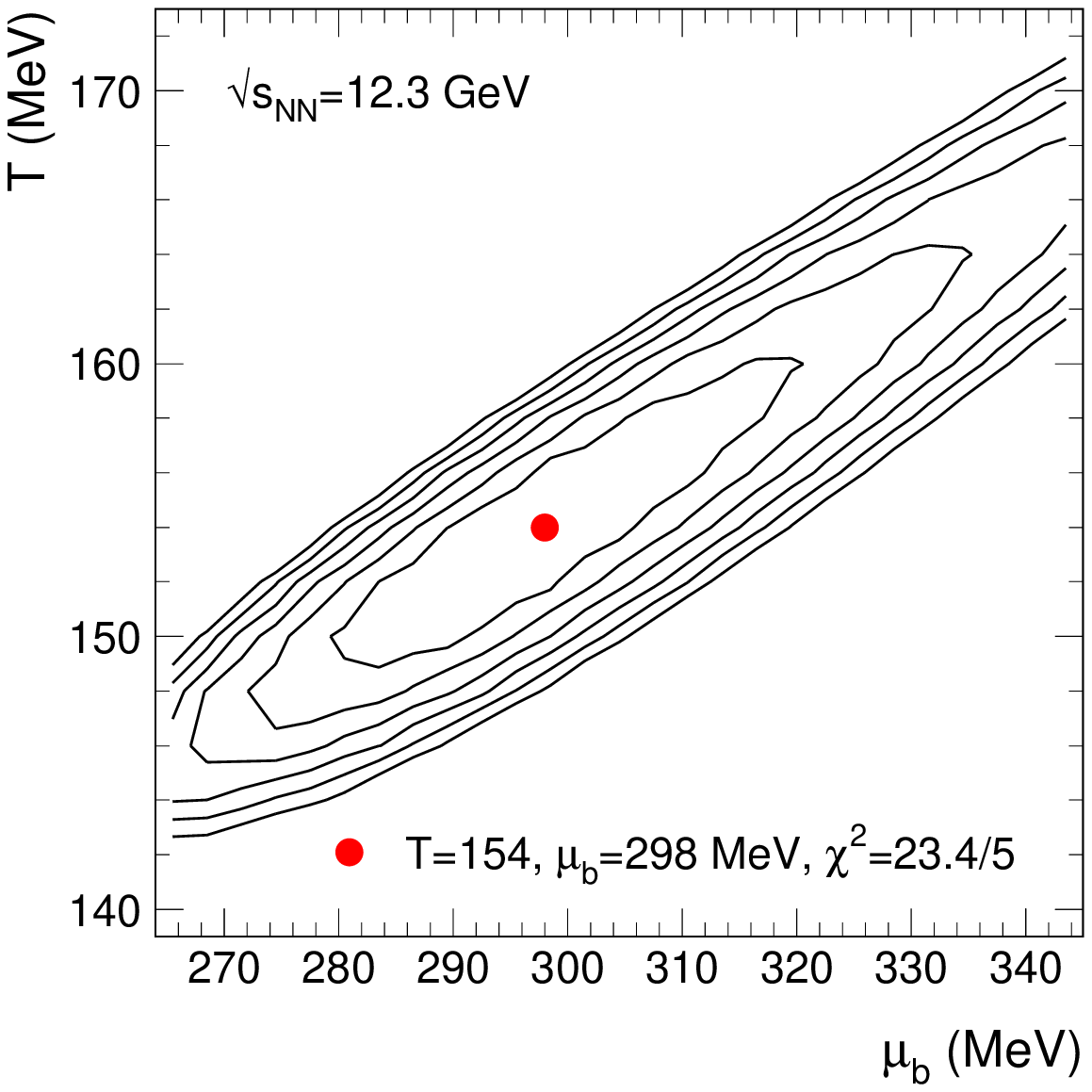}
\end{minipage} &\begin{minipage}{.58\textwidth}
\hspace{-.7cm}
\includegraphics[width=1.15\textwidth,height=.93\textwidth]{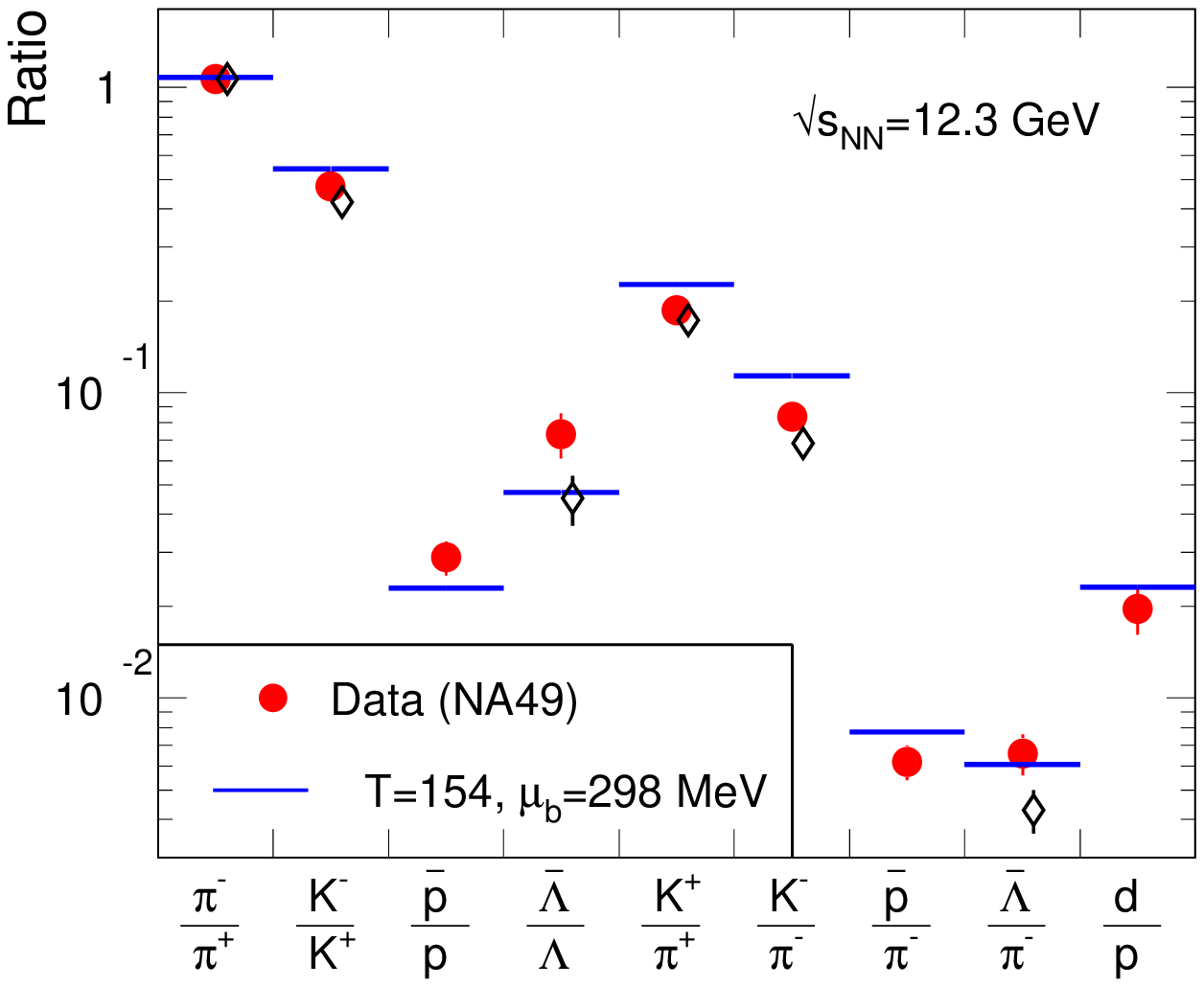}
\end{minipage} \end{tabular} 
\caption{Distribution of $\chi^2$ and hadron yield ratios with best fit at 
the beam energy of 80 AGeV using the NA49 data (the ratios $K^-/\pi^-$ and 
$d/p$ were not included in the fit). 
The 4$\pi$ yields (represented with diamonds) are plotted for comparison.} 
\label{fig5}
\end{figure}

The $\chi^2$ distribution and the comparison of the measured and
calculated ratios for the beam energy of 80 AGeV ($\sqrt{s_{NN}}$=12.3 GeV)
are shown in Fig.~\ref{fig5}. 
The best fit is achieved for $T$=154 MeV, $\mu_b$=298 MeV,
with 1~$\sigma$ errors of 6 and 20 MeV, respectively
($T$=152 MeV, $\mu_b$=271 MeV, with minimum $\delta^2$=0.23). 
The quality of the fit is not good, as seen from the value of
$\chi^2/N_{df}$=23.4/5.  
We have noticed that the outcome of the fit is sensitive to the inclusion
of different (statistically equivalent) hadron ratios. For instance,
if the ratios $p/\pi^-$ and $\Lambda/\pi^-$ are considered instead of
$\bar{p}/\pi^-$ and $\bar{\Lambda}/\pi^-$ the resulting temperature from
$\chi^2$ minimization is lower by about 10 MeV, while the temperature
from $\delta^2$ minimization is unchanged. This shows the effect of the
errors on the outcome of the $\chi^2$ minimization and indicates that the
errors of the data are not entirely consistent. We have incorporated
these results into the systematic errors of the thermal parameters.

\vspace{-.5cm}
\begin{figure}[hbt]
\centering\includegraphics[width=.9\textwidth]{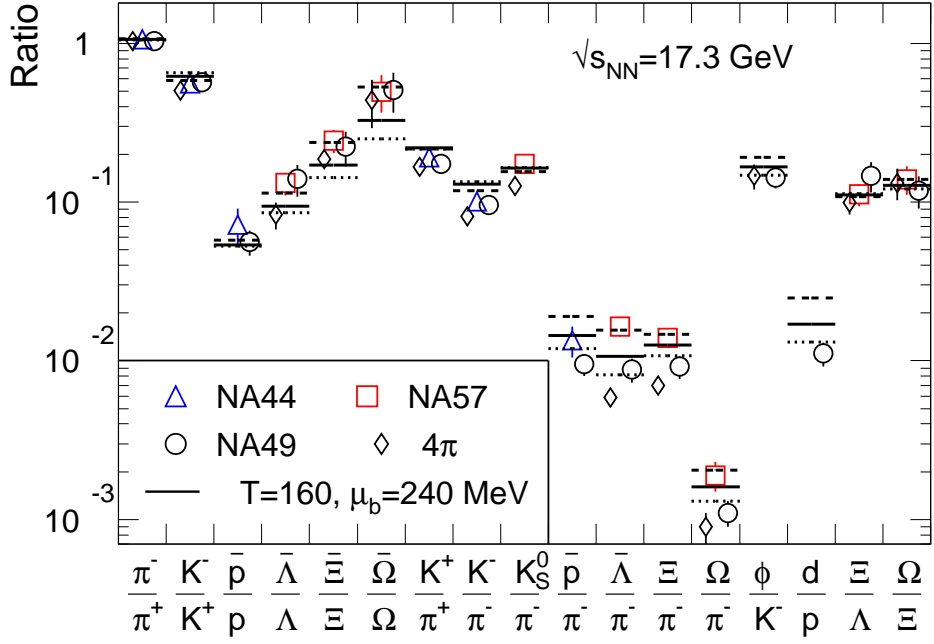}
\caption{The ratios of hadron yields at mid-rapidity with best fits at the 
top SPS energy of 158 AGeV.
For the ratios of hyperons and $K^0_S$ from NA57 relative to pions, we have 
used the $\pi^-$ yields measured by NA44. 
The 4$\pi$ ratios with NA49 data (diamonds) are plotted for comparison.
The full lines are for the combined fit, the dashed ones for the NA44+NA57 
data ($T$=180, $\mu_b$=268 MeV) and the dotted ones for the NA49 data
($T$=150, $\mu_b$=266 MeV).
The ratios $K^-/\pi^-$, $d/p$, $\Xi/\pi^-$ and 
$\Omega/\pi^-$ are not included in the fit.}
\label{fig6}
\end{figure}

For the top SPS energy of 158 AGeV ($\sqrt{s_{NN}}$=17.3 GeV), there are 
mid-rapidity measurements available from the experiments
NA49 \cite{na49pik,na49pd,na49l,na49xi,na49o}, 
NA57 \cite{na57,na57b} (hyperons and $K^0_S$) and 
NA44 \cite{na44} ($\pi$, $K$, $p$).
Note that the NA44 pion yields are not explicitly corrected for the 
contribution from weak decays, but this is estimated to be small if one 
takes into account the reconstruction procedure in the NA44 spectrometer
\cite{na44b}.
We have assumed an efficiency of reconstruction of 40\% for pions from
$K^0_S$ (considered the average yield of $K^+$ and $K^-$ measured by NA44) 
and of 20\% from $\Lambda$ (considered the average of the NA57
and NA49 yields) decays. In this way, 5.9 $\pi^+$ and 7.8 $\pi^-$ 
($\ud N/\ud y$) are estimated to originate from weak decays.

We have used in the fits the ratios $\Xi/\Lambda$ and $\Omega/\Xi$.
This choice leads to better values of $\chi^2$ compared to the case
when using the ratios $\Xi/\pi^-$ and $\Omega/\pi^-$ instead.
The ratio $d/p$ was not included in the fits.
The measured and calculated ratios for the best fit are shown in 
Fig.~\ref{fig6}.
A fit to all the combined NA49, NA57 and NA44 data gives  
$T=160\pm 5$ MeV and $\mu_b=240\pm 18$ MeV, with $\chi^2/N_{df}$=56/22
(from the least square deviation minimization, 
$T$=172 MeV, $\mu_b$=243 MeV, $\delta^2$=0.86). 
For instance, if the ratios $\Xi/\pi^-$ and $\Omega/\pi^-$ are considered 
instead of  $\Xi/\Lambda$ and $\Omega/\Xi$,
$T=155\pm 2$ MeV and $\mu_b=224\pm 6$ MeV, with $\chi^2/N_{df}$=63/22 
($T$=170 MeV, $\mu_b$=237 MeV, $\delta^2$=1.03).

\begin{figure}[hbt]
\hspace{-.5cm}
\begin{tabular}{lcr} \begin{minipage}{.31\textwidth}
\centering\includegraphics[width=1.25\textwidth]{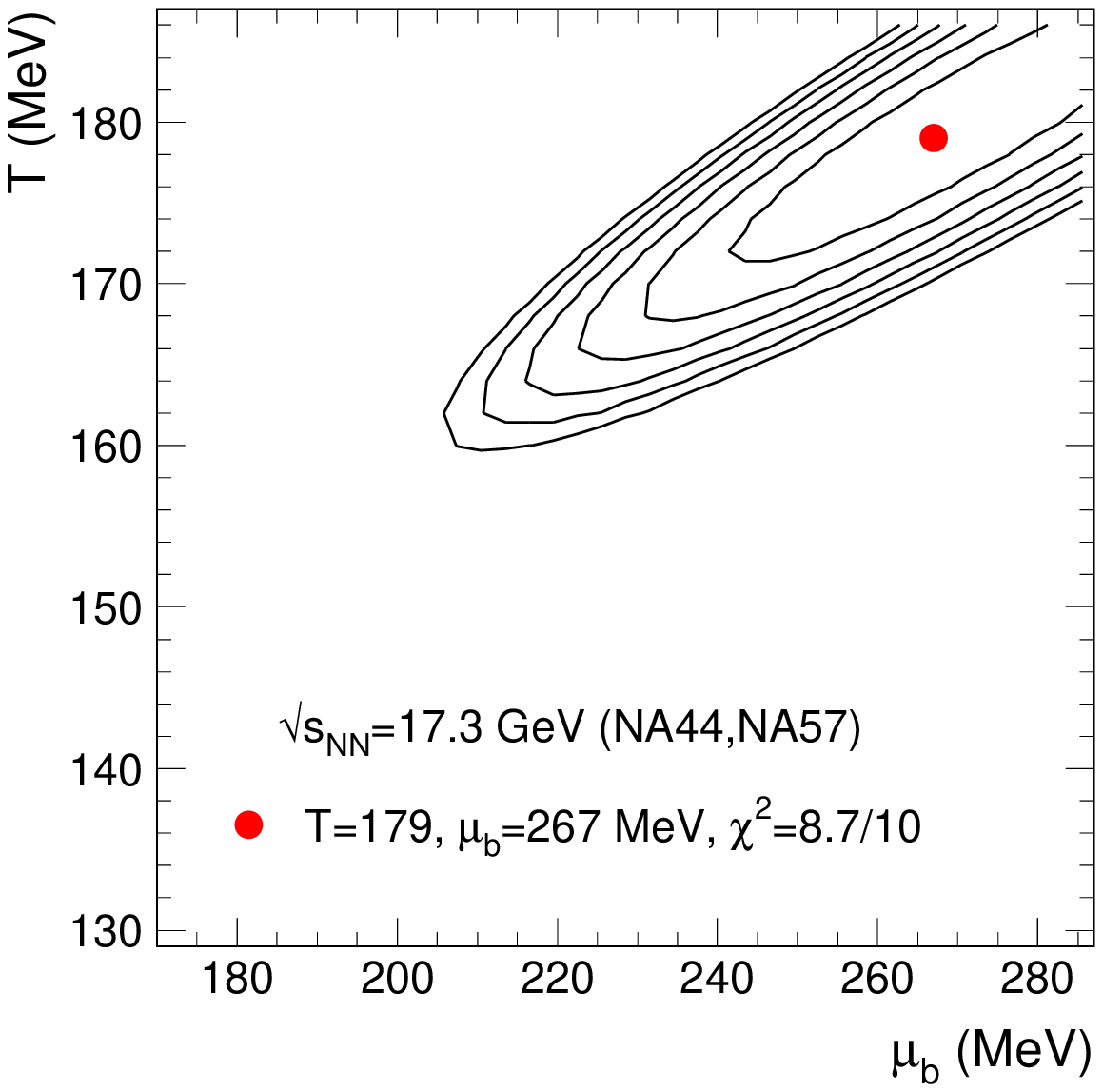}
\end{minipage} &\begin{minipage}{.31\textwidth}
\centering\includegraphics[width=1.25\textwidth]{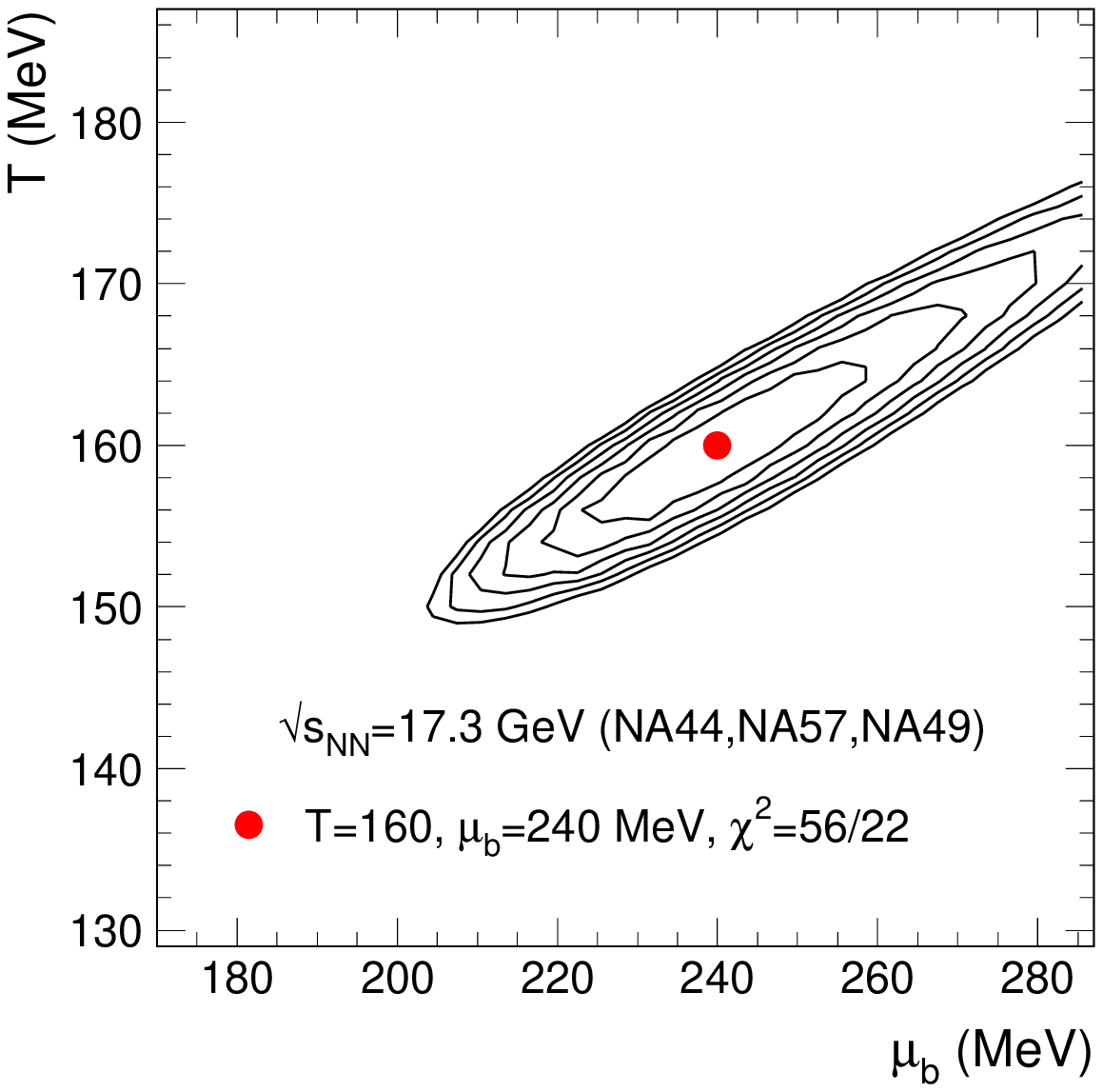}
\end{minipage} &\begin{minipage}{.31\textwidth}
\centering\includegraphics[width=1.25\textwidth]{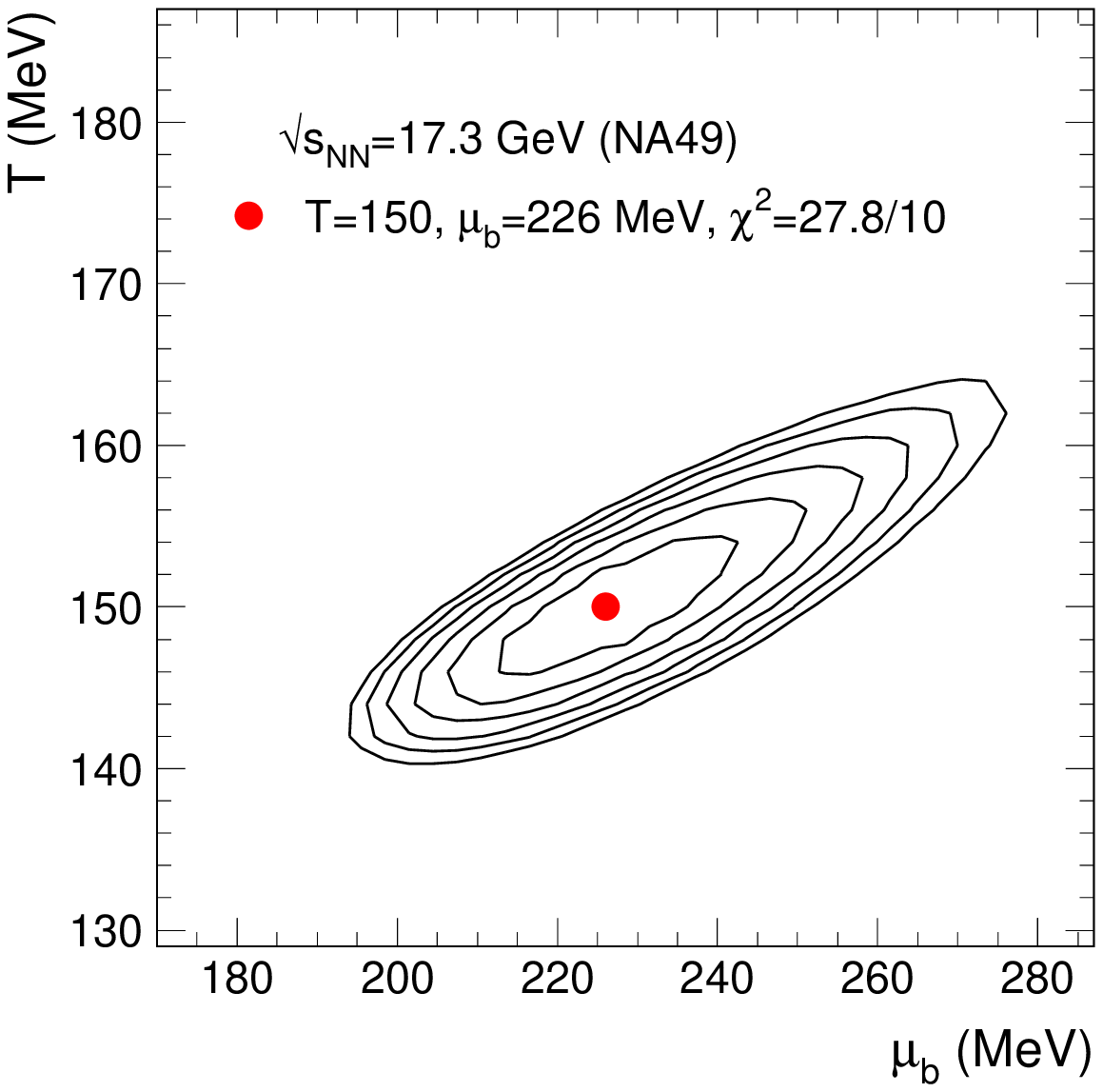}
\end{minipage} \end{tabular} 
\caption{The $\chi^2$ distributions at the top SPS energy of 158 AGeV for 
three cases (see text).}
\label{fig6b}
\end{figure}

To investigate the contribution to these large values of $\chi^2/N_{df}$, 
we have considered separatly two sets of hadron ratios: \\
i) the combined NA57 and NA44 data\footnote{
In combining data from different experiments we neglect the possible 
difference in centralities at the same quoted fraction of the inelastic 
cross section which may arise from different centrality measures.}
give $T=179\pm 7.5$ MeV and $\mu_b=267\pm 26$ MeV.
With $\chi^2/N_{df}$=8.7/10, the fit quality is good.\\
ii) using the NA49 data we extract 
$T=150\pm 4.5$ MeV, $\mu_b=226\pm 15$ MeV, with $\chi^2/N_{df}$=27.8/10.
The fit quality is much worse compared to the NA57 and NA44 data set.

The $\chi^2$ contours for the combined fit and for the two separate
data sets are shown in Fig.~\ref{fig6b}. The outcome of the fits
is summarized in Table ~\ref{tab1a}.
While $\mu_b$ is compatible within the errors for all the three cases,
the value of the temperature differs significantly, well beyond the errors,
between the NA49 and the NA44+NA57 data sets.
This discrepancy imposes large systematic errors for the extracted thermal 
parameters (see next section).
Remarkably, the values extracted from the $\delta^2$ minimization
are very similar for all three cases, $T\simeq$170 MeV, $\mu_b\simeq$240 MeV.
This indicates that the outcome of the combined fit is driven by the 
NA49 data and in particular by their small errors.
Although unsatisfactory thermal fits may signal a possible change
in physics in connection to the critical (end)point of a first order 
phase transition from hadrons to QGP \cite{nona}, the experimental situation
just discussed allows unfortunately no conclusion on this issue.
The apparent inconsistency of the results obtained using two independent 
data sets is not alleviated if one uses in the fit hadron yields instead 
of ratios. The results for this case (presented in Table~\ref{tab1b} in 
the Appendix) show a similar pattern as in case of the fits using ratios.
We note that poor quality of the fits at SPS energies characterizes other
recent analyses within the thermal model \cite{bec1,zsch}. 

\begin{table}[hbt]
\caption{Summary of the results of the fits using mid-rapidity data 
at top SPS energy ($\sqrt{s_{NN}}$=17.3 GeV).
The first three and next three columns show the results of the $\chi^2$ 
(with their 1 $\sigma$ error) and $\delta^2$ minimization, respectively.}
\label{tab1a}
\begin{tabular}{l|ccc|ccc}
data set & $T$ (MeV) & $\mu_b$ (MeV)  & $\chi^2/N_{df}$ &
$T$ (MeV) & $\mu_b$ (MeV)   & $\delta^2$ \\ 
\hline
NA44+NA57 &179$\pm$7.5 & 267$\pm$26 & 8.7/10 & 174 & 243 & 0.15 \\
NA49      &150$\pm$4.5 & 226$\pm$15 & 27.8/10 & 168 & 240 & 0.66 \\
combined  &160$\pm$5 &240$\pm$18 & 56/22 & 172 & 243  &0.86 
\end{tabular}
\end{table}

\subsection{RHIC}

\begin{figure}[hbt]
\begin{tabular}{lr} \begin{minipage}{.37\textwidth}
\centering\includegraphics[width=1.25\textwidth]{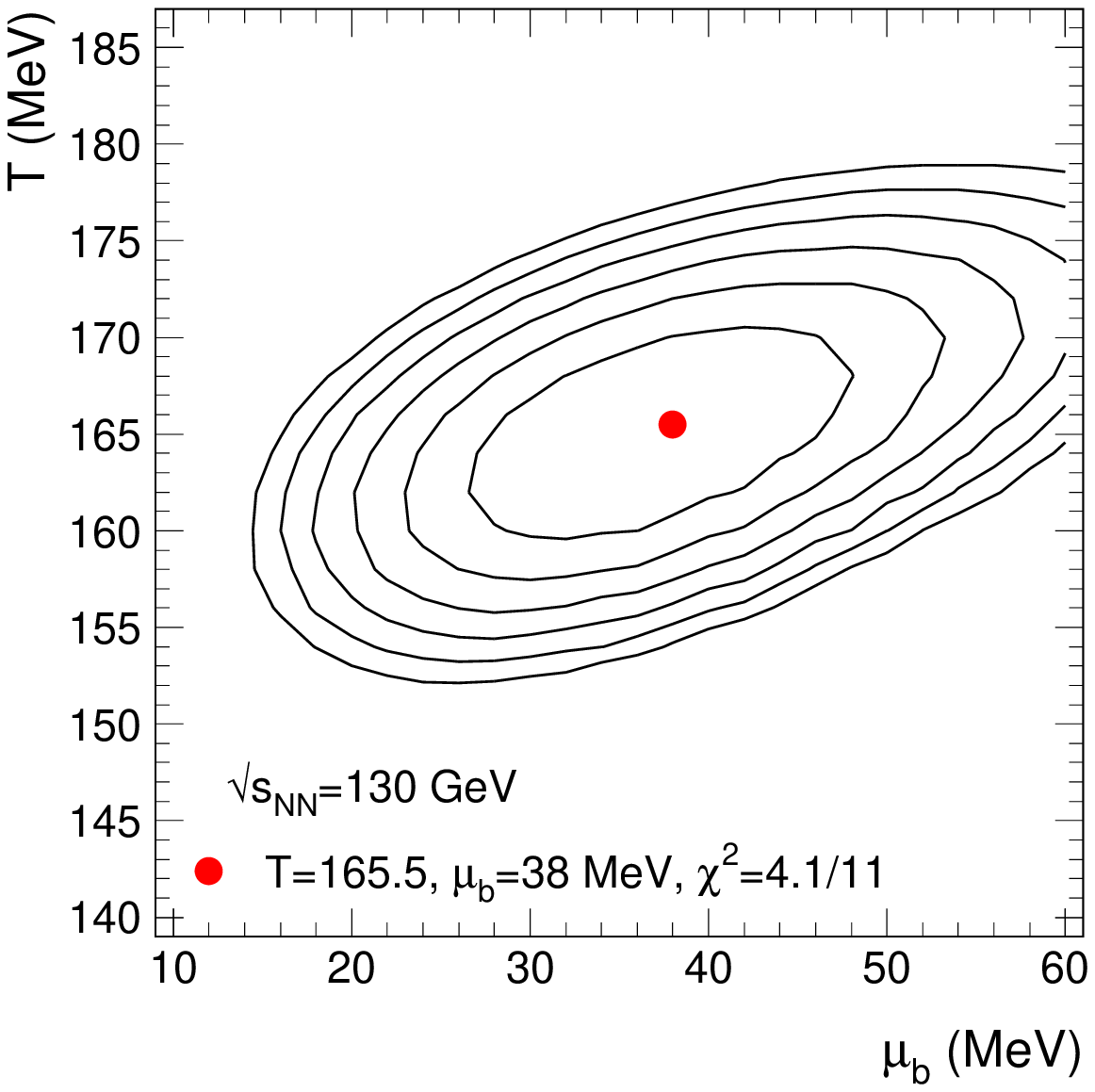}
\end{minipage} &\begin{minipage}{.58\textwidth}
\centering\includegraphics[width=1.15\textwidth]{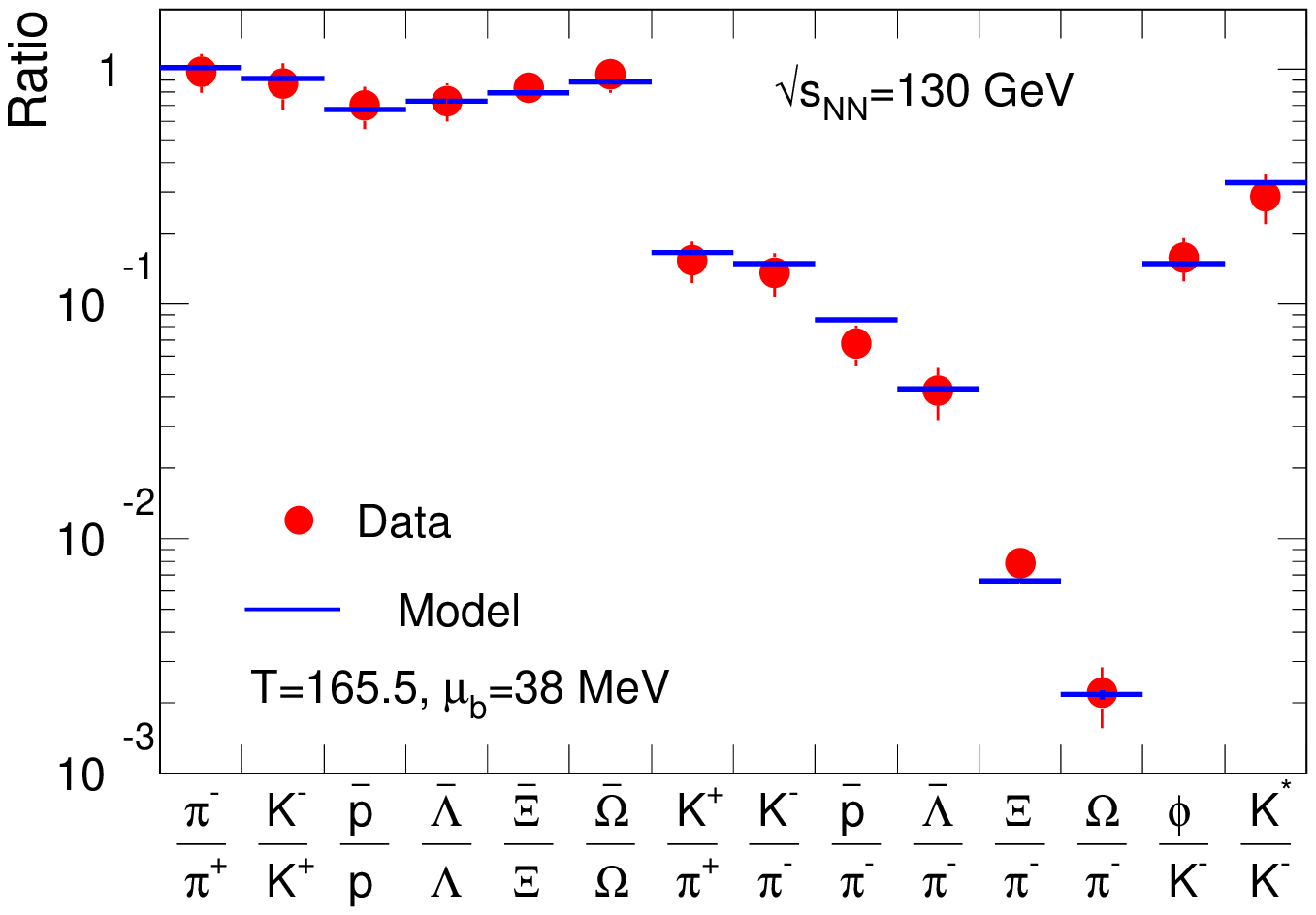}
\end{minipage} \end{tabular} 
\caption{The $\chi^2$ distribution and hadron yield ratios with best fit 
at $\sqrt{s_{NN}}$=130 GeV. 
Here, the $\Omega$ yield includes both $\Omega$ and $\bar{\Omega}$.
} 
\label{fig8}
\end{figure}

A complete set of final data for Au-Au collisions at RHIC is available 
at $\sqrt{s_{NN}}$=130 GeV.
We use for our analysis the yields of pions, kaons, (anti)protons \cite{p130}
and $\Lambda$, $\bar{\Lambda}$ \cite{pl}
measured by PHENIX, while for the other hadron species, 
$\Xi$, $\Omega$ \cite{sxio,srat}, $\phi$ \cite{sphi}, 
$K^*$ \cite{skst}, the data are from STAR.
Most of the available data are not corrected for feed-down from weak decays, 
but these contributions are properly taken into account in the model 
calculations.
One exception are the pions, for which the weak decay feed-down contribution
is not clearly specified by the experiment (PHENIX).We have considered three cases: \\
i) Assuming that all the pions from feed-down are included in the data,
the values of the thermal parameters are $T=170.5\pm 6$ MeV and 
$\mu_b=38\pm 15$ MeV, with a minimum $\chi^2/N_{df}$ value of 3.6/11.
\\
ii) At the other extreme, if one considers that the pion data do not contain 
any contribution from weak decays, the resulting values are
$T=162.5\pm 5.5$ MeV, $\mu_b=35\pm 11$ MeV, $\chi^2/N_{df}$=4.6/11.
\\
iii) Assuming that the data contain 30\% of the pions from weak decays,
$T=165.5\pm 5.5$ MeV, $\mu_b=38\pm 11$ MeV, $\chi^2/N_{df}$=4.1/11.

Although the fit is good for all cases (and even if the minimal
$\chi^2$ is obtained for case i), we consider the intermediate case iii) 
as the most likely situation, as it could be an implicit result 
of the reconstruction in the experiment.
The outcome of the fit is shown in Fig.~\ref{fig8}.
The very good fit is also apparent in the comparison 
of the hadron ratios: essentially all the experimental ratios are well 
reproduced by the model, including those involving $\phi$ and $K^*$ 
resonances.

\begin{figure}[hbt]
\centering\includegraphics[width=.87\textwidth]{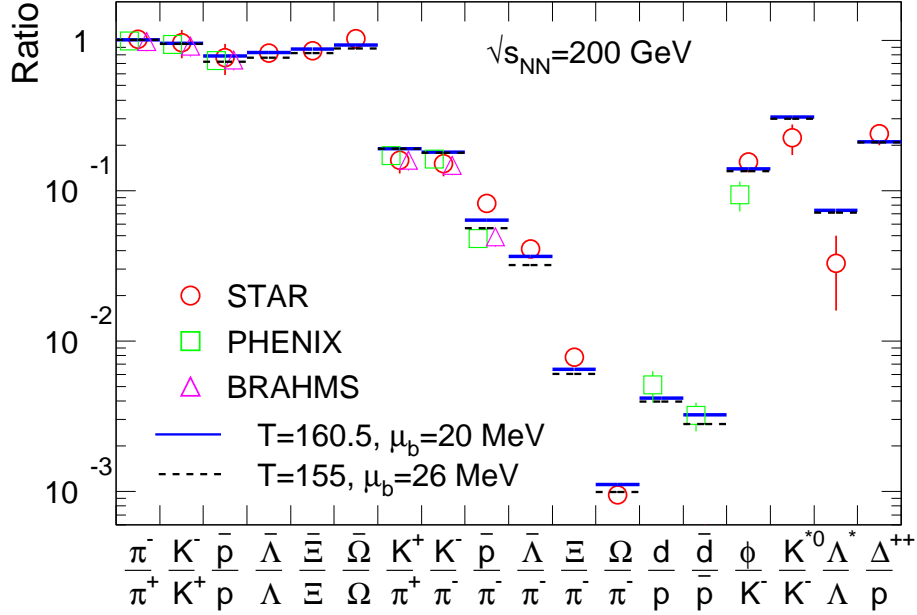}
\caption{Hadron yield ratios with best fits at $\sqrt{s_{NN}}$=200 GeV
(see text). The last three ratios, involving resonances, were not included
in the fits.} 
\label{fig8a}
\end{figure}

At $\sqrt{s_{NN}}$=200 GeV only a limited set of yields of identified 
hadrons is available to date. The yields of $\pi^\pm$, $K^\pm$, $p$, and 
$\bar{p}$ are published by PHENIX \cite{p200}, STAR \cite{s200}, and
BRAHMS \cite{b200} and the values agree within the quoted systematic errors.
In addition, available are the yields of $\phi$ \cite{s200phi,p200phi},
$K(892)^{*}$ \cite{s200k}, $d$ and $\bar{d}$ \cite{p200d}.
We use the hadron ratios with the corresponding errors
whenever provided by the experiments \cite{p200,s200phi,p200phi,s200k}
or otherwise calculate the ratios using the published yields
quoted above.
 Further ratios are available as preliminary data on strange 
hyperon ratios \cite{s200h}, $\Delta^{++}/p$ \cite{s200d}, 
$\bar{p}/\pi^-$, $\bar{\Lambda}/\pi^-$, 
$\Xi/\pi^-$, $\Omega/\pi^-$, and $\Lambda^*/\Lambda$\cite{s200s}. 
Unless specified, our fits do not include the ratios involving
resonances ($\Delta^{++}/p$, $K(892)^{*}$, and $\Lambda^*$).

We consider the following cases for the fits: \\
i) a combined fit of all available data, with the exception of strongly 
decaying resonances:
$T=155\pm2$ MeV, $\mu_b=26\pm5$ MeV, $\chi^2/N_{df}=34.1/23$
(with $\delta^2$ minimization: $T$=164 MeV, $\mu_b$=24 MeV, $\delta^2$=0.40).
If we include in the fit the three ratios involving resonances 
($K^*/K^-$, $\Lambda^*/\Lambda$, and $\Delta^{++}/p$), 
the results are the same within the errors, $T=155\pm2$ MeV, 
$\mu_b=25\pm5$ MeV, but with a worse $\chi^2/N_{df}=41.8/26$
($T$=162 MeV, $\mu_b$=22 MeV, $\delta^2$=0.82). \\
ii) as i), but excluding from the fit the ratios $\bar{p}/\pi^-$ and 
$\phi/K^-$ from PHENIX. The resulting parameters are: $T=160.5\pm2$ MeV, 
$\mu_b=20\pm4$ MeV, with $\chi^2/N_{df}$=16.0/21
($T$=166 MeV, $\mu_b$=26 MeV, $\delta^2$=0.19).
Including the resonances the outcome of the fit is identical, 
$T=160\pm2$ MeV, $\mu_b=20\pm4$ MeV, with a reasonable $\chi^2/N_{df}=25.2/24$ 
($T$=164 MeV, $\mu_b$=20 MeV, $\delta^2$=0.61). \\

\begin{figure}[hbt]
\hspace{-.7cm}
\begin{tabular}{lr} \begin{minipage}{.49\textwidth}
\centering\includegraphics[width=1.12\textwidth]{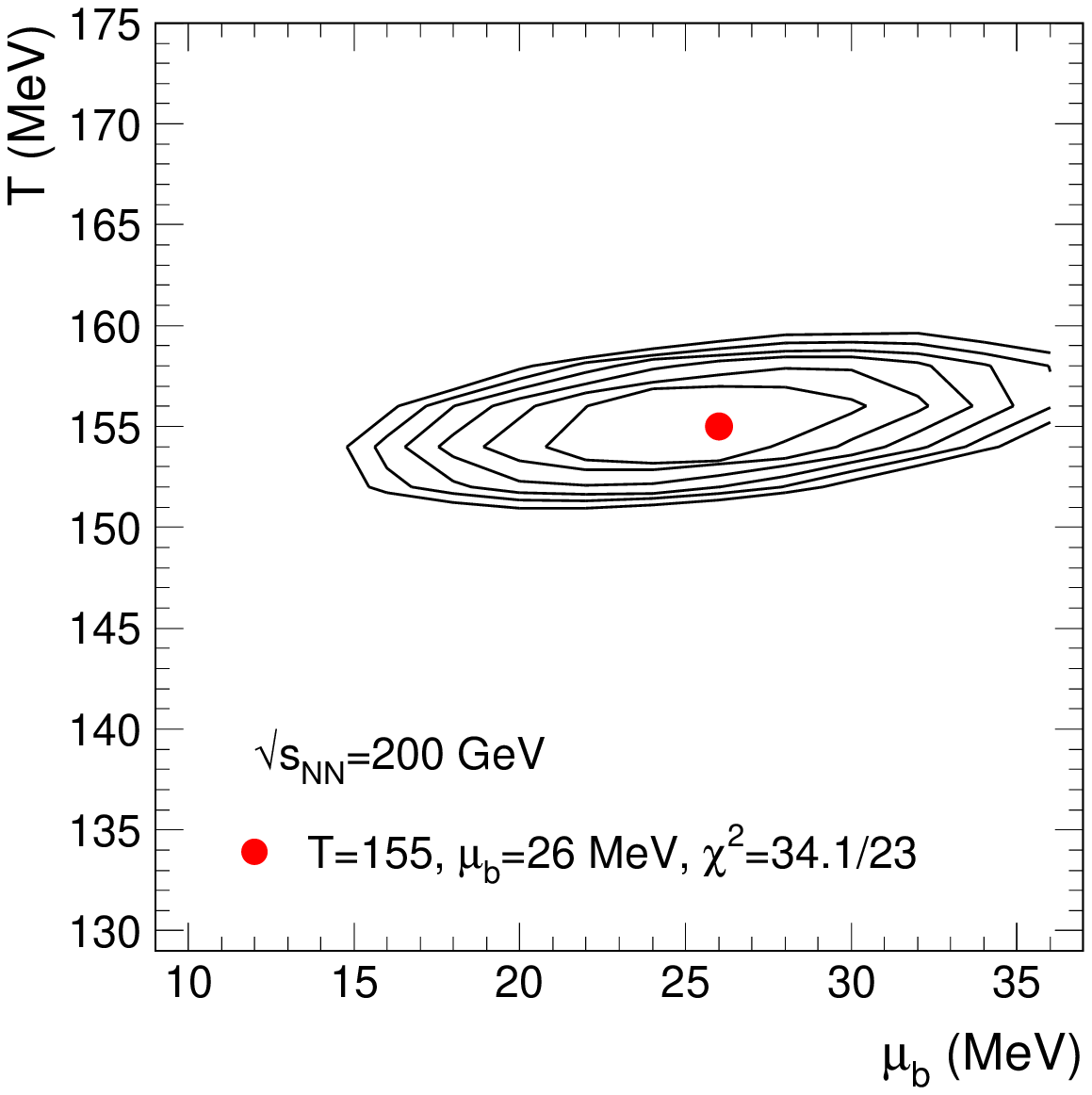}
\end{minipage} &\begin{minipage}{.49\textwidth}
\hspace{-.7cm}\includegraphics[width=1.12\textwidth]{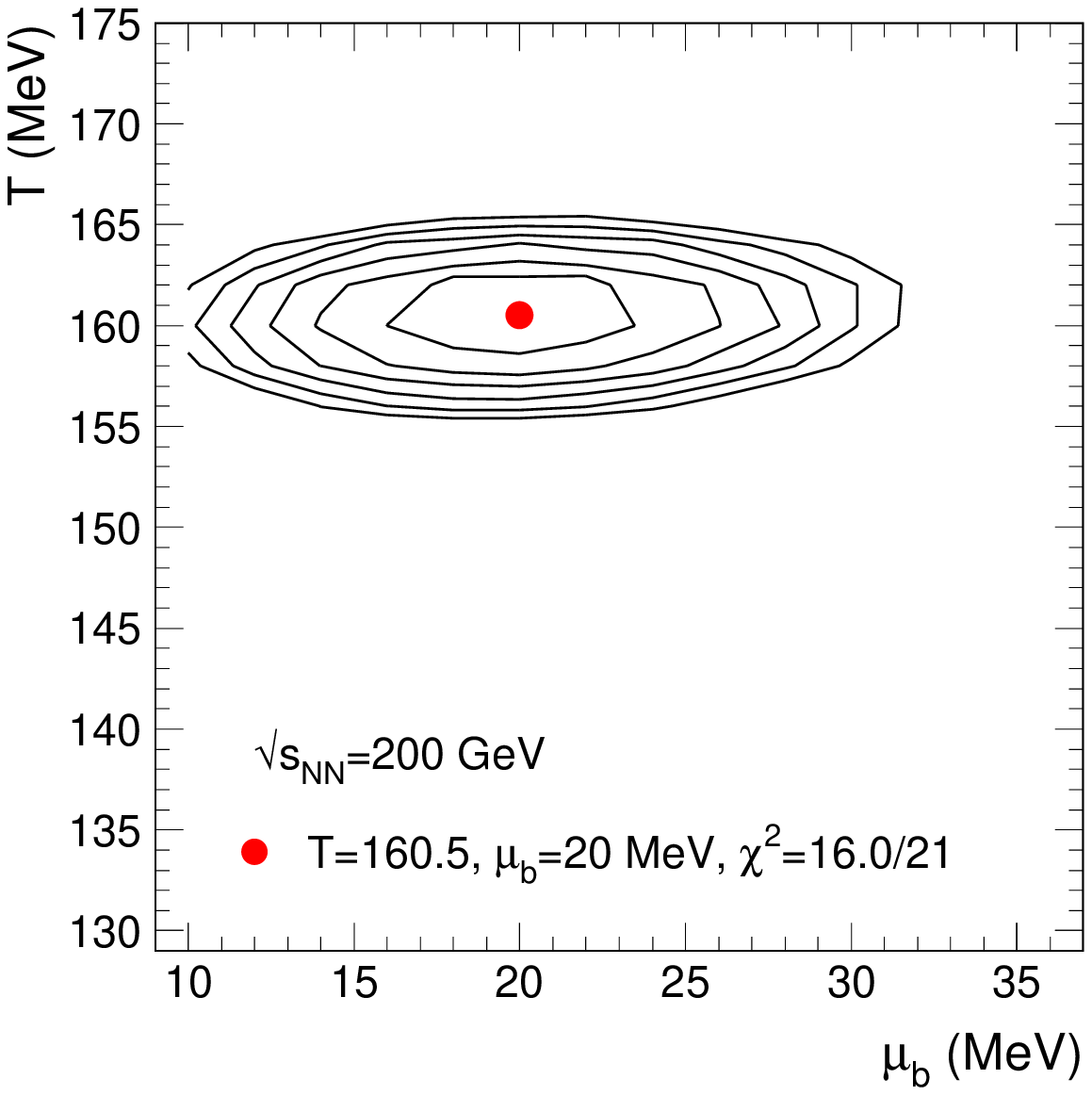}
\end{minipage} \end{tabular}
\caption{Distributions of $\chi^2$ at $\sqrt{s_{NN}}$=200 GeV for the
two cases in Fig.~\ref{fig8a}: 
a combined fit of all available data (left panel) and excluding the ratios
$\bar{p}/\pi^-$ and $\phi/K^-$ from PHENIX (right panel).} 
\label{fig8b}
\end{figure}

Obviously, the outcome of the fit (lower $T$) for case i) is 
determined by two contributions: 
a) the 6\% systematic error assigned by PHENIX to the ratio $\bar{p}/\pi^-$,
which is much smaller than the values 14\% and 18\% which we 
calculate from the systematic errors of the absolute yields 
in case of BRAHMS and STAR data, respectively;
b) the ratio $\phi/K^-$ from PHENIX \cite{p200phi}, which is much lower 
compared to the value from STAR \cite{s200phi}.
Using all data from STAR (without resonances) and the $d/p$ and 
$\bar{d}/\bar{p}$ ratios from PHENIX gives
$T=160.5\pm2$ MeV, $\mu_b=15\pm7$ MeV, with $\chi^2/N_{df}=11.6/13$
(from $\delta^2$ minimization: $T$=166 MeV, $\mu_b$=22 MeV, $\delta^2$=0.17).
Taking into account that, for all cases, the $\delta^2$ 
minimization results lead to $T>$160 MeV, and despite the fact that the 
$\chi^2/N_{df}$ is reasonable even for the global fit, we consider the fit 
without the ratios $\bar{p}/\pi^-$ and $\phi/K^-$ from PHENIX (case ii) 
as the nominal case for $\sqrt{s_{NN}}$=200 GeV. The systematic errors 
are then estimated as the differences to the case when all measured 
hadrons (but no resonances) are included in the fit and to the values 
from the $\delta^2$ minimization.

The experimental ratios and the best fit calculations for the two  
cases, i) and ii) are shown in Fig.~\ref{fig8a}.
Their respective $\chi^2$ distributions are shown in Fig.~\ref{fig8b}.
We note that, comparing the resonances, only the $\Lambda^*$ ratio
is clearly deviating from the model fits in our case, at variance with other
claims in this respect \cite{s200r}.
Despite this deviation, which needs further support from the experimental
side, our results on resonances do not point to a sizeable hadronic 
rescattering effect after chemical freeze-out \cite{s200r}.

\begin{figure}[hbt]
\hspace{-.7cm}
\begin{tabular}{lr} \begin{minipage}{.49\textwidth}
\centering\includegraphics[width=1.12\textwidth]{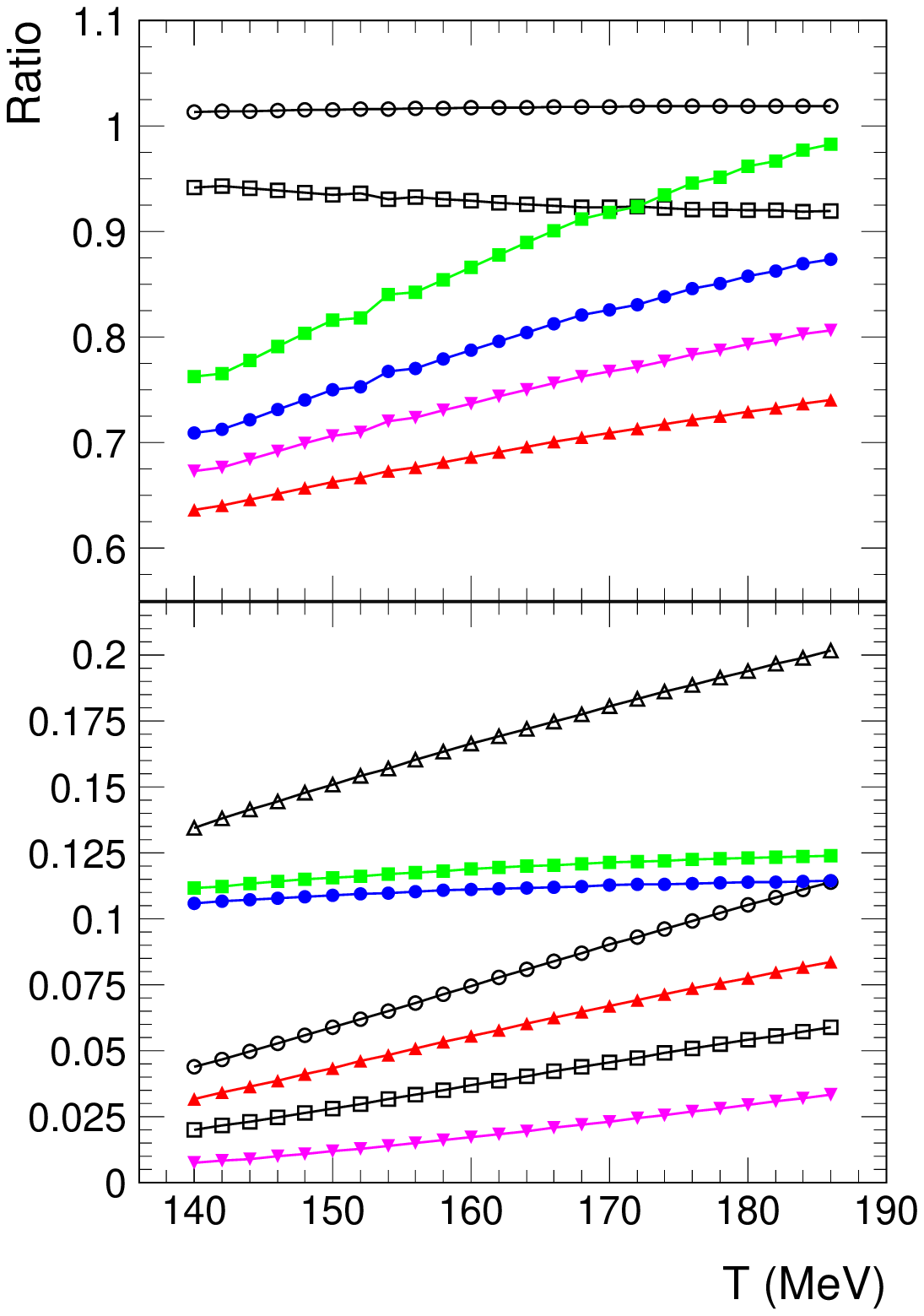}
\end{minipage} &\begin{minipage}{.49\textwidth}
\hspace{-.7cm}\includegraphics[width=1.12\textwidth]{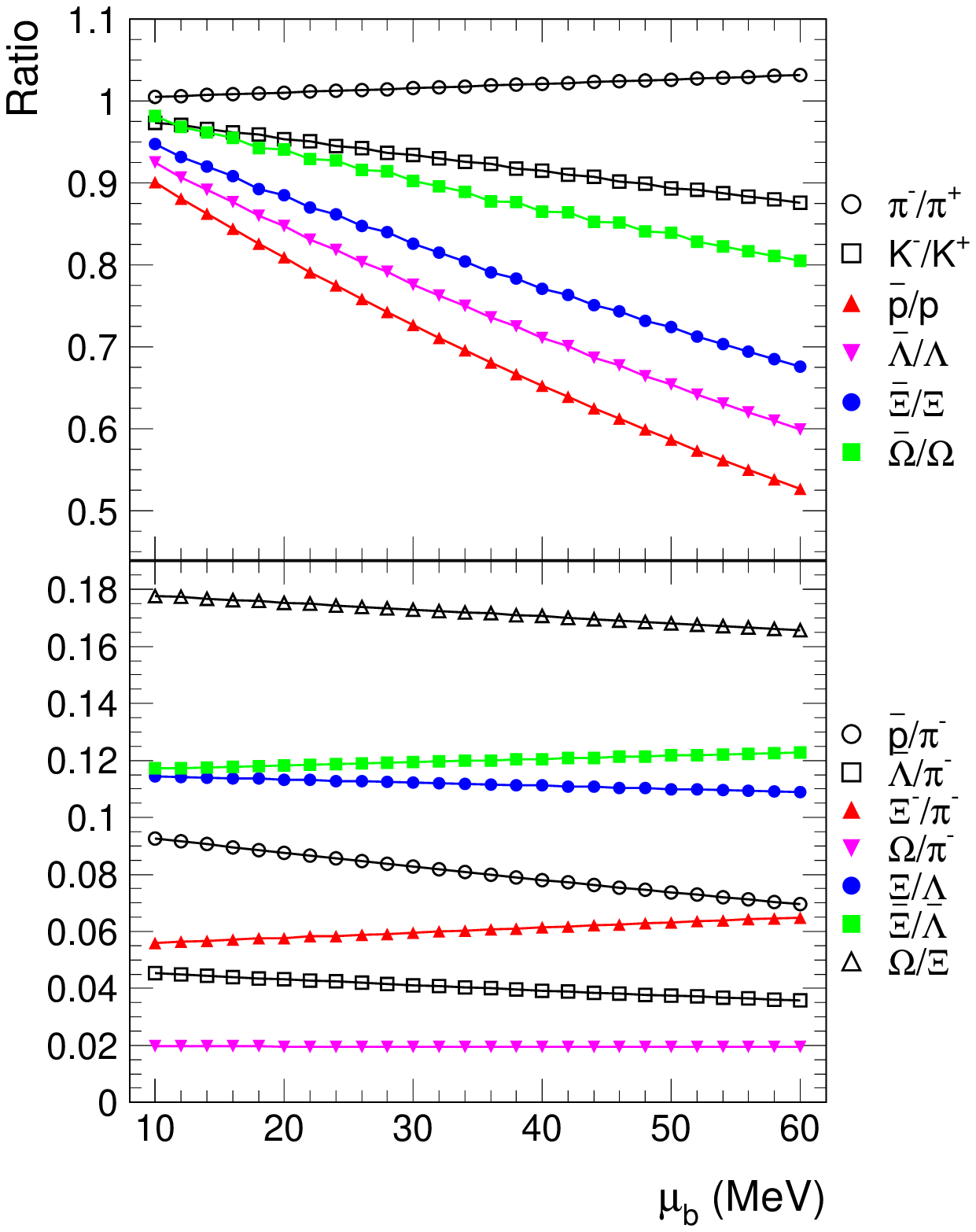}
\end{minipage} \end{tabular}
\caption{The dependence of various hadron ratios on $T$ and $\mu_b$
for the range of parameters relevant for the RHIC energies.
Note that the ratios $\Xi^-/\pi^-$ and $\Omega^-/\pi^-$ are scaled up
by a factor 10.} 
\label{fig9}
\end{figure}

In Fig.~\ref{fig9} we show the sensitivity of various hadron ratios
on $T$ and $\mu_b$ for the range of values relevant for the RHIC energies
\cite{pbm3}.
The importance of particular ratios in constraining $T$ and/or $\mu_b$
is evident. The antihyperon to hyperon ratios as well as the ratios of heavy
hadrons to pions are of special value in this respect.

We note that thermal models have also been used to describe hadron production 
in e$^+$e$^-$ and hadron-hadron collisions \cite{bec0,bec3}, 
leading to temperature parameters in the range 160-170 MeV. 
We mention that our model does not fit well the e$^+$e$^-$ data.
Our fits to the data at $\sqrt{s_{NN}}$=29 GeV compiled in ref. \cite{bec3}
were performed constraining $\mu_b$=0 and with three free parameters: 
the temperature, $T$, the strangeness suppression factor, $\gamma_S$, and 
the volume, $V$ (used to calculate the canonical suppression factor). 
The fits yield $T=154\pm 3$ MeV, $\gamma_S=0.74\pm 0.06$ and 
$V=42\pm 14$ fm$^3$, with a $\chi^2/N_{df}$=56/11.
Obviously, the fit quality is very poor.
For comparison, in the analysis of ref. \cite{bec3} $T=159\pm 2.6$ MeV, 
$\gamma_S=0.71\pm 0.05$, $\chi^2/N_{df}$=29.3/12.
Both $T$ and $\mu_b$ are compatible within the errors, but
our fit shows a much worse quality.
This may be due to the approximations concerning canonical strangeness 
suppression which are employed in the present approach.
However, even with the appropriate canonical treatment as done in 
ref.~\cite{bec3}, the quality of the fits (both in e$^+$e$^-$ and in
hadron-hadron collisions) is not good, raising questions about the 
applicability of the statistical model for elementary collisions.

\section{Energy dependence of the thermal parameters} 
\label{edep}

Our results for the best fits are summarized in Table~\ref{tab2}.
The upper part in Table~\ref{tab2} shows the results of the fits using 
mid-rapidity data, the lower part is for the data integrated over 4$\pi$.
One notices that in general $\chi^2/N_{df}$ is close to (and often below) 
unity, being significantly above unity only for the SPS energies.
The fits 4$\pi$-integrated data result in comparable temperature values as 
for the mid-rapidity data in case of the AGS energies and sizably lower values 
for the SPS energies. For the higher SPS energies, the fits of 4$\pi$ data 
are characterized by even poorer $\chi^2$ values. We would like to stress 
that, although we have performed the fits of 4$\pi$ data up to the top SPS 
energy (in order to compare to similar analyses in other works), we consider 
the extracted thermal parameters in this case adequate only up to the energy 
of 40 AGeV.
Also included in Table~\ref{tab2} are the results of the $\delta^2$ 
minimization, which differ sometimes from those obtained from the
$\chi^2$ minimization.
The central values used in the following are those from the $\chi^2$ 
minimization. 

\begin{table}[hbt]
\caption{Summary of the results of the thermal fits using ratios. 
The upper part shows the results obtained using mid-rapidity data,
the lower part is for data integrated over 4$\pi$.
The first three and next three columns show the results of the $\chi^2$ 
(with their 1~$\sigma$ error) and $\delta^2$ minimization, respectively.
The quantities $\Delta T^{syst}$ and $\Delta \mu_b^{syst}$ (in MeV) are 
the assigned systematic errors (see text).}
\label{tab2}
\begin{tabular}{l|ccc|ccc|cc}
$\sqrt{s_{NN}}$ (GeV) & $T$ (MeV) & $\mu_b$ (MeV)  & $\chi^2/N_{df}$ &
$T$ & $\mu_b$   & $\delta^2$ & $\Delta T^{syst}$ &$\Delta \mu_b^{syst}$\\ 
\hline
2.70 & 64$\pm$1   & 760$\pm$20   & 1.0/2   & 64  & 760 & 0.012 &+10,-2 &23\\
3.32 & 78$\pm$2   & 670$\pm$16   & 0.43/3 & 78  & 670 & 0.005 &+12,-3 &20\\
3.84 & 86$\pm$2   & 615$\pm$8    & 1.15/4  & 86  & 615 & 0.018 &+13,-3 &19\\ 
4.30 & 93$\pm$3   & 580$\pm$11   & 1.14/3 & 94  & 580 & 0.021 &+14,-3 &17\\
4.85 & 124$\pm$3  & 537$\pm$10   & 6.9/7   & 130 & 540 & 0.34 & 7  &12\\
8.76 &156$^{+4}_{-3}$ & 403$^{+18}_{-14}$& 15.4/13 & 166 & 434 & 0.61 &11 &32\\
12.3 &154$\pm$6   & 298$\pm$19    & 23.4/5  & 152 & 271 & 0.23  &12  &33\\
17.3 &160$\pm$5 & 240$\pm$18  & 55/22 & 172 & 243  &0.86 & +19,-12 & 24 \\
130  &165.5$\pm$5.5 &38$\pm$11   & 4.1/11  & 168  & 42  & 0.12 &6 &11\\
200 &160.5 $\pm$2 & 20$\pm$4 & 16.0/21  & 166 & 22 & 0.19 &6 & 7\\ 
\hline
3.84 & 86.5$\pm$2   & 631$\pm$11  & 6.7/4  & 88  & 640 & 0.14 &12 &19\\ 
4.85 & 115$\pm$4  & 535$\pm$11  & 9.0/5   & 116 & 530 & 0.15 &5  &15\\
6.27  & 132.5$\pm$3 & 518$\pm$15 & 4.4/4 & 136 & 516 & 0.13 &5 &16\\
7.62  & 138$\pm$3 & 459$\pm$12   & 22/4 & 140 & 449 & 0.22 &7 &19 \\
8.76  & 135$\pm$5 & 380$\pm$19   & 32/5 & 142 & 387 & 0.21 &9 &22\\
12.3 & 146.5$\pm$3.3 & 316$\pm$17  & 30/4  & 144 & 277 & 0.13  &4  &39\\
17.3 &151$\pm$9 & 256$\pm$41  & 36/9 & 170 & 285  &0.39 & +18,-5 & 41 \\

\end{tabular}
\end{table}

The "systematic" errors $\Delta T^{syst}$ and $\Delta \mu_b^{syst}$
shown in Table~\ref{tab2} are obtained by quadratically adding 
the 1~$\sigma$ error from the $\chi^2$ minimization with the difference 
between the values obtained by $\chi^2$ and $\delta^2$ minimization .
Whenever redundant sets of data are available from different experiments, 
which is the case at top SPS and top RHIC energies, our "nominal" values 
are those resulting from the combined fit of all data.
We have assigned as systematic error the quadratic sum of 1~$\sigma$ fit
errors and the deviation between the ($T$,$\mu_b$) values of this case 
and those resulting from the fit of the data which results in the smallest
value of $\chi^2$ (see previous section and Table~\ref{tab2}).
For the SPS energy of 80 AGeV ($\sqrt{s_{NN}}$=12.4 GeV) the systematic 
errors are evaluated based on the sensitivity of the fit on the inclusion
of different hadron ratios.
As discussed in the previous section, for lower AGS beam energies
of 2-8 AGeV, we have estimated 14\% and 3\% systematic errors for 
$T$ and $\mu_b$, respectively, based on the sensitivity of the fit
using ratios at top AGS energy. 
For the AGS energy of 6 AGeV, we have used the available 4$\pi$ yields of 
pions \cite{e895pi}, protons \cite{e895p}, $\Lambda$ and $K^0_S$ \cite{e895l}.
We note that, at this energy, there are no $K^+$ and $K^-$ yields measured 
in full phase space. Based on a comparison of the ratio 
$K^-/K^+$ at mid-rapidity and in 4$\pi$ at top AGS energy \cite{e802k},
we have assumed that at 6 AGeV the ratio $K^-/K^+$ in 4$\pi$ is 10\% smaller
than at mid-rapidity. At the top AGS energy, some of the 4$\pi$ yields do not
correspond to complete coverage \cite{e802k}; also, the ratio $\bar{p}/p$ is 
obtained with a rather large extrapolation.

\begin{figure}[hbt]
\begin{tabular}{lr} \begin{minipage}{.49\textwidth}
\centering\includegraphics[width=1.15\textwidth]{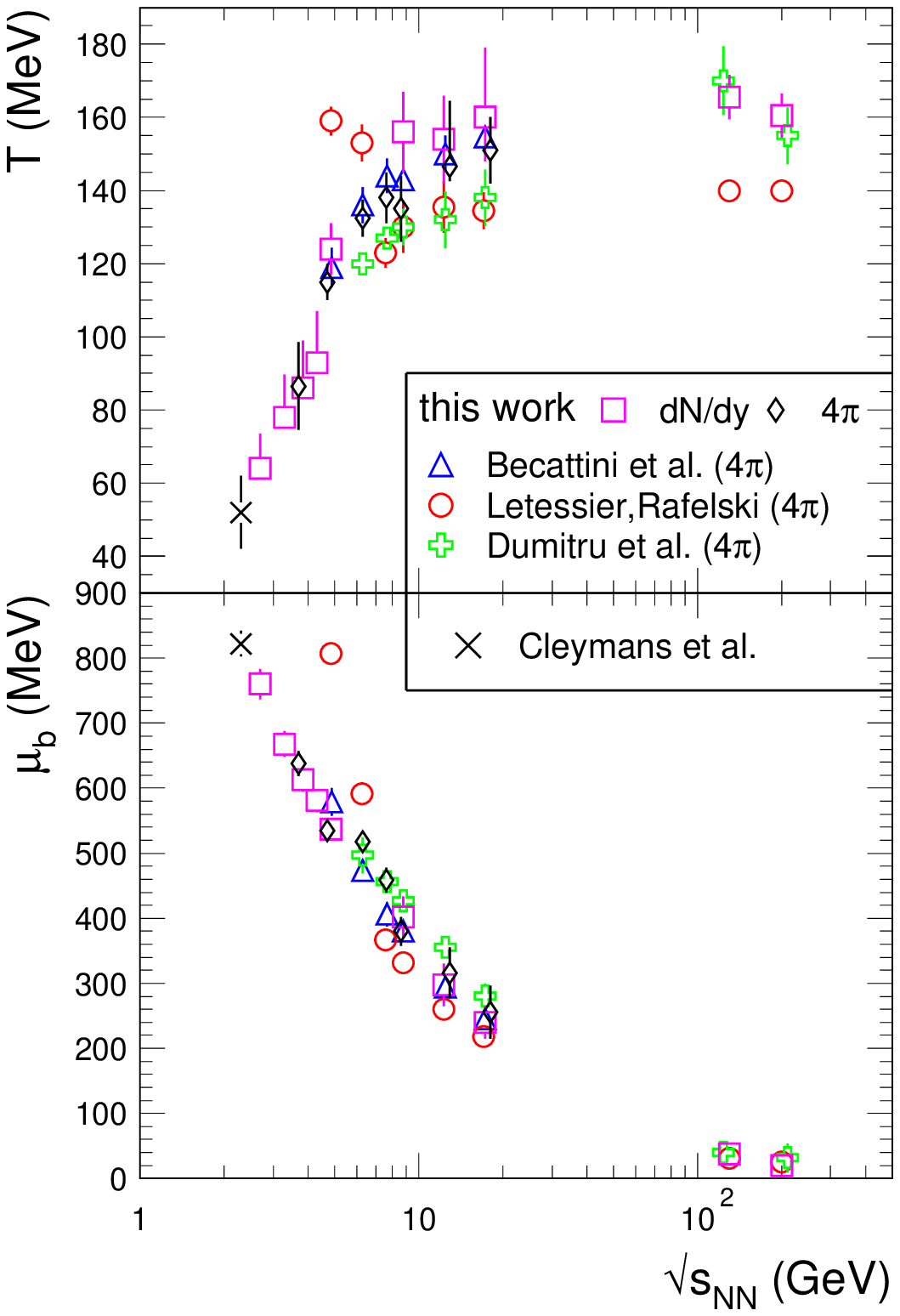}
\end{minipage} &\begin{minipage}{.49\textwidth}
\centering\includegraphics[width=1.15\textwidth]{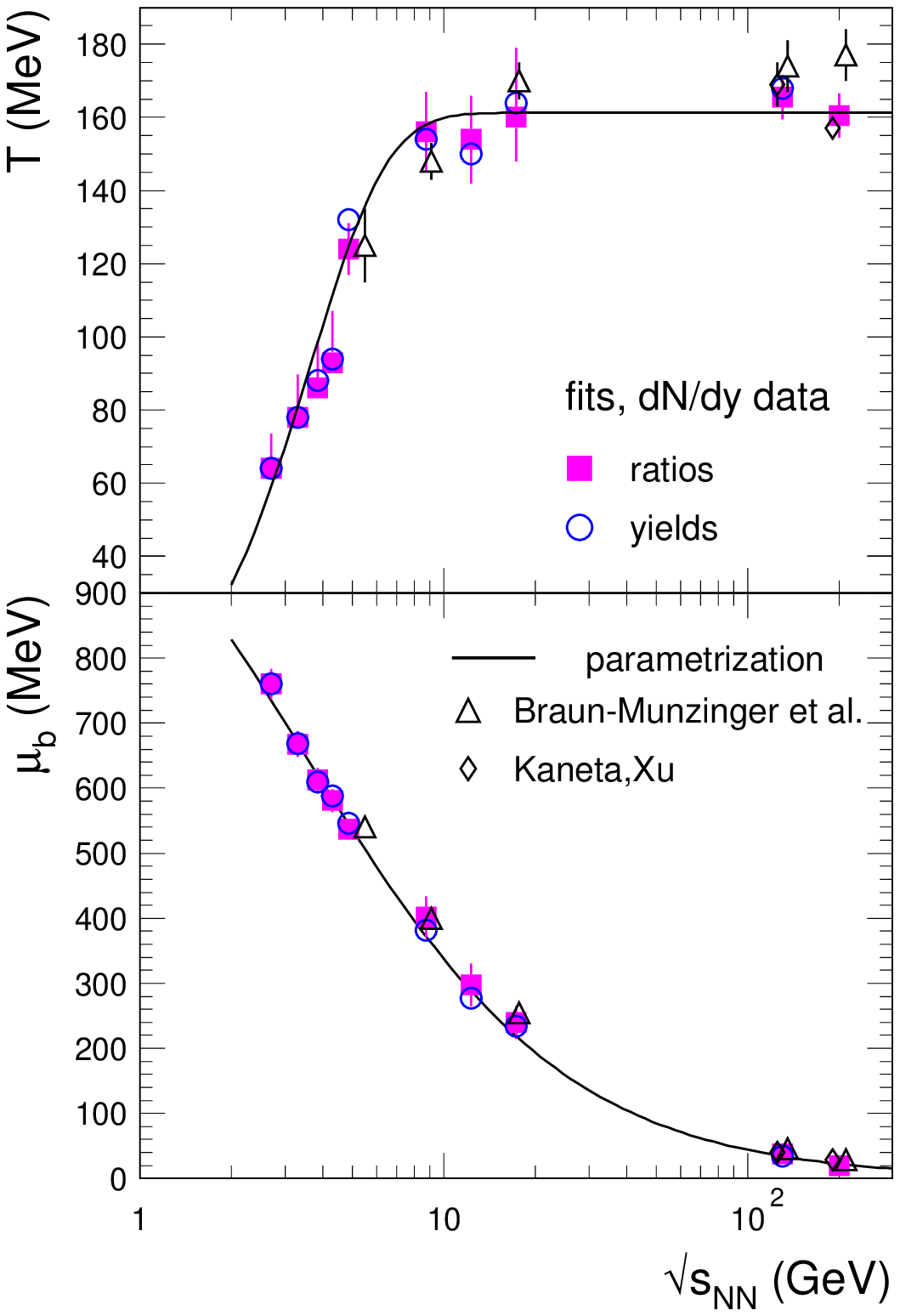}
\end{minipage} \end{tabular} 
\caption{The energy dependence of temperature and baryon chemical potential.
Left panel: the results of the present work are compared to the values
obtained in earlier studies (see text);
right panel: the results of our fits using $\ud N/\ud y$ data, both
with ratios and yields. 
The lines are parametrizations for $T$ and $\mu_b$ (see text).} 
\label{fig10}
\end{figure}

In Fig.~\ref{fig10} we show the energy dependence of $T$ and $\mu_b$ 
extracted from our thermal fits. 
The temperature $T$ exhibits a sharp rise up to $\sqrt{s_{NN}}\simeq$7-8 GeV,
while $\mu_b$ sharply decreases all the way up to RHIC energies.
Our results are compared with results from other studies 
\cite{rev,cle1,kan,bec1,raf1,zsch}, which exhibit similar trends. 
One notable exception are the results of Letessier and Rafelski \cite{raf1};
in this case, the observed difference may arise from the usage in their
work of seven free parameters, including, besides strangeness fugacity 
and suppression factor ($\gamma_S$), 
a light quark occupancy factor ($\gamma_q$) and an isospin fugacity.
The non-monotonic change in $\gamma_q$ and $\gamma_S$ as a function of
energy determines the temperature extracted in ref. \cite{raf1}.
To alleviate the poor quality of the fits at SPS energies, the model of
Dumitru et al. \cite{zsch}, introducing an inhomogeneous freeze-out scenario,
goes beyond other thermal models.

Not surprisingly, our results support those obtained earlier within 
the same model implementation \cite{rev}.
In detail, there are differences between our results and the rest of
the other results. Our values for the analysis of the 4$\pi$ data are 
significantly different from those of Becattini et al. \cite{bec1},
obtained within a model employing the strangeness suppression factor.
A remarkable agreement between our results and the analyses in ref.
\cite{pbm3,kan} is seen in case of the RHIC data, in particular at 
$\sqrt{s_{NN}}$=130 GeV. The results at $\sqrt{s_{NN}}$=200 GeV 
\cite{kan,star} are also in agreement with our values.
The higher $T$ value at $\sqrt{s_{NN}}$=200 GeV in ref.~\cite{pbm3}
is due to preliminary data.

We have parametrized our results from the fits of mid-rapidity data 
(left panel in Fig.~\ref{fig10} and Table~\ref{tab2}) as a function 
of $\sqrt{s_{NN}}$ (in GeV) with the following expressions:
\be
T \mathrm{[MeV]}=T_{lim}\left(1-\frac{1}{0.7+(\exp(\sqrt{s_{NN}}\mathrm{(GeV)})-2.9)/1.5}\right)
\label{pt}
\ee
\be
\mu_b \mathrm{[MeV]}=\frac{a}{1+b\sqrt{s_{NN}}\mathrm{(GeV)}}, 
\label{pm}
\ee
where the parameters $a=1303\pm$120 MeV and $b=0.286\pm$0.049 GeV$^{-1}$ 
are the results of a fit ($\chi^2/N_{df}$=0.48/8).
Our $\mu_b$ parametrization is the one proposed in ref. \cite{pbm4}, but
with different parameters to better fit the newly obtained $\mu_b$ values
of the present analysis.
In Eq.~\ref{pt} we assume a "limiting" temperature $T_{lim}$, which was 
obtained by fitting the five points for the highest energies (SPS and RHIC). 
The result of the fit is $T_{lim}=161\pm4$ MeV, with $\chi^2/N_{df}$=0.3/3.

\vspace{-.7cm}
\begin{figure}[hbt]
\centering\includegraphics[width=.75\textwidth]{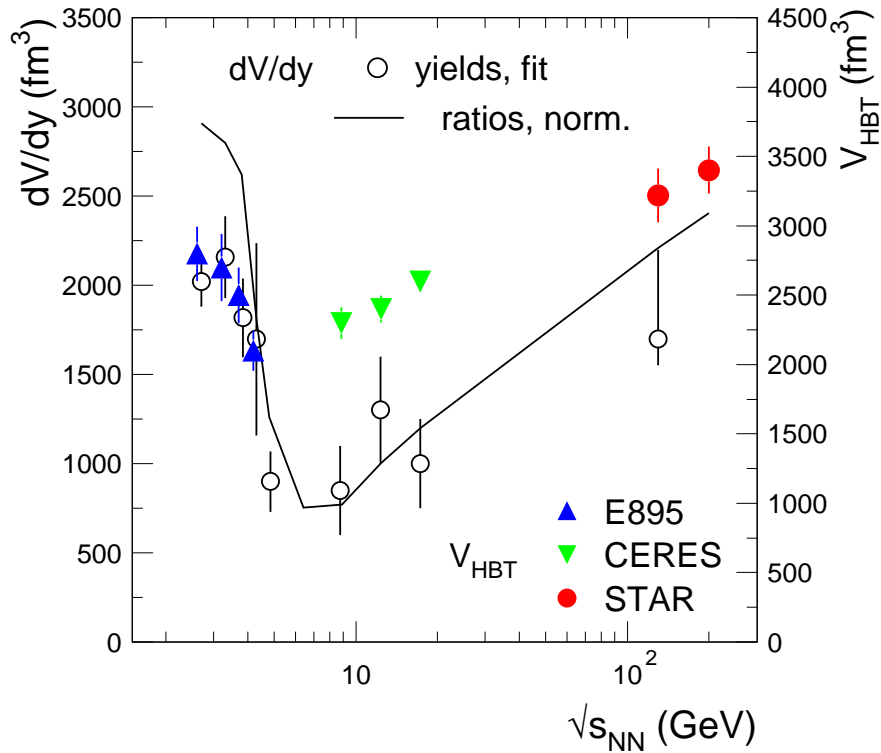}
\caption{Energy dependence of the volume for central collisions 
($N_{part}$=350). The chemical freeze-out volume for one unit of rapidity, 
$\ud V/\ud y$, is compared to the kinetic freeze-out volume from 
HBT measurements, $V_{HBT}$ \cite{cer}. Note that the scales are different 
for the two observables.}
\label{figv}
\end{figure}

We now briefly turn to another interesting parameter which is (either
implicitly or explicitly) determined in the course of thermal model
analyses. 
The volume at chemical freeze-out (corresponding to a slice of one unit 
of rapidity, $\ud V/\ud y$) is shown in Fig.~\ref{figv} as a function of 
energy.
The values extracted directly from the fits of particle yields (see Appendix) 
are compared to the values obtained by dividing measured charged particle 
yields with calculated densities (based on the above parametrization of 
$T$ and $\mu_b$; note that for the AGS energies of 2-8 AGeV the values of
$T$ corresponding to the upper limit of the systematic error were used 
instead). As expected, the two methods give identical results, with the 
exception of a small discrepancy for the lowest energies. 
The chemical freeze-out volume is compared to the kinetic freeze-out volume 
extracted from Hanbury Brown and Twiss (HBT) measurements, $V_{HBT}$ 
\cite{cer}.
We note here (see also the Appendix) that to determine a volume from thermal
model analyses one needs to know absolute densities. This implies an explicit
dependence of the volume on whether or not excluded volumes are implemented 
in the calculations, and, if included, on details of the implementation. 
We follow here the procedure developed in \cite{pbm2}. 

While the bias towards unphysically large volumes seen at the energies 
of 2-8 AGeV (see Appendix) is not completely understood, it is clear 
that the volume at kinetic freeze-out should not be exceeded.
Note that, to relate quantitatively the magnitude of the rapidity density 
of the chemical freeze-out volume $\ud V/\ud y$  to the volume at kinetic 
freeze-out determined from HBT measurements, one needs to map rapidity 
onto space (see the discussion in \cite{cer}). 
Here we are mainly interested in the energy dependence of the two volumes.
It appears that the volume at chemical freeze-out does exhibit a similar 
non-monotonic behavior as the volume at kinetic freeze-out, a remarkable 
result considering that the latter is determined by a completely different 
procedure. 
A minimum is observed for the top AGS energy, followed by a logarithmic 
increase as a function of $\sqrt{s_{NN}}$.

\section{The energy dependence of hadron ratios} 
\label{sdep}

Using the parametrizations of the chemical freeze-out parameters derived
from the fits of experimental data at mid-rapidity, as shown in Eq.~\ref{pt} 
and \ref{pm}, we calculate the energy dependence of various hadron ratios.
No contribution from weak decay feed-down is included in the model 
calculations as the published data are mostly corrected in this respect. 
Whenever not the case, the measurements are scaled appropriately (see below).
The finite widths of resonances are taken into account in the model.
This contributes about 10\% of pions at low energies and 4\% at higher 
energies.
Within the smoothing hypothesis implied by the parametrizations of
Eq.~\ref{pt} and \ref{pm}, the model has interpolative and extrapolative 
predictive power. 
In this sense, we can provide quantitative predictions for LHC energy,
as well as for the low RHIC energy of 62.4 GeV, for which the experimental
data are becoming available \cite{s62}. 
This also applies to the low energy SPS data \cite{fri}.

\begin{figure}[hbt]
\begin{tabular}{lr} \begin{minipage}{.49\textwidth}
\centering\includegraphics[width=1.1\textwidth]{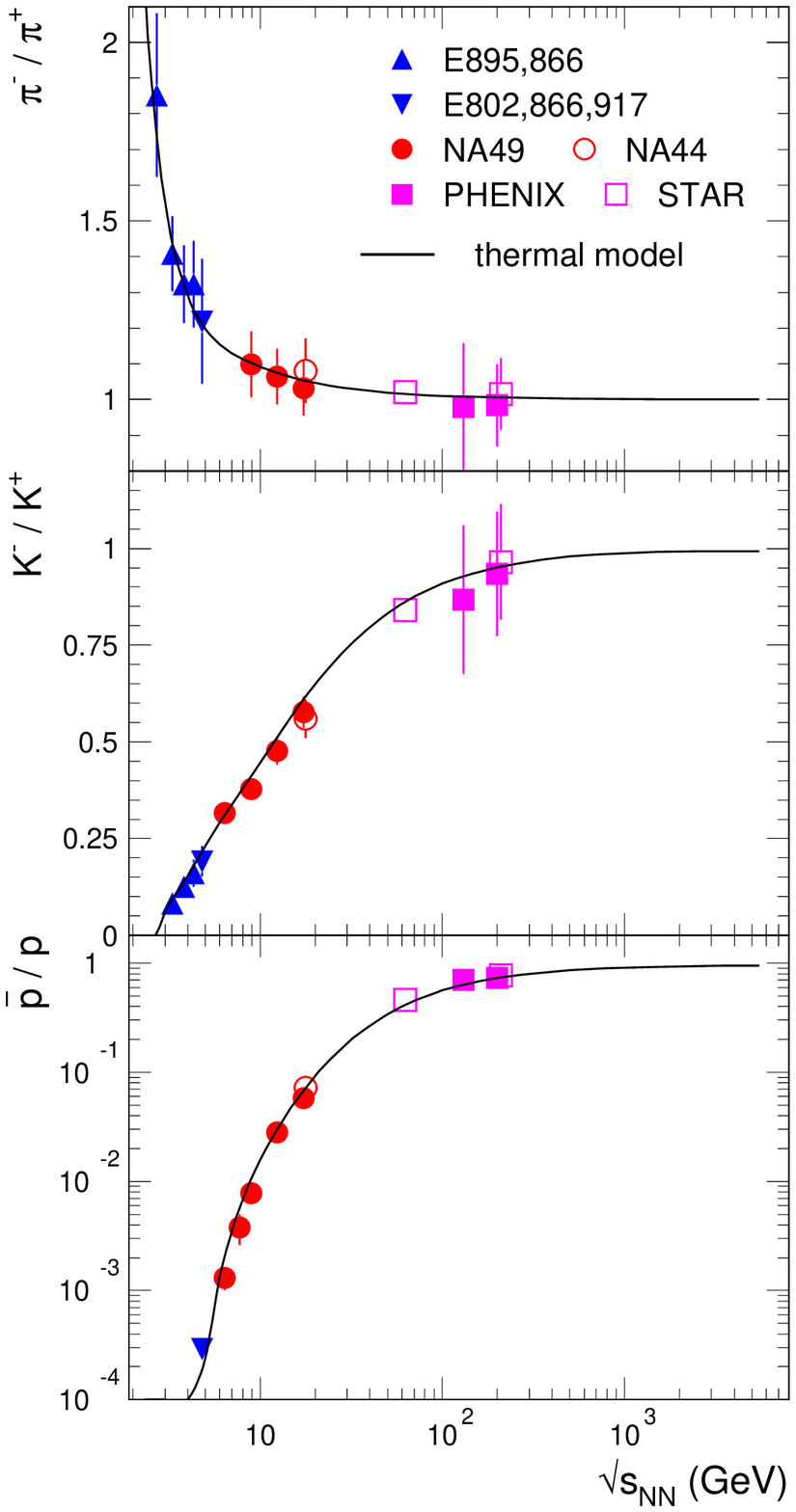}
\end{minipage} &\begin{minipage}{.49\textwidth}
\centering\includegraphics[width=1.1\textwidth]{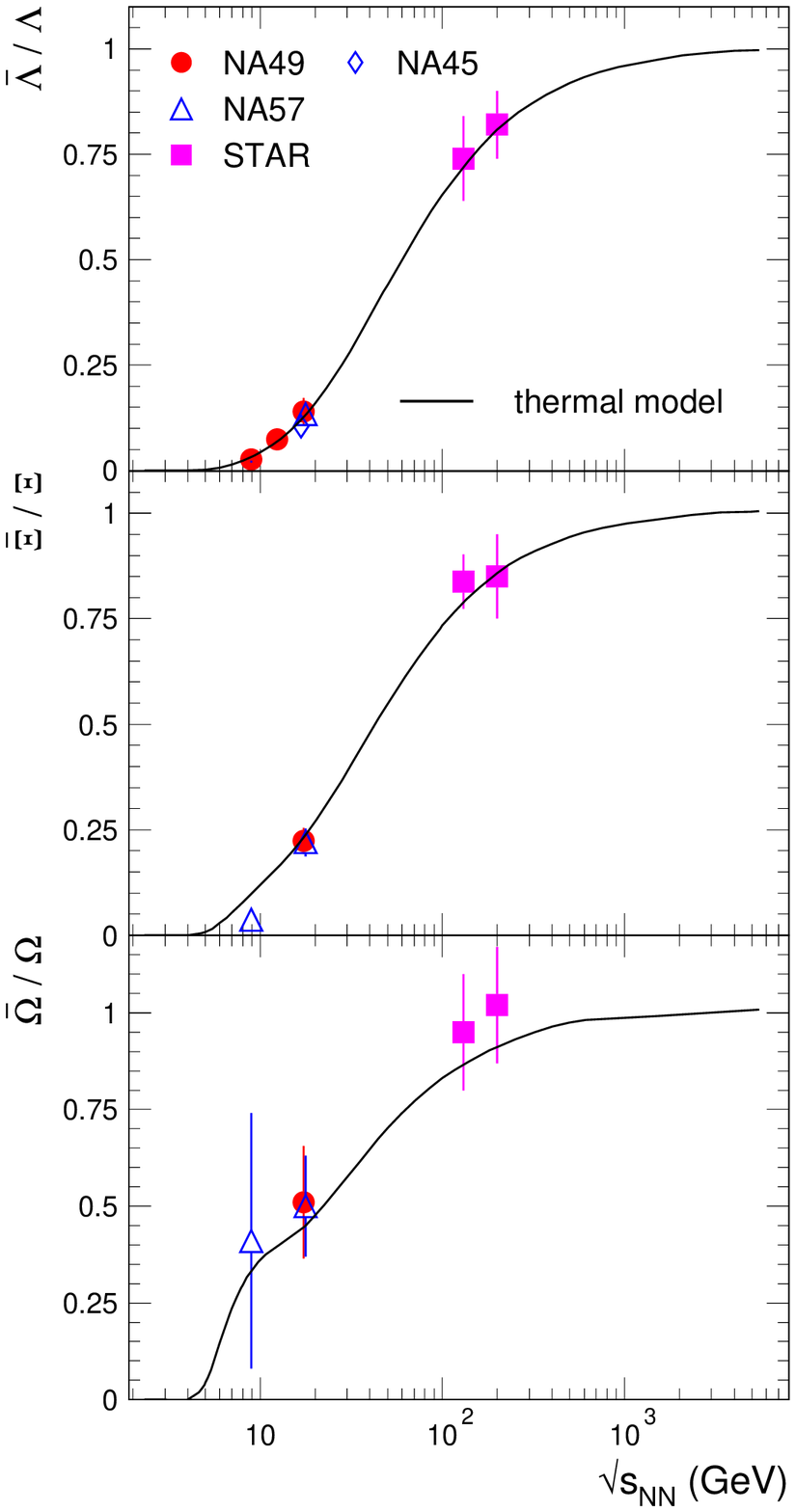}
\end{minipage} 
\end{tabular}
\caption{The energy dependence of antiparticle/particle ratios.}
\label{fig13}
\end{figure}

In Fig.~\ref{fig13} we show the energy dependence of antiparticle 
to particle ratios.
Calculations employing the global $T(\sqrt{s_{NN}})$ and 
$\mu_b(\sqrt{s_{NN}})$ from Eq.~\ref{pt} and \ref{pm}
reproduce very well all these experimental ratios, including
the preliminary values for $\sqrt{s_{NN}}$=62.4 GeV \cite{s62}.
The $\pi^-/\pi^+$, $K^-/K^+$, and $\bar{p}/p$ ratios illustrate the 
decreasing importance of isospin, the increasing importance of strangeness
production and the dramatic evolution as a function of energy of the balance
between the incoming and newly produced protons. The ratio $\bar{p}/p$ has 
the steepest dependence on energy. From the lowest measured value of 
3$\cdot10^{-4}$ at AGS, it reaches a value of about 0.7 at RHIC and 
is predicted to reach 0.95 at LHC.
The dominance of the initial nucleons at low energies changes to the newly
created hadrons at RHIC and beyond. 
At LHC, it is expected that the fireball will consist practically completely
of created hadrons ($\mu_b\simeq$1 MeV).

The energy dependence of antihyperon/hyperon ratios (right panel in 
Fig.~\ref{fig13}) follows a mass hierarchy: the saturation value of 1 
is achieved the earlier the more massive the hyperon species. 
This is a fingerprint of the preferred abundance of hyperons over antihyperons
containing 2 ($\Lambda$) or 1 ($\Xi$) light valence quarks, which may be 
remnants from the incoming nucleons, while in case of $\Omega$, all valence
quarks are newly produced.\footnote{
The small structure of the excitation function at $\sqrt{s_{NN}}$=10-20 GeV,
most prominently seen in Fig.~\ref{fig13} for the ratio $\bar{\Omega}/\Omega$,
but also for $\bar{\Xi}/\Xi$, is an artifact of our parametrization of $\mu_b$.
}
Remarkably, as already seen in Fig.~\ref{fig6}, there is a very good
agreement between the NA49 \cite{na49l,na49xi,na49o} and the NA57 \cite{na57}
data. 
The ratio $\bar{\Lambda}/\Lambda$ from the NA45 (CERES) experiment 
\cite{na45l} is as well in agreement.
The generally poor quality of thermal fits at SPS energies is not caused by 
ratios involving (multi)strange baryons.

\begin{figure}[htb]
\begin{tabular}{lr} \begin{minipage}{.49\textwidth}
\centering\includegraphics[width=1.1\textwidth]{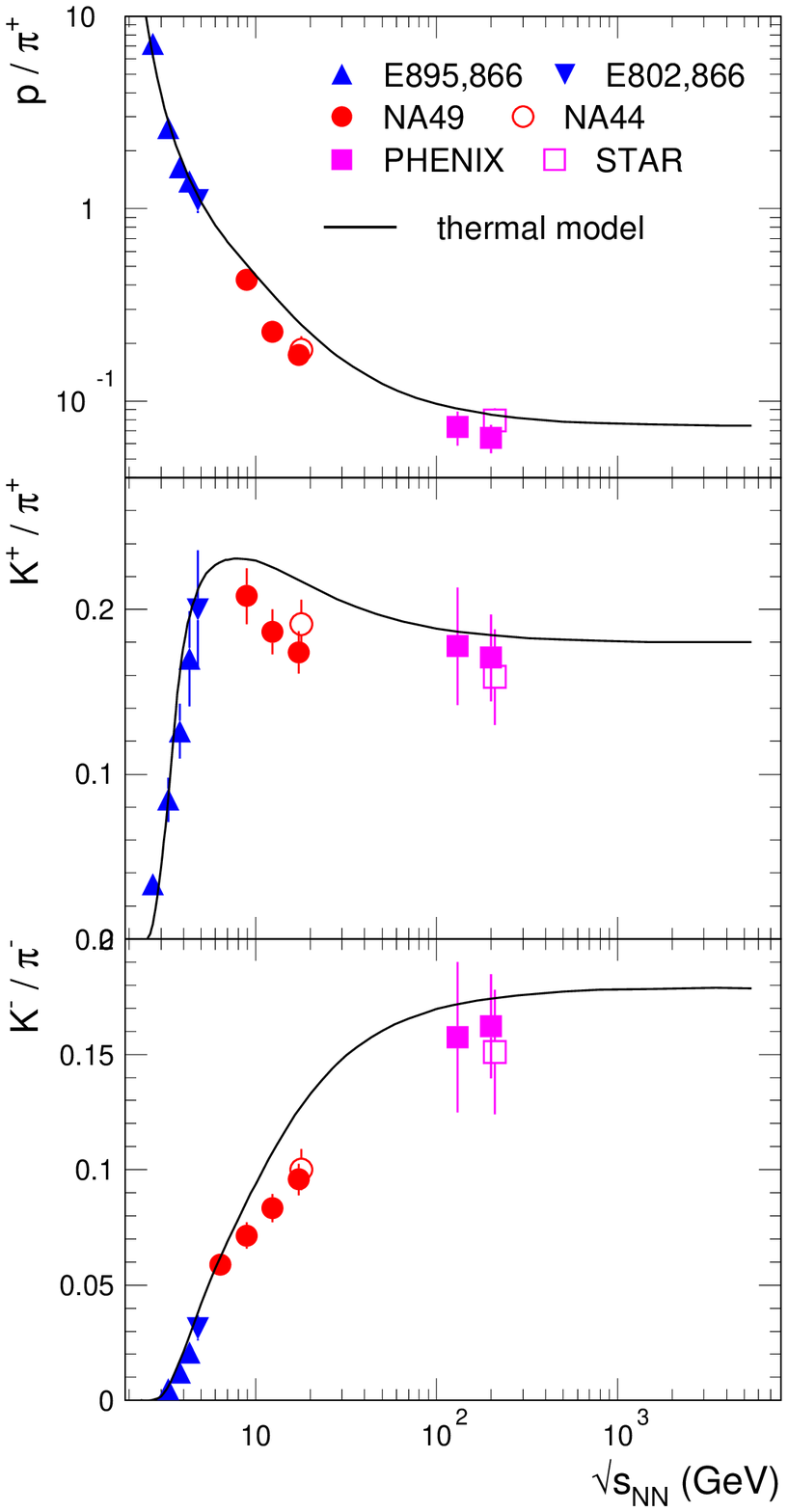}
\end{minipage} &\begin{minipage}{.49\textwidth}
\centering\includegraphics[width=1.1\textwidth]{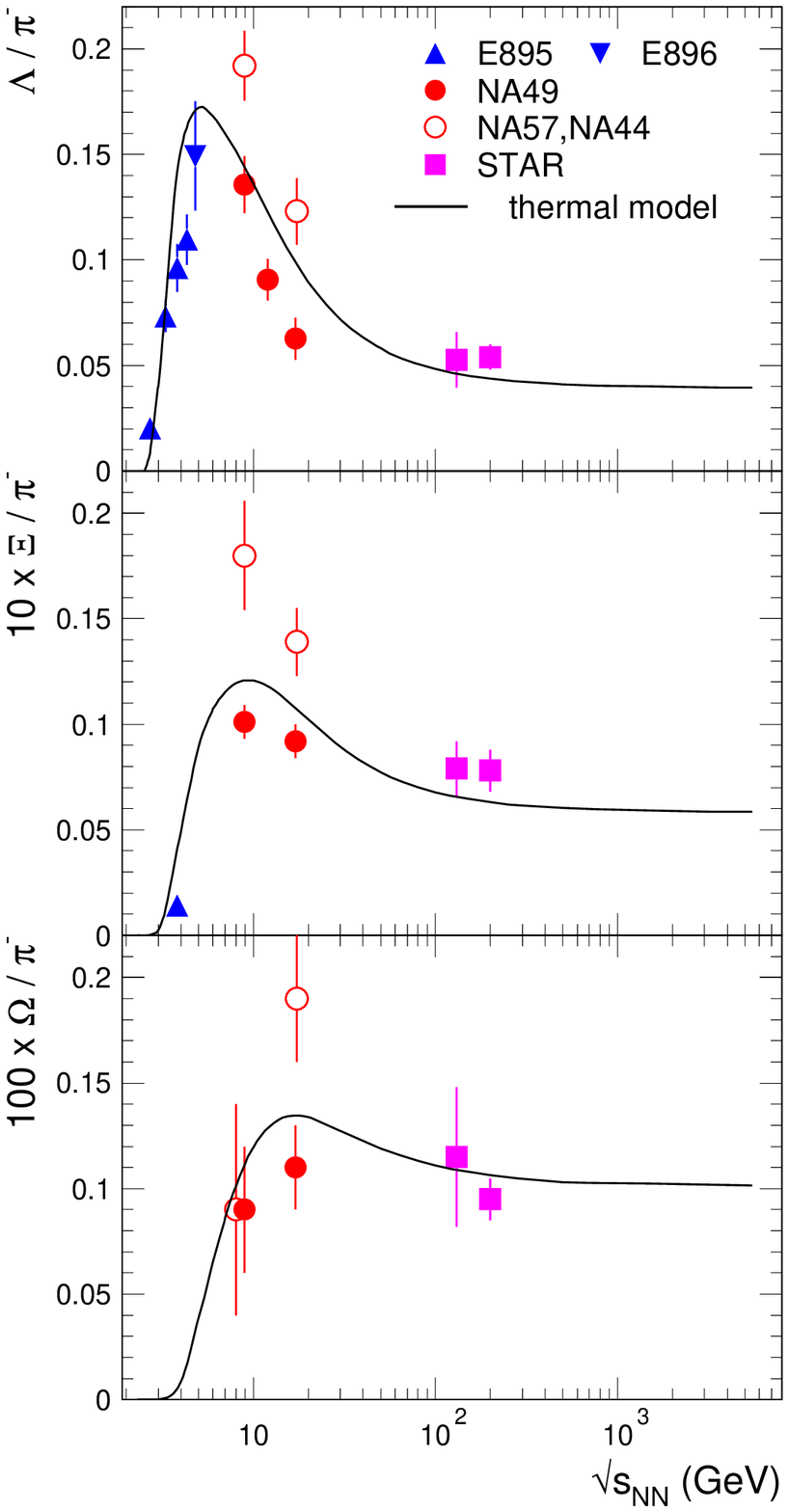}
\end{minipage} \end{tabular}
\caption{The energy dependence of hadron yields relative to pions. 
In the right panel, note the scaling factors of 10 and 100 for the 
$\Xi^-/\pi^-$ and $\Omega^-/\pi^-$ ratios, respectively.}
\label{fig14}
\end{figure}

The comparison of the measured and calculated excitation functions for
hadron abundances with respect to pions is shown in Fig.~\ref{fig14}.
For a consistent comparison to the other measurements and to the calculations,
the measured pion yield at $\sqrt{s_{NN}}$=130 GeV \cite{p130} was scaled 
assuming that 30\% of the weak decays contribution was contained in the
data (see above).
The ratios shown in Fig.~\ref{fig14} have a very different energy 
dependence compared to those in Fig.~\ref{fig13}, reflecting the evolution
of the fireball composition at freeze-out as a function of energy.
The steep decrease of the $p/\pi^+$ ratio directly reflects the decrease 
as a function of energy of stopping of the incoming protons, implying a
decrease of $\mu_b$. 
The increase of pion production also plays a role in this ratio.
Beyond $\sqrt{s_{NN}}\simeq$100 GeV, the flattening is a consequence of
the dominance of newly created baryons. 
The steep variation of the $K^+/\pi^+$ and $K^-/\pi^-$ ratios at the lowest
energies reflects the threshold for strangeness production, determined
in the model by the steep increase of the temperature.
The canonical suppression plays an important role as well.
While the ratio $K^-/\pi^-$ shows a monotonic increase with energy, 
followed by a saturation, essentially determined by the temperature
(as both particles are newly created), the ratio $K^+/\pi^+$, 
which is much discussed as a signature for the onset of QGP 
\cite{gaz,gaz2,bra,vk2},
shows a characteristic broad peak around $\sqrt{s_{NN}}\simeq$8 GeV. 
As the $K^+$ contains a $u$ valence quark, which may come from the initial 
nucleons, its yield is the convolution of two competing contributions as 
a function of energy: 
i) the decreasing net light quark content and 
ii) the increasing production of quark-antiquark pairs.
The peak in the $K^+/\pi^+$ ratio occurs naturally in the thermal model
\cite{pbm4,cle2}, but is broader and has to be seen in the context of other 
strange hadron yields \cite{cle2} (right panel of Fig~\ref{fig14}).
It appears that, at the SPS energies of 80 and 158 AGeV, the yields
of kaons and protons relative to pions are systematically below the 
thermal model predictions.
We note that recent transport model calculations \cite{bra} grossly deviate
from the measured $K^+/\pi^+$ ratio, while earlier results reproduced 
the data well \cite{wan}.
We emphasize the good agreement between the two data sets (NA49 \cite{na49pik}
and NA44 \cite{na44}) available at the top SPS energy.

The energy dependence of the relative hyperon yields, $\Lambda/\pi^-$, 
$\Xi^-/\pi^-$ and $\Omega^-/\pi^-$, shown in the right panel of 
Fig.~\ref{fig14}, reveals the presence of characteristic peak structures,
already noted in \cite{pbm4}.
Their strength and location follows a mass hierarchy, recently discussed 
by Cleymans et al. \cite{cle2}.
The peaks are less pronounced and located at larger energies 
($\sqrt{s_{NN}}\simeq$ 5, 10 and 20 GeV) the more massive the hyperon species.
This results from an interplay between the baryochemical potential
(presence of the light quarks from the initial nucleons) and temperature.
The agreement between the model and the data is good at AGS and RHIC 
energies, while at SPS the situation is more complex. 
The two available sets of data, NA49 \cite{na49l,na49xi,na49o}  
and NA57 \cite{na57}, are in a clear disagreement: generally, the NA49 data
are below the NA57 data. 
As a consequence of our adopted fitting procedure (see above) the model 
prediction is between the two data sets.
We notice that the discrepancy of the data is manifest in the absolute
yields: for instance, $\ud N/\ud y$ of $\Lambda$ at top SPS (5\% central 
collisions) for NA49 and NA57 are 11.0$\pm$1.6 and 18.5$\pm$2.2, 
respectively. 
The disagreement is well beyond the errors (statistical and systematic, 
added in quadrature) quoted by both experiments.

\begin{figure}[hbt]
\begin{tabular}{lr} 
\begin{minipage}{.49\textwidth}
\centering\includegraphics[width=1.17\textwidth]{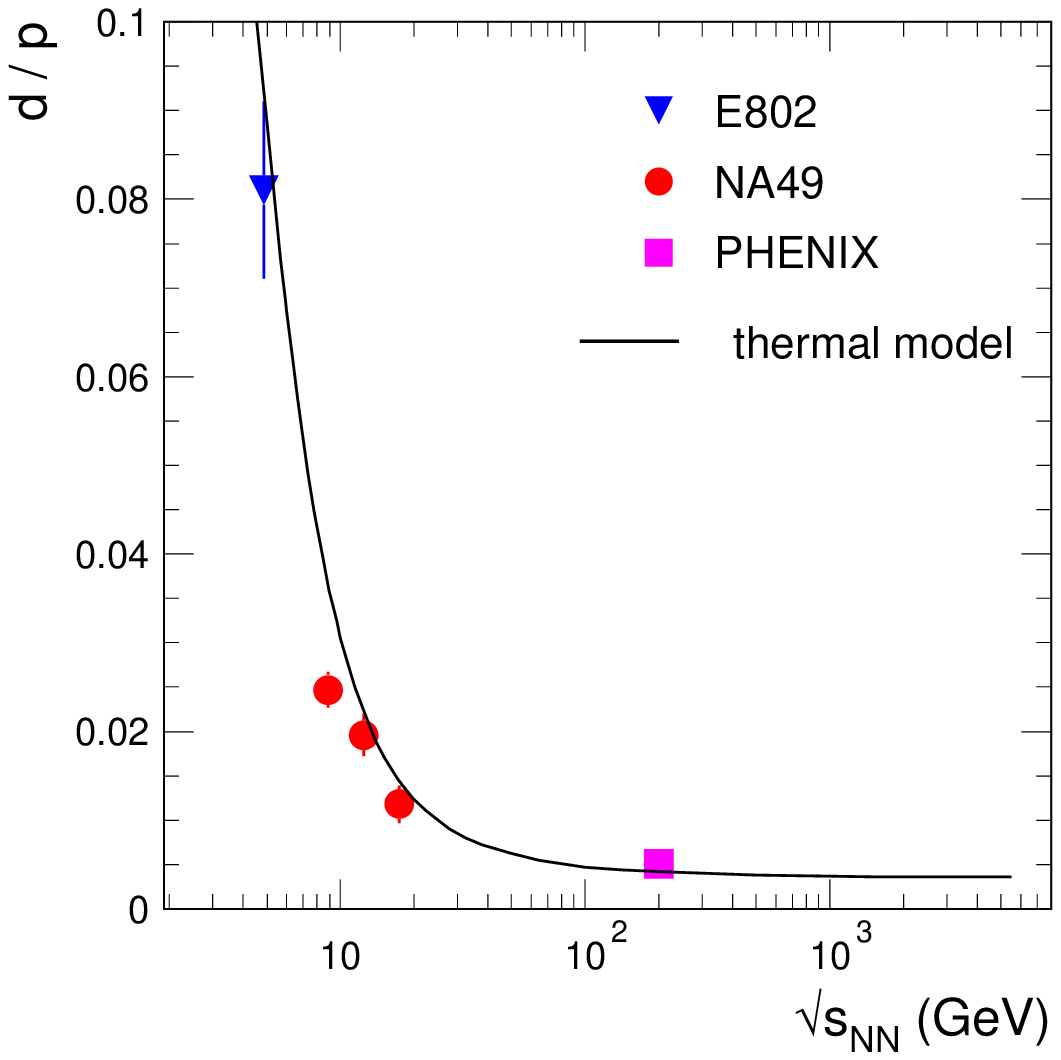}
\end{minipage} &\begin{minipage}{.49\textwidth}
\centering\includegraphics[width=1.17\textwidth]{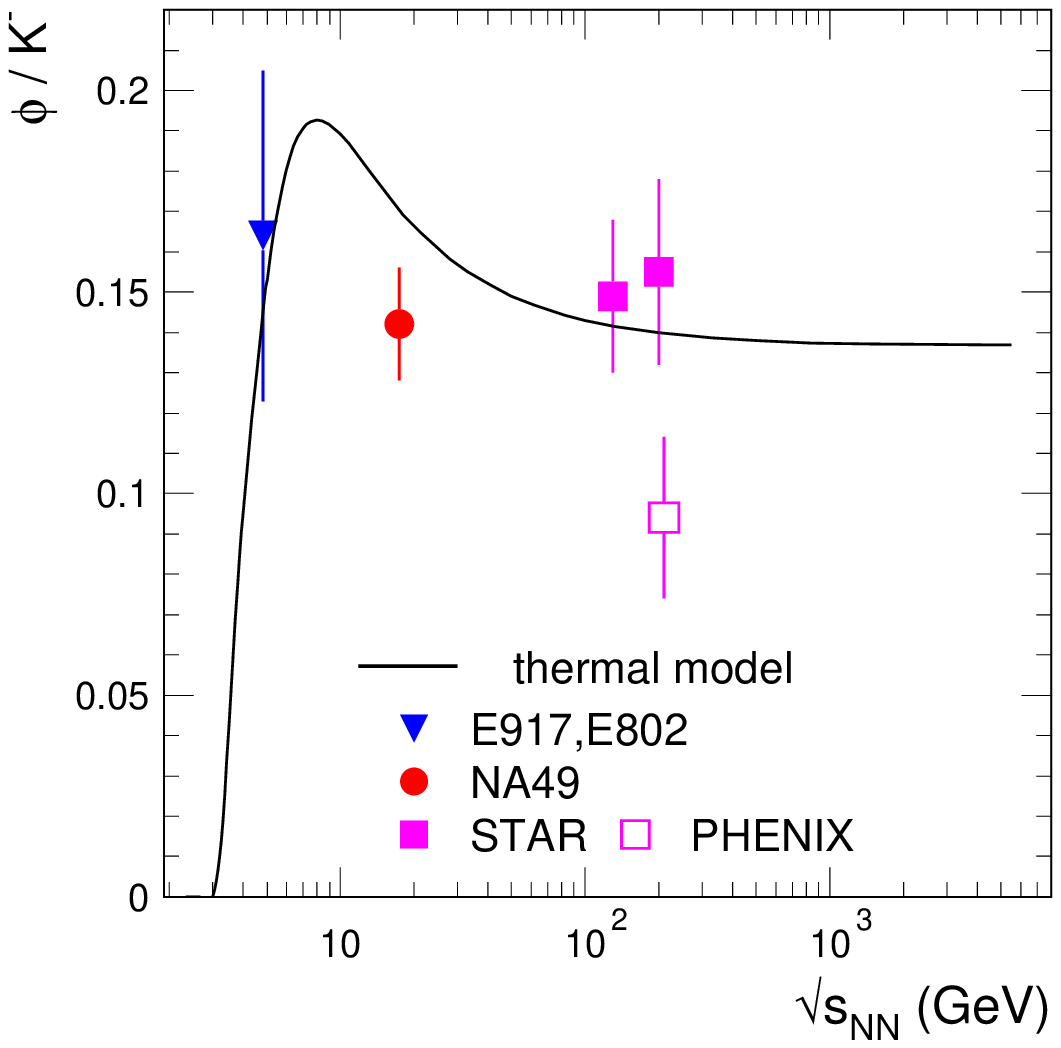}
\end{minipage} \end{tabular}
\caption{The energy dependence of $d/p$ and $\phi/K^-$ ratios.}
\label{fig15}
\end{figure}

In Fig.~\ref{fig15} we show the excitation function of the ratios
$d/p$ and $\phi/K^-$.
The ratio $d/p$ shows a monotonic decrease as a function of energy,
well explained by the model.
The ratio $\phi/K^-$ shows a peak in the calculations, centered in this case 
at $\sqrt{s_{NN}}\simeq$8 GeV. 
The model reproduces well the AGS measurement \cite{e917phi}, while 
at SPS there is a disagreement.
At RHIC the experimental situation is unclear; the model is in agreement
with the STAR data.

Given the accuracy of the description of hadron ratios presented 
in the previous section, it is not surprising that, based on a  
parametrized energy dependence of $T$ and $\mu_b$, the model does reproduce 
the experimental values well over a broad energy range.
However, systematic discrepancies between the model calculations and data 
are evident for the SPS data, in particular concerning the ratios involving 
strange to non-strange hadrons.

\begin{figure}[hbt]
\centering\includegraphics[width=.8\textwidth,height=.73\textwidth]{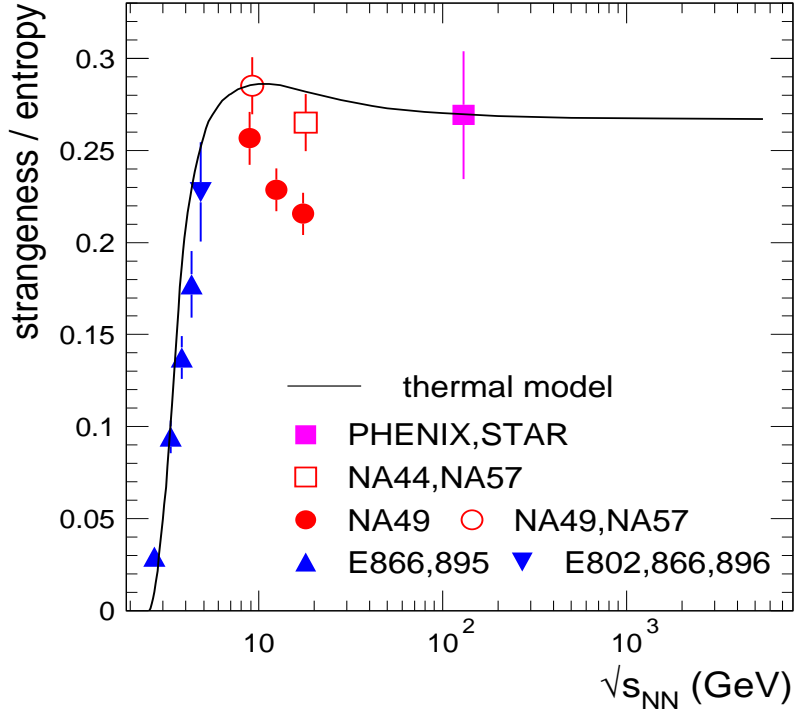}
\caption{Energy dependence of strangeness to entropy ratio (see text).} 
\label{fig16}
\end{figure}

A more global way to represent the ratio of strange to non-strange hadrons
is the strangeness ($\sigma$) to entropy ($S$) ratio.
Its excitation function is presented in Fig.~\ref{fig16}. 
We adopt an experiment-oriented construction of the two quantities, 
which we consistently employ for the model as well, calculated from
the yields at mid-rapidity as:
\be
\sigma=2\times(K^++K^-)+1.54\times(\Lambda+\bar{\Lambda}) , \quad
S=1.5\times(\pi^++\pi^-)+2\times\bar{p}
\ee

The strangeness has in principle to be complemented with the 
yields of $\phi$, $\Xi$ and $\Omega$ (and $\bar{\Xi}$ and $\bar{\Omega}$), 
but, since the measurements for these yields are scarce (as seen in 
Fig.~\ref{fig14} and \ref{fig15}), we have chosen to leave them out for 
the strangeness count.
The factor 2 multiplying the kaon yields takes into account $K^0$, while
the factor 1.54 for $\Lambda$ hyperons accounts for the contribution of 
$\Sigma^\pm$ and was deduced from the model calculations.\footnote{The yield 
of $\Sigma^0$, which decays with 100\% branching ratio into $\Lambda\gamma$,
is always included in the $\Lambda$ yield.}
The factor 1.5 for the pion yields accounts for the $\pi^0$ yield, while 
in case of $\bar{p}$ yields the factor 2 is used to account for the produced
protons.
As expected from the individual particle ratios studied above, the ratio
strangeness/entropy is well reproduced by the model, with the exception
of the data at SPS, where the NA49 data exhibit a sharper peak than predicted
by the model. This feature is not supported by the NA57 data, which are in
agreement with the model.

\begin{table}[hbt]
\caption{Thermal model prediction for hadron ratios at LHC ($T$=161 MeV, 
$\mu_b$=0.84 MeV). No contribution from weak decays is included.}
\label{tab3}
\begin{tabular}{cccccc}
$\pi^-/\pi^+$ & $K^-/K^+$ & $\bar{p}/p$ & $\bar{\Lambda}/\Lambda$ & 
$\bar{\Xi}/\Xi$ & $\bar{\Omega}/\Omega$ \\ \hline
1.00 & 0.99 & 0.95 & 1.00 & 1.00 & 1.00 \\ \hline\hline
$p/\pi^+$ & $K^+/\pi^+$ & $K^-/\pi^-$ & $\Lambda/\pi^-$ & 
$\Xi^-/\pi^-$ & $\Omega^-/\pi^-$ \\ \hline
0.074 & 0.180 & 0.179 & 0.039 & 0.0058 & 0.00106 \\ \hline\hline
$\phi/K^-$ & $K^{*0}/K^0_S$ & $\Delta^{++}/p$ & $\Sigma(1385)^+/\Lambda$ & 
$\Lambda^*/\Lambda$ & $\Xi(1530)^0/\Xi^-$ \\ \hline
0.136 & 0.312 & 0.216 & 0.140 & 0.075 & 0.396
\end{tabular}
\end{table}

The results discussed above allow a rather confident prediction for
hadron ratios in central Pb-Pb collisions at LHC energy \cite{pbm6}, 
presented in Table~\ref{tab3}.
The antiparticle/particle ratios are all very close to unity, with the
exception of $\bar{p}/p$, reflecting the small, but nonzero, $\mu_b$
value obtained from Eq.~\ref{pm}.
Within the extrapolation scenario for the thermal parameters, the model 
predicts that the yields relative to pions are very similar to the values 
measured at the RHIC energy.

\section{The phase diagram of hadronic matter}
\label{spha}

\begin{figure}[htb]
\begin{tabular}{lr} \begin{minipage}{.65\textwidth}
\hspace{-.5cm}
\includegraphics[width=1.18\textwidth]{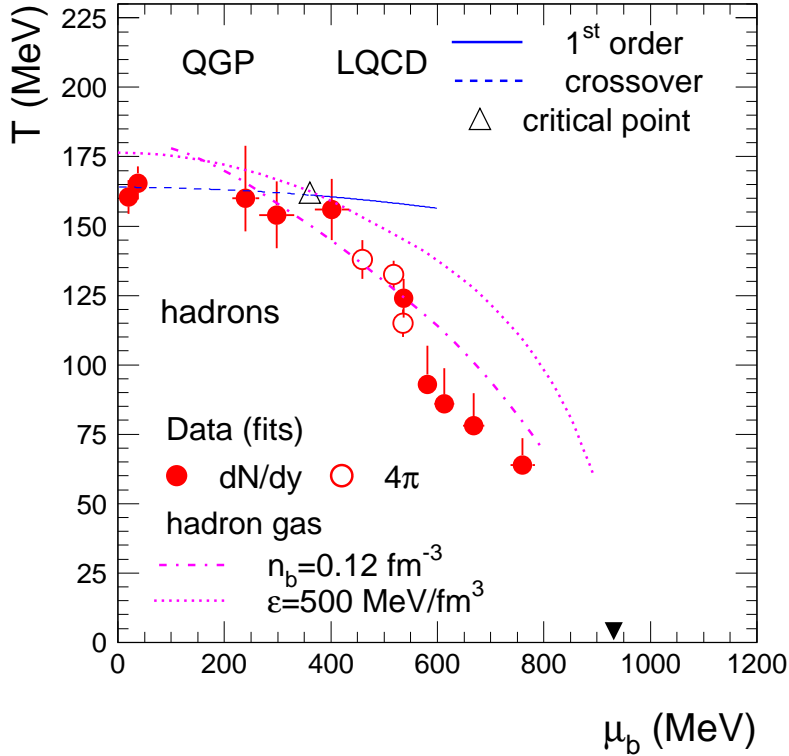}
\end{minipage} &\begin{minipage}{.32\textwidth}
\caption{The phase diagram of hadronic and quark-gluon matter in the 
$T$-$\mu_b$ plane. 
The experimental values for the chemical freeze-out 
are shown together with results of lattice QCD calculations \cite{lqcd}.
The predicted \cite{lqcd} critical point is marked by the open triangle.
Also included are calculations of freeze-out curves for a hadron gas 
at constant energy density ($\varepsilon$=500 MeV/fm$^3$) and at constant 
total baryon density ($n_b$=0.12~fm$^{-3}$).
The full triangle indicates the location of ground state nuclear matter 
(atomic nuclei).} 
\label{fig17}
\end{minipage} \end{tabular}
\end{figure}

The values of $T$ and $\mu_b$ obtained from our fits of the experimental
data are shown in a phase diagram of hadronic and quark-gluon matter in 
Fig.~\ref{fig17}. Full points are from fits of ratios of yields at 
midrapidity, open points are from fits of 4$\pi$ ratios.
An important observation about the phase diagram is that, for the 40 AGeV 
SPS energy and above, the thermal parameters agree with the phase 
boundary calculations from lattice QCD (LQCD) \cite{lqcd,lqcd1,lqcd2}, 
implying that hadron yields are frozen at the phase boundary.
The LQCD calculation \cite{lqcd} shown in Fig.~\ref{fig17} is for two
light quarks ($u, d$) with realistic (close to physical) masses and a heavy 
strange quark.
The critical temperature at $\mu_b$=0 from LQCD calculations is about 
165 MeV \cite{fk} (and refs. therein), with a scale uncertainty of the order 
of 10 MeV \cite{fk,lqcd2} and with comparable systematic errors \cite{fk}.

Also included in Fig.~\ref{fig17} are calculations of freeze-out curves for 
a hadron gas at constant energy density ($\varepsilon$=500 MeV/fm$^3$) and 
at constant total baryon density ($n_b$=0.12~fm$^{-3}$) \cite{pbm5}.
The LQCD phase boundary calculated in \cite{lqcd} and shown in 
Fig.~\ref{fig17} apparently does not follow the expectation \cite{lqcd1} 
(see also \cite{taw1}) of a constant energy density dependence.
However, depending on the method, the errors for the critical line
from LQCD calculations can be large for finite $\mu_b$ \cite{lqcd1}.

The freeze-out points which are departing from the LQCD phase 
boundary are approximately described by the curve of a hadron gas at 
constant baryon density.
An earlier proposed freeze-out criterion corresponds to an average energy 
per average number of hadrons of approximately 1~GeV \cite{1gev}. 
It has also been argued recently that the freeze-out points can be 
described by a constant entropy density divided by $T^3$ \cite{taw}.
This is a measure of the degrees of freedom; in a hadron gas with 
realistic masses the effective number of degrees of freedom depends 
explicitly on temperature and is not constant. 
As Cleymans et al. \cite{cle2} have recently pointed out, for realistic 
excluded volume corrections, this criterion grossly deviates from the 
freeze-out points.
A comparison of various freeze-out criteria was recently done in 
ref. \cite{cle2b}, which showed that all are identical except for
large and small $\mu_b$ values. All are smooth curves and consequently
not consistent with the new results presented here, which exhibit a rather 
steep trend at intermediate $\mu_b$ values.
However, the errors need to be improved before one can confidently
rule out any (smooth) universal freeze-out criterion.
An exciting possibility is that the rather abrupt turn-over in the
freeze-out points near $\mu_b$=400~MeV is caused by the approach to
the QCD phase boundary.

The underlying assumption of the thermal model used to extract the 
($T$,$\mu_b$) values is equilibrium at chemical freeze-out.
A natural question then is how the equilibrium is achieved? 
The answer obviously cannot come from within the framework of the thermal 
model.
Considerations about collisional rates and timescales of the hadronic 
fireball expansion imply that at SPS and RHIC the equilibrium cannot be 
established in the hadronic medium below the critical temperature $T_c$ 
\cite{stock,pbmx}.
In a recent paper \cite{pbmx} many body collisions near $T_c$ were
investigated as a possible mechanism for the equilibration within the
hadronic stage. There it is argued that because of the rapid density change 
near a phase transition such multi-particle collisions provide a natural 
explanation for the observation of chemical equilibration at RHIC energies 
and lead to $T=T_c$ to within an accuracy of a few MeV. 
While this argument is expected to be valid also for SPS energies, 
the situation at lower energies needs further consideration.

The critical temperature determined from RHIC and SPS data assuming 
$T\approx T_c$ coincides well with lattice estimates \cite{lqcd,lqcd2,fk}.
These arguments buttress the conclusion that the hadrochemical freeze-out 
parameters probe experimentally in a unique manner the critical line of 
the QCD phase transition between hadrons and QGP \cite{js}.
Despite the rather large systematic errors of the extracted temperature
for the SPS energies, our results imply that the phase boundary is reached 
for beam energies around 40 AGeV.
The experimental data at 20 AGeV \cite{fri} will constrain further this
energy range.

Based on the LQCD results of Fodor and Katz \cite{lqcd} shown in 
Fig.~\ref{fig17}, the experimental freeze-out points at SPS are located
in the vicinity of the critical point.
It was pointed out recently \cite{op} that the existence of a critical 
point for $\mu_b<$500 MeV requires a fine tuning of the (light) quarks
masses within 5\%. However, it is important to recognize that serious 
open problems of LQCD \cite{op} need to be solved before one could address
quantitatively such a delicate possibility.
Nevertheless, it is interesting to speculate whether the deviations from
the thermal model (including rather poor-quality fits) which we have 
encountered for the SPS energies are a hint for the critical point.
It is expected that, in the (broad) vicinity of the critical (end)point 
the thermal model would not work \cite{nona}.
Thermal fits including fluctuations have been already performed for the
top SPS energy \cite{zsch}.
Unfortunately, the present experimental situation, namely the level of
disagreement between data, does not allow any firm conclusion on the
interesting issue of the critical point.

\section{Summary}

We have analyzed the experimental hadron yields ratios over a broad
energy range ($\sqrt{s_{NN}}$=2.7-200 GeV) in comparison with thermal model 
calculations. The fits of the experimental data with the model
provide the temperature and baryo-chemical potential, at hadrochemical 
freeze-out.
The quality of the fits is very good for most cases, providing support
for the validity of the approach.
From our analysis we have established parametrizations of $T$ and $\mu_b$ 
as a function of energy.
This allows to assess and understand trends visible in the experimental
data. In particular, we have discussed for the yields of strange and 
multi-strange hadron yields relative to pions the resulting non-monotonic
energy dependence, which describes the main features observed in the data,
but not the details.
Furthermore, we have provided quantitative predictions for hadron ratios
at $\sqrt{s_{NN}}$=62.4 GeV and at LHC energy, where first data are expected
soon.
The analysis presented here has also led to an updated version of the 
chemical freeze-out curve.
The results buttress previous observations that, from $\sqrt{s_{NN}}\approx$
10 GeV on, the chemical freeze-out points coincide with the phase
boundary between hadrons and the QGP as predicted by solving QCD on the 
lattice.
We have provided chemical freeze-out points, for the first time, at the 
4 low AGS energies. In this energy range the freeze-out curve exhibits a
peculiar and intriguing structure which needs further explanation.

\section*{Appendix}

In this Appendix we discuss technical aspects related to the thermal 
model analysis. First, we concentrate on the technical aspects of fitting 
yields rather than ratios, and the demonstration that both methods produce 
equivalent results. Secondly, we briefly address the issue of the role 
of $\gamma_S$ in the thermal model fits. Finally, we comment on the issue 
of weak feeding.
The absolute yields (and consequently the values for the volume) quoted 
here are for $N_{part}$=350.

\subsection*{A. Analysis of particle yields}

\begin{figure}[hbt]
\centering\includegraphics[width=.8\textwidth]{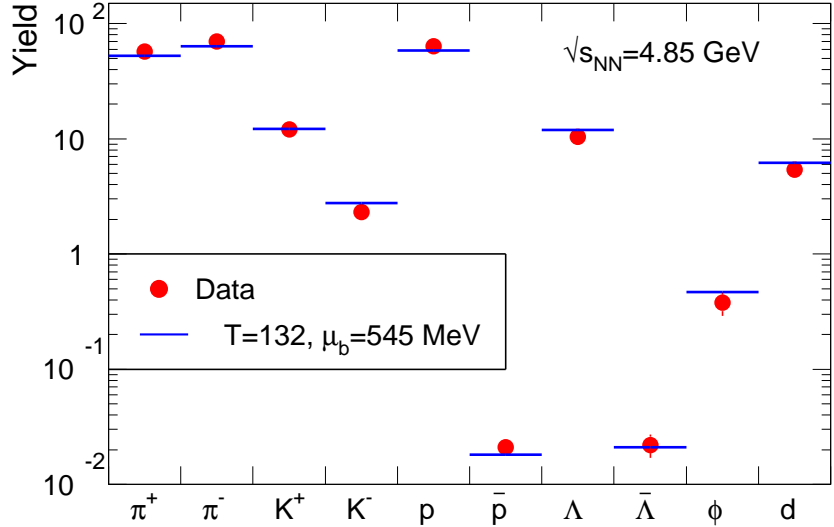}
\caption{Hadron yields at mid-rapidity and the best fit for the 
top AGS energy.} 
\label{figx}
\end{figure}

The fit of mid-rapidity yields for the top ASG energy is shown in 
Fig.~\ref{figx}.
The resulting parameters are: $T$=132 MeV, $\mu_b$=546 MeV, 
$V$=900 fm$^3$. With $\chi^2/N_{df}$=10.3/7, the fit is good.
The temperature is somewhat higher than the temperature obtained
when fitting ratios.
Excluding from the fit the yields of $\bar{p}$, $\bar{\Lambda}$, $\phi$, 
and $d$ one obtains: $T$=110 MeV, $\mu_b$=550 MeV, $V$=2620 fm$^3$, 
$\chi^2/N_{df}$=1.2/3.
The resulting change in $T$ is 20\%, larger than the difference derived 
from the fit of ratios (15\%). This lower temperature is "compensated" by 
an unphysically large volume, leading to a good fit in this case too.
The main contribution to this effect arises from the absence of $\bar{p}$
and $\bar{\Lambda}$ (when fitting without $\bar{p}$ and $\bar{\Lambda}$: 
$T$=114 MeV, $\mu_b$=551 MeV, $V$=2100 fm$^3$, $\chi^2/N_{df}$=2.8/5).

Our fits of (mid-rapidity) yields at lower AGS energies result in identical
values for $T$ and $\mu_b$ as when using ratios. 
The extracted volume is much larger than for the top AGS energy.
This originates in the bias of the temperature due to the limited set of 
particle yields available, as discussed above and as also observed when 
fitting ratios.
Based on the above results at top AGS energy, we conclude that the 
extracted volume at the lower energies is overestimated by a factor 
of 2.9.

\begin{table}[hbt]
\caption{Summary of the results of the fits using mid-rapidity yields 
at top SPS energy.}
\label{tab1b}
\begin{tabular}{l|cccc|cccc}
data set & $T$ (MeV) & $\mu_b$ (MeV) & $V$ (fm$^3$) & $\chi^2/N_{df}$ &
$T$ (MeV) & $\mu_b$ (MeV) & $V$  & $\delta^2$ \\ \hline
NA44+NA57 &174 & 240 & 750 & 24.2/10 & 174 & 240 & 750 & 0.28 \\
NA49      &158 & 231 & 1250 & 56.2/10 & 160 & 225 & 1150 & 0.44 \\
combined  &164 & 234 & 1000 & 105/23 & 168 & 234 & 900 &1.10  \\
\end{tabular}
\end{table}

\begin{figure}[hbt]
\centering\includegraphics[width=.8\textwidth]{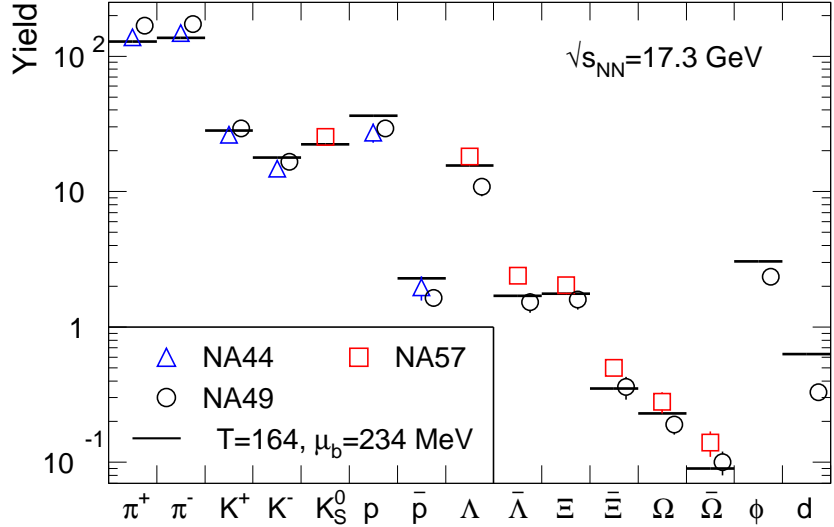}
\caption{Mid-rapidity yields and best fit (combined data) at the SPS 
beam energy of 158 AGeV ($d$ was not included in the fit).}
\label{fig6x}
\end{figure}

\begin{figure}[hbt]
\centering\includegraphics[width=.8\textwidth]{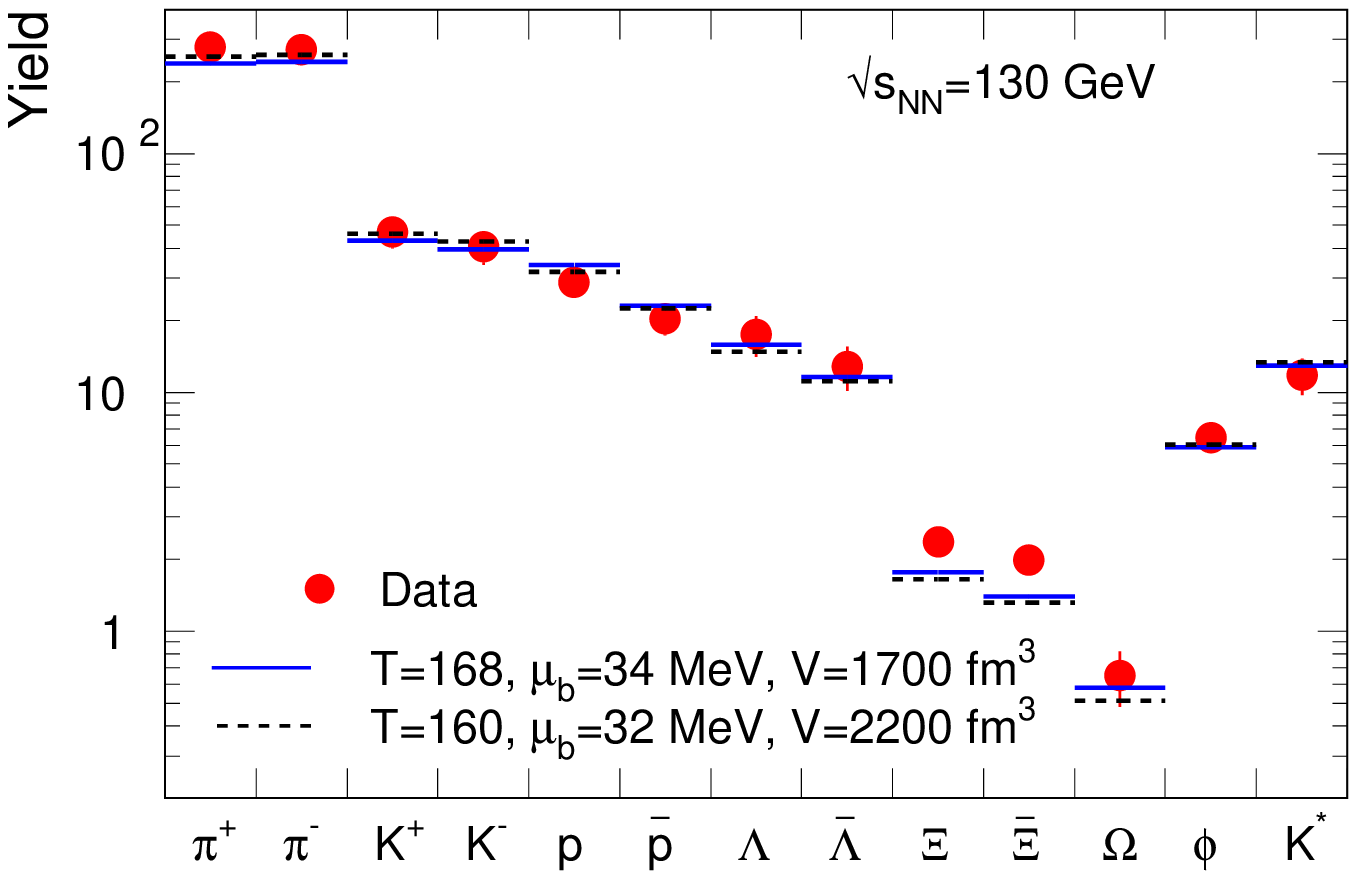}
\caption{Hadron yields with best fit at $\sqrt{s_{NN}}$=130 GeV.
The dashed lines are for the best fit excluding the $\Xi$ hyperons.
The $\Omega$ yield includes both $\Omega$ and $\bar{\Omega}$.} 
\label{fig8x}
\end{figure}

The comparison of mid-rapidity yields and the best fit for the top SPS energy
is shown in Fig.~\ref{fig6x}. The temperature is slightly higher than in case 
of ratios, but this is within 1-$\sigma$ errors. 
The discrepancy between different data sets noted in case of ratios
is evident also for the yields. As for the ratios, this leads to 
contradictory results when fitting the data separately, see Table~\ref{tab1b}.
The quality of the fits is for all SPS energies significantly poorer than 
in case of ratios. 

At the RHIC energy of $\sqrt{s_{NN}}$=130 GeV, Fig.~\ref{fig8x}, the resulting 
temperature from the fits of absolute yields is slightly higher
than in case of ratios. This, as well as a poorer $\chi^2/N_{df}$=15.3/10,
is mainly due to an apparent experimental upward bias of the yield of 
the $\Xi$ (and, to a lesser extent, $\Lambda$) hyperons. 
Without $\Xi$'s in the fit: $T$=160 MeV, $\mu_b$=32 MeV, 
$V$=2200 fm$^3$, $\chi^2/N_{df}$=4.3/8.

The main results for the best fits of absolute hadron yields are 
summarized in Table~ \ref{tab2x}.
The upper part in Table~\ref{tab2x} is for the fits using mid-rapidity data, 
the lower part is for data integrated over 4$\pi$.
The fits of yields give similar temperatures as for ratios, but larger 
values for baryo-chemical potentials and larger $\chi^2$ values.
One notices that, in general, $\chi^2/N_{df}$ is close to or below unity, 
being significantly above unity only for the SPS energies, as for the case 
using ratios of yields.
For all energies, the  $\chi^2$ values are larger than in case of the
fits of ratios.
This is expected, as part of the discrepancy between data and model is 
cancelled in the ratios, for instance in case of antiparticle/particle 
ratios.
These results demonstrate that the use of ratios is less prone to biases.
This is one of the reasons why we prefer to fit particle ratios rather
than yields.

\begin{table}[hbt]
\caption{Summary of the results of the thermal fits of yields. 
The upper part shows the results obtained using mid-rapidity data,
the lower part is for data integrated over 4$\pi$.}
\label{tab2x}
\begin{tabular}{l|cccc|cccc|ccc}
$\sqrt{s_{NN}}$ (GeV) & $T$  & $\mu_b$  &$V$ & 
$\chi^2/N_{df}$ & $T$ & $\mu_b$   &$V$  & $\delta^2$ & 
$\Delta T$ &$\Delta \mu_b$ &$\Delta V$\\ 
\hline
2.70 & 64  & 760 &5880 & 1.19/2 & 64  & 760 & 5900 & 0.013 &+13,-2 &23 &140\\
3.32 & 78  & 669 &6280 & 0.54/3 & 78  & 670 & 6200 & 0.003 &+16,-3 &20 &230\\
3.84 & 88  & 610 &5290 & 1.62/3 & 88  & 610 & 5200 & 0.020 &+18,-3 &19 &220\\ 
4.30 & 94  & 588 &4940 & 1.5/3  & 96  & 575 & 4400 & 0.026 &+19,-3 &17 &540\\
4.85 & 132 & 546 & 900 & 10.3/7 & 132 & 540 & 900  & 0.13  & 4     &11 &170\\
8.76 & 154 & 382 & 850 & 51/13  & 160 & 418 & 700  & 0.69  & 6     &36 &250\\
12.3 & 150 & 277 &1300 & 47/5   & 152 & 265 & 1300 & 0.18  & 5     &12 &300\\
17.3 & 164 & 234 &1000 & 105/23 & 168 & 234 & 900  & 1.10  & 6     &11 &250\\
130  & 168 & 34  &1700 & 15.3/10& 170 & 34  & 1600 & 0.24  & 4     &6  &220\\
\hline
6.27  &128 & 489 & 3850 &15.9/4 &128 & 453 & 4200 & 0.10 &3 &36 &450\\
7.62  &138 & 471 & 3450 & 27/4 & 142 & 474 & 2900 & 0.22 &4 &13 &550\\
8.76  &136 & 426 & 4600 & 56/6 & 140 & 366 & 4100 & 0.60 &5 &60 &500\\
12.3  &146 & 373 & 5400 & 61/4 & 148 & 298 & 4800 & 0.30 &6 &75 &750\\
17.3  &148 & 228 & 6100 & 64/9 & 152 & 210 & 5700 & 0.51 &5 &22 &650\\
\end{tabular}
\end{table}

\subsection*{B. Inclusion of $\gamma_S$ into the fits}

In Table~\ref{tab2y}, we present results obtained with mid-rapidity 
and 4$\pi$-integrated yields with a thermal model including a strangeness 
suppression factor $\gamma_S$. 
The results for the mid-rapidity data demonstrate that the key quantities, 
namely $T$ and $\mu_b$, are very little influenced by whether or not 
$\gamma_S$ is included as a fit parameter. Also, the quality of the fits 
is not significantly improved by including $\gamma_S$. 

\begin{table}[hbt]
\caption{Results of the fits of yields using $\gamma_S$.
The upper part is for mid-rapidity data, the lower part is for 4$\pi$ data.}
\label{tab2y}
\begin{tabular}{l|ccccc|ccccc}
$\sqrt{s_{NN}}$ (GeV) & $T$  & $\mu_b$  &$V$ & $\gamma_S$ &
$\chi^2/N_{df}$ & $T$ & $\mu_b$   &$V$  & $\gamma_S$ & $\delta^2$ \\ 
\hline
8.76  &156 & 386 & 800  & 0.94 & 48/12 & 160 & 414 & 700 & 0.95 & 0.59 \\
12.3 &144 & 265 & 1900 & 0.84 & 32/4  & 152 & 265 & 1300 & 0.93 & 0.18 \\
17.3 &162 & 231 & 1100 & 0.96 & 102/23 &168 & 231 & 900  & 1.0  & 1.12 \\
\hline
6.27  &132 & 480 & 3600 & 0.83 & 9.0/3 & 128 & 450 & 4300 & 0.94 & 0.10 \\
7.62  &140 & 448 & 3850 & 0.81 & 10.6/3 & 142 & 444 & 3550 & 0.75 & 0.08 \\
8.76  &152 & 418 & 3000 & 0.74 & 7.0/5 & 152 & 418 & 3200 & 0.69 & 0.04 \\
12.3 &162 & 340 & 3400 & 0.71 & 5.9/3 & 158 & 322 & 4000 & 0.70 & 0.023 \\
17.3 &162 & 267 & 4600 & 0.76 & 17.6/8 & 162 & 258 & 4300 & 0.84 & 0.85 \\
\end{tabular}
\end{table}

\begin{figure}[hbt]
\hspace{-.7cm}
\begin{tabular}{lcr} \begin{minipage}{.49\textwidth}
\centering\includegraphics[width=1.18\textwidth]{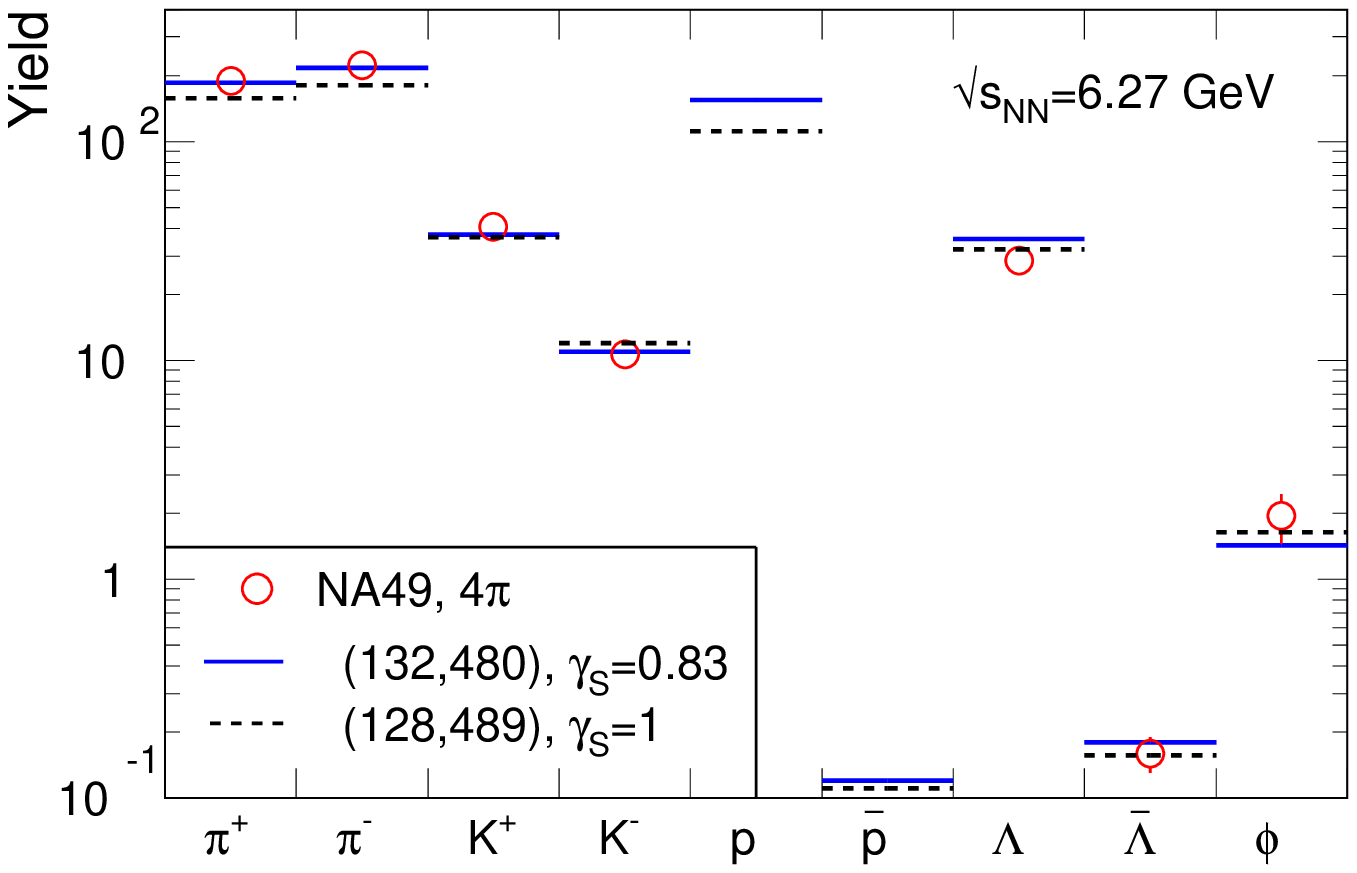}
\end{minipage} &\begin{minipage}{.49\textwidth}
\centering\includegraphics[width=1.18\textwidth]{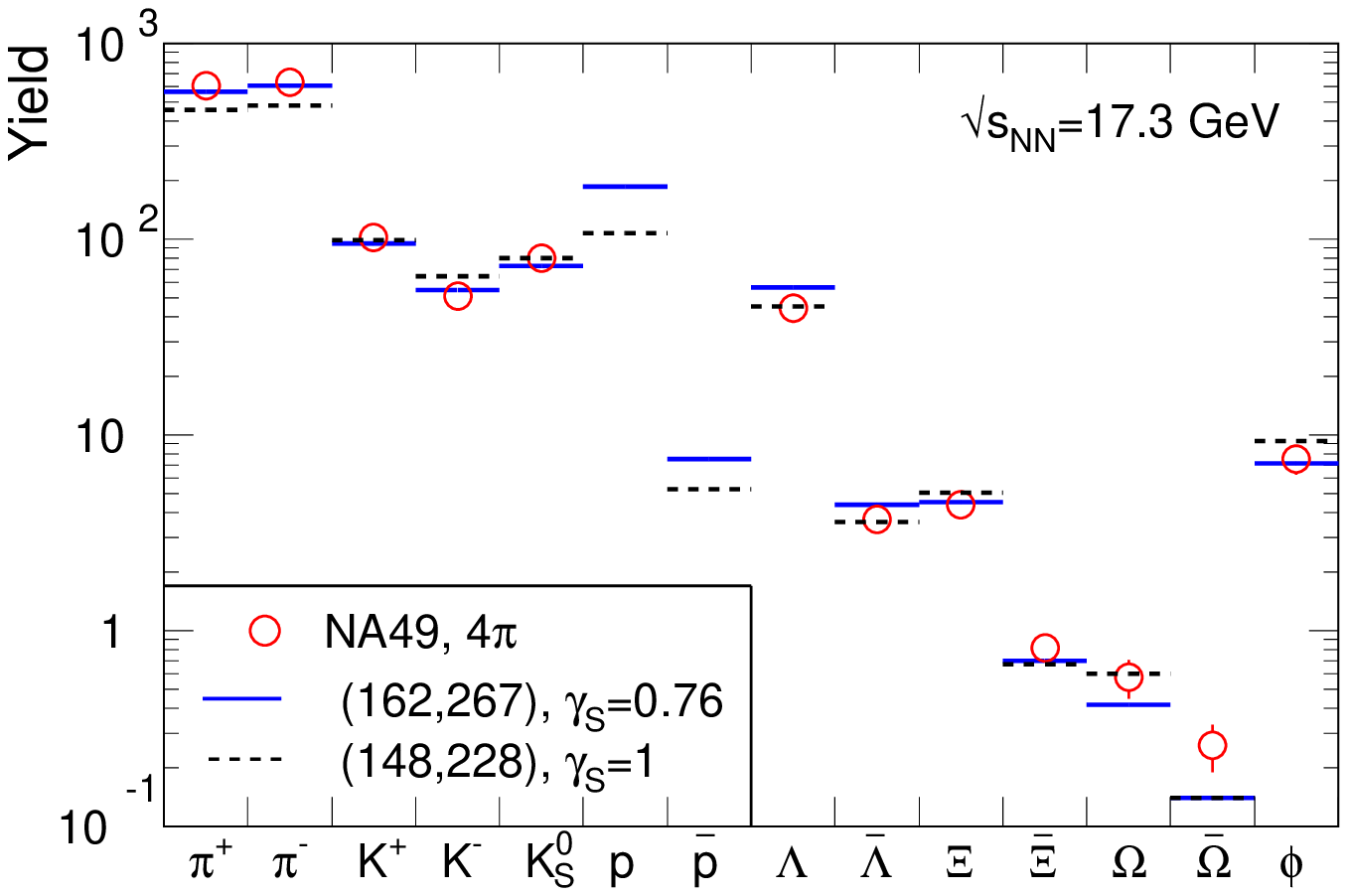}
\end{minipage} \end{tabular} 
\caption{Hadron yields integrated over 4$\pi$ with best fits with (continuous
lines) and without (dashed lines) $\gamma_S$ at the SPS energies of 20 and 
158 AGeV.}
\label{fig6y}
\end{figure}

The fits of 4$\pi$ yields including $\gamma_S$ leads to a substantial 
improvement of the fit quality only for the higher SPS energies, but still 
the fit seldom attains a good quality.
For these energies, the resulting $T$ (and also $\mu_b$) values are larger 
than in the case without $\gamma_S$ (Table~\ref{tab2x}).
This effect is negligible for the lowest SPS energy and increases gradually
as a function of energy, see Table~\ref{tab2y}. 
This is illustrated in Fig.~\ref{fig6y} for the lowest and highest SPS 
beam energies.
Contrary to expectations, the volume and $\gamma_S$ show a non-monotonic 
behavior as a function of energy (also apparent in the analysis of 
Becattini et al. \cite{bec1}).
In view of this, we regard $V$ and $\gamma_S$ as compensatory parameters 
for an inadequate fit of 4$\pi$ data.
As a consequence, the fact that the temperatures from the fits of 4$\pi$ 
data using $\gamma_S$ are very similar to those extracted from fits of
mid-rapidity data should be viewed as a coincidence.

\subsection*{C. Importance of feed-down from weakly decaying states}

\begin{figure}[htb]
\begin{tabular}{lr}\begin{minipage}{.55\textwidth}
\centering\includegraphics[width=1.15\textwidth]{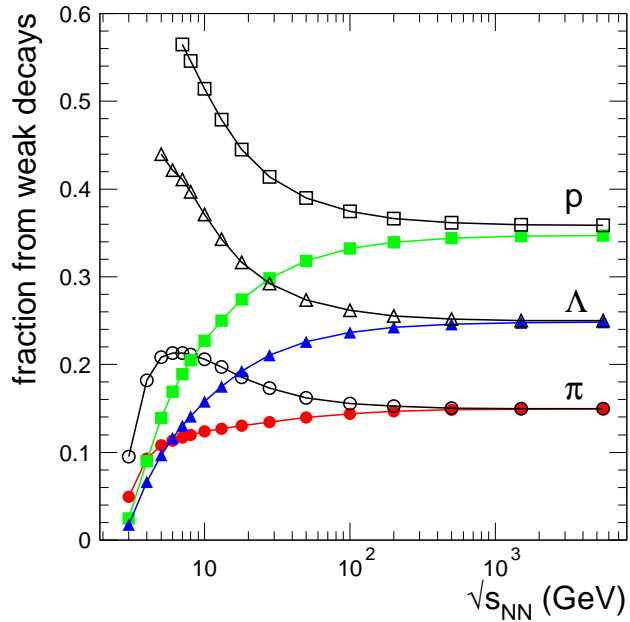}
\end{minipage} &\begin{minipage}{.42\textwidth}
\caption{The energy dependence of the fraction of total hadron yields 
originating from weak decays.
The full symbols are for particles ($\pi^+$, $p$, $\Lambda$), the open
ones for antiparticles ($\pi^-$, $\bar{p}$, $\bar{\Lambda}$).}
\label{fig11}
\end{minipage} \end{tabular}
\end{figure}

In Fig.~\ref{fig11} we show, as a function of energy, the contribution 
of weak decays to the yields of pions, protons and $\Lambda$ hyperons 
calculated with the parametrizations for $T$ and $\mu_b$ (Eqs. \ref{pt}
and \ref{pm} in Section~\ref{edep}).
Shown is the fraction of the total yield of these particles originating 
from weak decays.
This fraction reaches the asymptotic values of 15\%, 25\% and 35\% for 
$\pi$, $\Lambda$ and $p$, respectively, being significantly larger in case 
of antiparticles at AGS and SPS energies.
Since this contribution is sizeable, a consistent comparison between
data and model is required.

\end{document}